\documentclass[hidelinks]{article}

\usepackage{arxiv}

\usepackage[utf8]{inputenc} 
\usepackage[T1]{fontenc}    
\usepackage{amsmath}
\usepackage{amssymb}
\usepackage{graphicx}
\usepackage{lipsum}
\usepackage{siunitx}
\usepackage{physics}
\usepackage{xspace}
\usepackage{multirow}
\usepackage{hyperref}
\usepackage{lineno}
\usepackage[authoryear]{natbib}
\usepackage[nameinlink, capitalize]{cleveref}

\newcommand{\Fr}{\mathrm{Fr}}
\newcommand{\Ro}{\mathrm{Ro}}
\newcommand{\UBmag}{U}
\newcommand{\uBvec}{\vec{u}^B}
\newcommand{\uB}{u^B}
\newcommand{\vB}{v^B}
\newcommand{\wB}{w^B}
\newcommand{\wdir}{\psi}
\newcommand{\wdB}{\wdir}
\newcommand{\rews}{U_\mathrm{REWS}}

\newcommand{\Uhub}{U_h}
\newcommand{\zhub}{z_h}
\newcommand{\veer}{\alpha_\text{v}}
\newcommand{\shear}{\alpha_\text{s}}
\newcommand{\powerlaw}{\alpha}
\newcommand{\iea}{IEA~\SI{15}{\mega\watt}\xspace}

\newcommand{\wxB}{\omega_x^B}
\newcommand{\ti}{\mathrm{TI}}
\newcommand{\ud}{\langle u_d \rangle}
\newcommand{\an}{a_n}

\author{
 Kirby S. Heck$^1$, Storm A. Mata$^1$, and Michael F. Howland$^{1*}$ \\
 \\
  $^{1}$Civil and Environmental Engineering, Massachusetts Institute of Technology, Cambridge, MA 02139, USA \\
  $^*$Corresponding author: \texttt{mhowland@mit.edu} \\
}

\title{Influence of wind shear and veer on power, thrust, and induction of an actuator disk}

\begin{document}

\maketitle


\begin{abstract}
Wind shear and wind veer (gradients of wind speed and direction, respectively) are ubiquitous in the atmospheric boundary layer (ABL) due to the combined effects of pressure gradient forcing, wall friction, and planetary rotation. 
Therefore, wind turbines, which extract energy from the ABL, routinely operate in sheared and veered conditions. 
Previous field campaigns have observed statistically significant variations in power production efficiency (quantified by a power coefficient), upwards of $\pm 15\%$, due to the effects of shear and veer in the ABL. 
However, it is not yet clear how non-uniform inflow conditions alter rotor aerodynamics and drive the mechanisms of these power and efficiency variations. 
In this study, we perform concurrent-precursor large-eddy simulations of an actuator disk-modeled wind turbine across neutral and stable ABL conditions to demonstrate that shear and veer can reduce wind power efficiency by more than $20\%$. 
To further support these ABL simulations, we perform simplified inflow simulations where we can independently control the inflow properties. 
Using these controlled simulations, we demonstrate that the effects of shear and veer can be decomposed into two components: (1) geometric effects, due to changes in mass and energy flux upwind of the rotor, which are straightforward to model through a rotor-equivalent wind speed, and (2) inductive effects, where the aerodynamics of the rotor and induced velocities are altered. 
We identify that inductive effects of wind shear modulate the power coefficient through changes to the local induction, which varies over the rotor disk, while inductive effects of wind veer reduce the power coefficient by generating an adverse pressure gradient at the rotor scale. 
The geometric and inductive effects of shear and veer can be approximately linearly superimposed, summing to efficiency losses exceeding $20\%$ for a $\SI{240}{\meter}$-diameter turbine operating in stably stratified conditions, relative to the same turbine operating in uniform inflow, with increasing losses as shear and veer magnitudes increase. 
Inductive effects account for a significant fraction of the total observed efficiency loss, and we show that the induction of a turbine is affected by the thrust coefficient, wall proximity, shear, and veer through processes that are neglected in existing momentum theory models. 
By revealing the distinct physical mechanisms through which shear and veer affect rotor performance, this work establishes a framework that can enable improved power prediction and rotor design in realistic atmospheric inflow conditions. 
\end{abstract}

\maketitle

\section{Introduction}
\label{sec:intro}


As the wind energy industry grows to meet the growing global demand for carbon-free energy, turbine manufacturers continue designing and deploying machines with increasing nameplate capacity and rotor diameters \citep{iea_renewables_2025, diaz_review_2020}. 
Contemporary-scale offshore wind turbines can reach over \SI{200}{\meter} in diameter \citep{musial_offshore_2022}, marking a $100\%$ increase in size between 2001--2021 \citep{jung_properties_2023}. 
As wind turbines grow in size, the region of the atmospheric boundary layer (ABL) that they operate in becomes increasingly complex \citep{lundquist_wind_2022}. 
Wind speed, wind direction, turbulence intensity, thermal stratification, and other properties of the ABL vary with height due to the presence of the ground. 
Consequently, the heterogeneity of the incident winds increases as wind turbine rotors increase in size and turbine blades sample a wider extent of heights in the ABL.

Despite the heterogeneity of ABL winds, wind turbines are traditionally designed and analyzed assuming simplified inflow conditions. 
For example, momentum theory is used ubiquitously to analyze the large-scale flow response and energy extraction by the rotor in industry and research design codes, but momentum theory is traditionally constrained to uniform, neutrally stratified inflow, assuming incident winds are spatially homogeneous and aligned with the rotor face \citep[c.f.,][]{jonkman_fast_2005, burton_wind_2011}. 
By contrast, winds in the ABL are almost always sheared (non-zero gradients of wind speed as a function of height). 
Furthermore, Coriolis effects in the ABL, which arise due to planetary rotation, interact with wind shear to create vertical gradients of wind direction, known as wind veer \citep{kelly_shear_2023}.
In the northern hemisphere, positive wind veer is defined as rightward (clockwise) turning wind direction with increasing height following the Ekman spiral \citep{ekman_influence_1905}. 
Both shear and veer depend on a variety of phenomena in the ABL, including stratification \citep{mahrt_stably_2014}, boundary layer turbulence \citep{turk_dependence_2010}, mesoscale dynamics \citep{munoz-esparza_bridging_2014}, complex terrain \citep{kustas_wind_1986}, and diurnal variability \citep{kumar_largeeddy_2006}, resulting in a wide range and high dimensionality of ABL characteristics that wind turbines operate in. 

Atmospheric inflow properties affect rotor performance, such as power production and fatigue loading (for a recent review, see \citet{kosovic_impact_2026}). 
Many field campaigns have documented statistically significant variations of wind turbine power generation as a function of inflow conditions, even for wind turbines with relatively small rotor diameters ($D \sim \SI{100}{\meter}$). 
However, the exact effects of inflow properties on power production vary between sites and studies. 
Most studies define wind shear based on a power law, 
\begin{equation}
    \label{eq:powerlaw}
    \frac{U}{U_h} = \left( \frac{z}{\zhub} \right)^\powerlaw, 
\end{equation}
where $\UBmag$ is the wind speed magnitude, $\Uhub$ is the wind speed at hub height $z=\zhub$ where $z$ is the wall-normal coordinate, and $\powerlaw$ is the power law exponent. 
While some field measurements have indicated that rotor power decreases with increasing inflow shear, i.e.,\ increasing $\powerlaw$ \citep{albers_influence_2007, vanderwende_modification_2012, vratsinis_impact_2026}, measurements at other field sites have observed negligible or increasing rotor power with increasing shear \citep{wharton_atmospheric_2012, wharton_assessing_2012, sanchezgomez_effect_2020}. 
\citet{albers_influence_2007} noted that the same strength of wind shear had differing effects between differing turbine geometries, noting that lower hub heights displayed higher losses in strongly sheared inflows. 
Thermal stability is often correlated with wind shear, and several studies have noted the importance in parsing ABL regimes based on multiple different variables representative of stably stratified conditions (binned using high shear values, $\powerlaw \gtrsim 0.3$, in conjunction with either low turbulence \citep{dorenkamper_atmospheric_2014} or bulk Richardson numbers greater than a critical value \citep{vanderwende_modification_2012, st.martin_wind_2016}) and unstable (convective) conditions (conversely, binned using low or near-zero shear values, $\powerlaw \lesssim 0.1$, and either high turbulence or negative bulk Richardson numbers). 
\citet{st.martin_wind_2016} found that filtering based on wind shear alone, without the additional stability criterion, did not yield statistically different turbine power production for a \SI{77}{\meter}-diameter turbine in Colorado, USA, but that filtering on shear and bulk Richardson number delineated regimes of overproduction in convective conditions and underproduction in stable conditions. 
Differences in power production conditioned on inflow conditions can exceed $15\%$ \citep[c.f.][]{wharton_atmospheric_2012, dorenkamper_atmospheric_2014}. 
Clearly, wind shear has a complex and significant influence on rotor power that depends on a wide variety of site- and turbine-specific characteristics. 

In addition to wind shear, previous studies have also focused on the effects of wind veer. 
In an analysis of offshore wind farm data for turbines exceeding \SI{7}{\mega\watt} ($\approx \SI{200}{\meter}$ diameter), \citet{vratsinis_impact_2026} found shear and veer to be negatively correlated with leading-row turbine power production. 
\citet{gao_effect_2021} observed power losses of $-6.5\%$ in veering conditions and slight gains of $+1.6\%$ in backing (negative wind veering, or leftward turning winds with increasing height) from a 5-year measurement campaign in Minnesota, USA. 
In contrast, \citet{tumenbayar_effect_2023} found opposite trends of veering and power generation in a similar analysis of turbines on Jeju Island, South Korea, citing site-specific dependencies such as local topography. 
Some studies have found that considering the effects of wind shear and wind veer individually better describes the variability of power production on ABL properties. 
For example, \citet{sanchezgomez_effect_2020} found turbine overperformance up to $+8\%$ with large shear and small veer, while strong veering and small shear led to underperformance of $-13\%$ for a field site in central Iowa, USA. 
Separately, \citet{mata_modeling_2024} found underperformance, relative to uniform inflow conditions, up to $-19\%$ in strongly veered and sheared conditions at a field site in northwest India. 
From these field campaigns, it is clear that further research is required to understand the influence of ABL conditions on wind turbine performance. 

Given the complexity of ABL winds, the rotor-equivalent wind speed $(\rews)$ has been proposed as an alternative to the hub height wind speed, $\Uhub$, for predicting turbine power. 
Here, we define $\rews$ as
\begin{equation}
\label{eq:rews}
    \rews = \frac{1}{\pi R^2} \int_{0}^{2\pi} \int_0^R 
    \left(\uBvec (z) \cdot \hat{n} \right) r \,dr \,d\theta,
\end{equation}
where $R = D/2$ is the rotor radius, $\uBvec(z)$ is the inflow wind velocity, $\hat{n}$ is the rotor normal vector, $r$ is a radial coordinate, and $\theta$ is the azimuthal coordinate \citep{wagner_influence_2009, wagner_accounting_2011}. 
One simple model to capture the influence of wind shear and veer on rotor power production is to use $\rews$ rather than $\Uhub$ to determine the turbine power production. 
This incorporates the geometric effects of non-uniform wind speed and direction on the mass flux through the rotor swept area upwind of the turbine (i.e., not considering the potential effects of shear and veer on turbine induction). 
We note that an alternative to \cref{eq:rews}, which we will refer to as the rotor-equivalent power model, integrates the cube of the local wind speed, then normalizes by $\pi R^2$, and computes the cube-root of the result, \citep[c.f.][]{wagner_influence_2009}. 
The rotor-equivalent models have been used in field-scale assessments to varied success. 
For example, \citet{murphy_how_2020} used a rotor-equivalent model to characterize ABL inflows, finding that $\rews$ is a more useful metric than $\Uhub$ for predicting turbine power. 
However, as shown by \citet{mata_modeling_2024}, the rotor-equivalent models capture only about $30\%$ of the variability in turbine power due to variability of the inflow conditions ($R^2 \approx 0.3$ for the rotor-equivalent models relative to wind turbine power production measurements).

More sophisticated models have also been proposed in the literature, based on numerical and experimental observations. 
Blade-element models, which use airfoil polars to estimate the forces and torque imparted on the flow by the blades using the velocities at the rotor plane, have been extended to arbitrary inflow geometries \citep{howland_influence_2020, tamaro_power_2024, mata_modeling_2024}.
In this way, the inflow velocity can impact rotor performance by changing the angle of attack and wind speed magnitude along the blades, as shown in \citet{howland_influence_2020}. 
However, blade-element models rely on velocities at the rotor plane, which are lower than the velocity upstream due to induction, and existing models of induction neglect the influence of shear, veer, and other ABL effects \citep{burton_wind_2011}.
Aeroelastic models, which account for fluid-structure interactions through structural turbine models, also depend on the same induction closure models \citep[c.f.][]{jonkman_fast_2005}. 
Therefore, higher-fidelity aeroelastic studies are also limited by induction models, which do not account for shear or veer \citep{wagner_simulation_2010, antoniou_wind_2009}. 
\citet{cheung_greens_2024} proposed a Greens function-based induction model that incorporates the effects of wind speed shear and wall effects into the turbine induction, but model predictions deviate from high-fidelity large-eddy simulations (LES) for moderate or high levels of thrust or shear. 
In parallel, a few LES studies have investigated the interaction between ABL inflows and rotor performance. 
For example, \citet{troldborg_actuator_2010} simulated actuator disk-modeled turbines in sheared inflow conditions with and without wall effects, finding a slight increase in power due to the interaction between inflow shear and the turbine wake. 
Later, \citet{parinam_exploring_2024} performed simulations of the \iea reference turbine \citep{gaertner_definition_2020}, finding a power increase in turbulent inflow conditions relative to laminar inflow conditions. 
Recently, \citet{ma_exploring_2026} performed LES of the \iea reference turbine in veered inflow conditions, showing non-monotonic deviations in power from uniform inflow conditions up to $+1.5\%$ gain and $-5\%$ power loss. 
They hypothesize that the observed trend is due to local changes in the angle of attack along the turbine blades, but do not perform further analysis. 
Therefore, the existing literature leaves the ABL parameter space largely unexplored, with a high degree of uncertainty on the different effects of wind shear and veer, both geometric and inductive. 

The underlying hypothesis of this study is that the aerodynamics of rotors in ABL inflow conditions differ significantly from uniform inflow conditions, and these differences are not explained solely by inflow geometry alone (e.g., $\rews$). 
That is, in addition to modifying the local wind speed, wind shear and wind veer have an additional aerodynamic impact on the rotor induction that affects rotor performance, which is poorly understood. 
Furthermore, larger turbines will experience greater magnitudes of shear and veer, leading to larger effects and deviations in turbine power than for smaller wind turbines. 
To this end, we perform LES of an actuator disk-modeled wind turbine in ABL inflows. 
Using a simplified inflow setup, we isolate the energy transport mechanisms that are modulated by ABL inflow conditions. 
With these simplified inflow simulations, we parse the aerodynamic and geometry effects of wind shear and veer.  

The rest of this paper is organized as follows. 
In \cref{sec:methods}, we present the large-eddy simulation numerical setup. 
Then, in \cref{sec:results_sbl_power}, we investigate the rotor performance of a $\SI{240}{\meter}$-diameter, actuator disk-modeled wind turbine across neutrally and stably stratified conditions using concurrent-precursor LES. 
Following in \cref{sec:shear_idealized} and \cref{sec:veer_ideal}, we parse the effects of shear and veer, respectively, on rotor performance and energy transport mechanisms, using LES with idealized inflows. 
Then, in \cref{sec:results_2}, we re-analyze the power losses arising from atmospheric boundary layer inflows with a new decomposition of shear and veer effects on turbine power production. 
Finally, we close with concluding remarks in \cref{sec:conclusions}.

\section{Methods}
\label{sec:methods}
We begin with the large-eddy simulation governing equations in \cref{ssec:les}.
Then, a description of the numerical solver is given in \cref{ssec:padeops}, followed by a description of the case setup for each set of simulations. 

\subsection{Large-eddy simulation governing equations}
\label{ssec:les}
We begin with the non-dimensional Navier--Stokes equations with rotation effects in the inviscid limit with the Boussinesq approximation for buoyancy, 
\begin{gather}
\label{eq:NSE_filtered}
\pdv{u_i}{t} + u_j \pdv{u_i}{x_j}
= 
-\pdv{p}{x_i}
-\pdv{\tau_{ij}}{x_j}
+\frac{\delta_{i3}}{\Theta_0 \Fr^2}(\Theta - \langle \Theta \rangle_{xy})
- \frac{2}{\Ro}\epsilon_{ijk}\Omega_j (u_k - G_k)
+ f_i, \\
\label{eq:continuity}
\pdv{u_i}{x_i} = 0,
\end{gather}
where $u_i = (u, v, w)$ is the filtered velocity vector and $x_i = (x, y, z)$ are the coordinates in the streamwise, lateral, and wall-normal directions. 
The potential temperature field is $\Theta$, $\Theta_0 = \SI{300}{\kelvin}$ is reference potential temperature, and $\langle \cdot \rangle_{xy}$ denotes horizontal averaging. 
The trace-free subgrid stresses are given by $\tau_{ij}$, and the filtered, modified pressure field $p$ absorbs the normal component of the subgrid stresses. 
The driving geostrophic pressure gradient is prescribed by a geostrophic wind velocity vector $G_k$. 
Coriolis forces are included with only the vertical component of Earth's rotation, such that the non-dimensional rotation vector is $\Omega_j = (0, 0, \sin(\phi))$, where $\phi$ is the latitude. 
The Rossby number $\Ro \equiv G/(\omega D)$ represents the ratio of inertial to Coriolis forces, where $G$ is the geostrophic wind speed magnitude, $\omega = \SI{7.29}{\radian\per\second}$ is the rotation rate of the earth, and $D$ is a length scale that we will define throughout this manuscript to be equal to the turbine diameter. 
The Froude number $\Fr = G/\sqrt{g D}$ is the ratio of inertial to gravitational forces, where $g = \SI{9.81}{\meter\per\second\squared}$ is gravitational acceleration. 
Finally, $f_i$ represents the force imparted by turbines on the flow. 

In simulations where the domain is thermally stratified, an additional prognostic equation for the filtered potential temperature is also solved: 
\begin{equation}
    \label{eq:theta}
    \pdv{\theta}{t} + u_j\pdv{\Theta}{x_j}
    = 
    -\pdv{q_j}{x_j}. 
\end{equation}
In \cref{eq:theta}, $q_j$ is the subgrid heat flux. 
A turbulent Prandtl number $\Pr = 0.5$ is used in all simulations. 

\subsection{Numerical setup}
\label{ssec:padeops}
We solve the filtered, incompressible Navier--Stokes equations with the open-source, pseudo-spectral LES code Pad{\'e}Ops \citep{klemmer_momentum_2024, heck_coriolis_2025, shin_addressing_2025}.
Fourier collocation is used in the horizontal directions, and a sixth-order accurate compact finite differencing scheme is used in the wall-normal direction \citep{lele_compact_1992}. 
In all cases, the problem domain is discretized into a regular grid, and time-stepping is performed with a fourth-order Runge-Kutta scheme using a Courant-Friedrichs-Lewy number of 1.0. 
Subgrid stresses are closed with the sigma model \citep{nicoud_using_2011}. 
A fringe region \citep{nordstrom_fringe_1999} is used when a turbine is present to break the periodicity in the streamwise direction and introduce unwaked inflow to the inlet of the domain. 

An irrotational actuator disk is used to represent the wind turbine. 
The actuator disk imparts a thrust force normal to its surface that is proportional to the square of the disk-averaged velocity $\ud$, $F_{T, i} = \tfrac 12 \rho A_d u_d^2 C_T' n_i$, where $\rho$ is the fluid density, $C_T'$ is the local thrust coefficient \citep{calaf_large_2010, shapiro_filtered_2019}, $A_d = \pi R^2$ is the rotor swept area, and $n_i$ is the normal vector to the disk. 
Throughout this study, we fix $C_T' = 4/3$ \citep{calaf_large_2010}. 
The actuator disk is represented numerically by an indicator function $I(\vec{x})$ that is computed by convolving a Heaviside function $H$ with a three-dimensional, isotropic Gaussian filter $G(\vec{x})$ of width $\Delta = 2.5h$, where $h=\sqrt{\Delta x^2 + \Delta y^2 + \Delta z^2}$ is the grid size: 
\begin{equation}
    \label{eq:adm_kernel}
    I(\vec{x}) = \frac{1}{\pi R^2}\iiint \delta(x') H(R-r') G(\vec{x} - \vec{x}') \,d\vec{x}'.
\end{equation}
In \cref{eq:adm_kernel}, $\delta(x)$ is the delta function and $r = \sqrt{y^2 + z^2}$ is the radial coordinate. 
A standard spherical Gaussian kernel $G(\vec{x})$ is used, following \citet{calaf_large_2010}: 
\begin{equation}
    \label{eq:gaussian_kernel}
    G(\vec{x}) = \left( \frac{6}{\pi \Delta^2} \right)^{3/2} 
    \exp \left( -\frac{6 |\vec{x}|^2}{\Delta ^2}\right). 
\end{equation}
Throughout this study, we use the correction factor introduced by \citet{shapiro_filtered_2019} to reduce the overprediction in actuator disk power and thrust introduced by the regularization function. 

We study the interaction between inflow heterogeneity and an actuator disk using different numerical setups, which vary in computational cost. 
In the most computationally intensive case, we perform concurrent-precursor simulations of a rotor immersed in a stably stratified atmospheric boundary layer (SBL). 
To complement the concurrent-precursor simulations, we also perform prescribed inflow profile simulations using a synthetic inflow method \citep{heck_unraveling_2025}. 
In the synthetic inflow simulations, the effects of wind shear, wind veer, wall proximity, and turbulence can be controlled independently to parse different effects of the inflow on the rotor and their underlying physical mechanisms. 
Details for the three simulation environments used in this study are given in \cref{sssec:setup_sbl}--\ref{sssec:setup_turbulent}.

\subsubsection{Concurrent-precursor ABL simulation setup}
\label{sssec:setup_sbl}
Winds in the true ABL are controlled by the interaction of Coriolis forces, thermal stratification, and wind shear. 
In the concurrent-precursor simulations, the wind shear and wind veer are varied by changing the cooling rate at the surface $C_r$ between $0 - \SI{0.5}{\kelvin\per\hour}$. 
For the stably stratified simulations ($C_r > 0$), we initialize the ABL in a spin-up simulation for \SI{10}{\hour} using the nocturnal stable boundary layer methodology described in \citet{shen_global_2024}, which we briefly summarize here. 
The grid resolution is $20 \times 12.5 \times 6.25$~\si{\meter} with $384 \times 256 \times 256$ grid points in the streamwise, lateral, and vertical directions. 
The latitude is set to $\SI{45}{\degree}$ and we prescribe a surface roughness of $z_0 = \SI{0.1}{\meter}$.
An unstratified (isothermal) initial condition of $\theta_0 = \SI{300}{\kelvin}$ is initialized with small temperature perturbations in the bottom $\SI{50}{\meter}$ of the domain to accelerate the development of a turbulent boundary layer.
A time-dependent Dirichlet boundary condition is prescribed to cool the bottom boundary condition with the prescribed cooling rate $C_r$, which is initially the same temperature as the domain. 
The geostrophic wind speed is fixed at $G = \SI{12}{\meter\per\second}$ and is initially aligned with the $x$-axis.

After the initial spin-up, we perform a domain rotation before starting the concurrent-precursor simulation \citep{stevens_concurrent_2014}. 
Between hours $10$--$11$, a wind angle controller is used \citep{sescu_control_2014} to rotate the domain such that the wind direction is aligned at the turbine hub height ($z_h = \SI{150}{\meter}$, matching the \iea reference turbine \citep{gaertner_definition_2020}). 
Finally,  the wind angle controller is turned off, and the inflow from the rotation phase is fed into a concurrent simulation with an actuator disk placed at $z=z_h$ and centered laterally in the domain.
Time averaging begins at hour 13 after a transient period of five flow-through times and continues until hour 20 ($\SI{7}{\hour}$ of time averaging) \citep{klemmer_momentum_2024}. 

The truly neutral boundary layer (TNBL, also known as the turbulent Ekman layer) has zero surface cooling ($C_r = 0$). 
As a result, the ABL depth grows unimpeded by stable stratification up to the Rossby-Montgomery height \citep{rossby_layer_1935, zilitinkevich_further_2007}.
Because stratification is not present to suppress turbulence, vertical gradients of wind speed and direction are smaller in the TNBL than in the SBL \citep{heck_coriolis_2025}.
Therefore, we expand the vertical domain height from \SI{1.6}{\kilo\meter} to \SI{6.4}{\kilo\meter}, but coarsen the vertical grid by a factor of two.
The TNBL simulation shares the same horizontal grid resolution as the SBL simulations. 
In separate sensitivity tests (not shown), we find that the coarsened vertical resolution has a minimal effect on the inflow profile, but the expansive vertical domain height is necessary to freely accommodate ABL growth. 
Inertial oscillations damp out slowly in neutrally stratified ABLs \citep{liu_geostrophic_2021, heck_coriolis_2025}, so the spin-up time in the TNBL is increased to $15/f_c \approx \SI{40}{\hour}$. 
In the concurrent-precursor stage, statistics are averaged over a full inertial period ($2\pi/f_c \approx \SI{15}{\hour}$) following the domain rotation and a transient period of five flow-through times \citep{heck_coriolis_2025}.

\subsubsection{Synthetic laminar inflow simulation setup}
\label{sssec:setup_laminar}
To independently explore the effects of wind shear, wind veer, and wall proximity on rotor aerodynamics, we introduce a synthetic inflow LES environment. 
Unlike the concurrent-precursor LES (\cref{sssec:setup_sbl}) where wind shear and wind veer are coupled to Coriolis forcing, stratification, turbulence, wall friction, and other ABL properties, we explicitly prescribe the inflow velocity profile in the synthetic inflow environment. 
The prescribed inflow profile is laminar ($\ti = 0\%$) with slip walls, preventing inflow adjustment upstream of the turbine. 
Thus, shear and veer can be prescribed independently through inflow profiles $\uB(z)$ and $\vB(z)$, respectively. 
In all laminar inflow simulations, the domain is isothermal, and the equation for the potential temperature is not solved. 

Our main consideration in the laminar inflow simulations is a sufficiently large domain such that strong wind veer does not cause the wake to interact with the lateral boundaries of the domain, which are periodic. 
We select a lateral domain size of $L_y 4\pi D$ for all simulations. 
Simulations run with twice the lateral domain size $L_y^\text{large} = 8 \pi D$ show that differences in rotor power between the small and large domains are within $0.5\%$ for veer angles up to $60^\circ$, the maximum veer that we present in this manuscript.
Therefore, we deem the domain size to be sufficiently large for the flow physics of interest.
The streamwise and vertical dimensions of the domain are chosen to be $L_x = 10\pi D$ and $L_z = 2\pi D$, respectively. 
Simulations are run with $(n_x, n_y, n_z) = (480, 192, 96)$ grid points in the streamwise, lateral, and wall-normal directions, resulting in an isotropic grid. 
The uniform, zero freestream turbulence inflow simulations converge to a quasi-steady state solution in approximately two flow-through times $L_x/\Uhub$ \citep{heck_modelling_2023}.
To add an extra buffer around the additional inflow heterogeneity from shear and veer, we increase the averaging time to four flow-through times, which we find sufficient across the parameters of wind shear and veer studied here. 
A transient time of four flow-through times is used before time-averaging begins to allow the flow to equilibrate. 

In the synthetic inflow setup with slip walls, the inflow profile can be prescribed exactly.
Throughout \cref{sec:shear_idealized} and \cref{sec:veer_ideal}, we prescribe idealized profiles for shear and veer to identify key mechanisms that modify the rotor aerodynamics.
For these idealized profile simulations, we switch off the effects of planetary rotation by setting $\Ro = 10^{10}$ (i.e., $\Ro \to \infty$). 
In \cref{ssec:results_sbl_cumulative}, we impose the mean inflow profiles from an ABL into the synthetic inflow setup, prescribing $\Ro$ to match the ABL flow. 
The specific profiles (forms of $\uB(z)$ and $\vB(z)$) imposed in the synthetic inflow LES are given in \cref{sec:shear_idealized}--\ref{sec:results_2}.

\subsubsection{Synthetic turbulent inflow simulation setup}
\label{sssec:setup_turbulent}
To investigate the effects of freestream turbulence on rotor aerodynamics in combination with wind shear and wind veer, we superimpose turbulent fluctuations on the prescribed mean profiles from \cref{sssec:setup_laminar}. 
The setup here exactly follows the setup described in \citet{heck_unraveling_2025}.  
Turbulent fluctuations are generated through a concurrently run homogeneous, isotropic turbulence box of size $(L_x^\text{HIT}, L_z^\text{HIT}, L_z^\text{HIT}) = (2\pi D, 4\pi D, 2\pi D)$. 
The magnitude of turbulence fluctuations is scaled using a proportional-integral turbulence controller to target a set level of turbulence intensity at the rotor, where $\ti_d \equiv \sqrt{\tfrac 23 \langle k\rangle_d} / \Uhub$ and $k \equiv \tfrac 12  \overline{u_i' u_i'}$ is the resolved turbulence kinetic energy. 
The domain size and grid resolution for the synthetic turbulent inflow LES match the synthetic laminar inflow LES from \cref{sssec:setup_laminar}. 

To compute converged time-averaged statistics, we increase the time averaging window in the turbulent synthetic inflow simulations. 
Specifically, we allow the wake and turbulence controller to equilibrate over a time window of 16 flow-through times. 
Then, the turbulence controller is shut off, and the flow is averaged over an additional 24 flow-through times. 
For a turbine of $D = \SI{240}{\meter}$ and $\Uhub = \SI{12}{\meter\per\second}$, this corresponds to $\approx \SI{9.5}{\hour}$ of time averaging, which is sufficient for converged statistics of the rotor and wake \citep{heck_unraveling_2025}. 

Similar to the synthetic laminar inflow setup, here we can prescribe idealized profiles of shear and veer or use realistic ABL profiles. 
When prescribing idealized velocity profiles, we do not solve the potential temperature equation. 
However, when prescribing the SBL profiles in \cref{sec:results_2}, we also prescribe the time- and horizontally-averaged potential temperature profile in the synthetic turbulent inflow setup, following \citet{heck_unraveling_2025}. 
This helps to suppress turbulence production in strongly veered inflows, mitigate inflow adjustment, and better match the SBL inflow velocity profiles at the rotor location.

\section{Rotor performance in stable boundary layers}
\label{sec:results_sbl_power}

\begin{figure}[htb]
    \centering
    \includegraphics[width=1\linewidth]{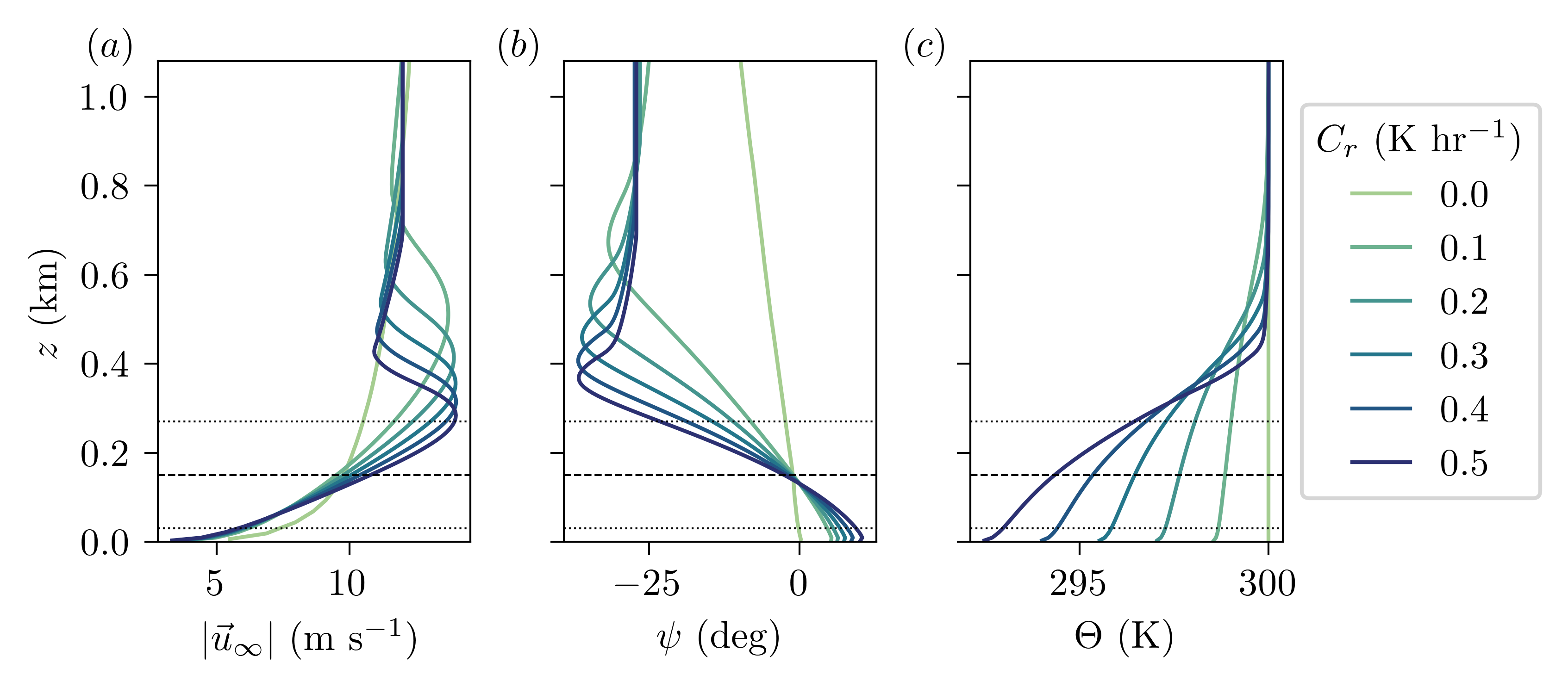}
    \caption{Inflow profiles of ($a$) wind speed magnitude, ($b$) wind direction, and ($c$) potential temperature as a function of cooling rate $C_r$ in the stably stratified ABL. Note that the full computational domain (\SI{1.6}{\kilo\meter} in the SBL and \SI{6.4}{\kilo\meter} in the TNBL) is not shown. Dashed lines denote the hub height of the \iea reference turbine ($\zhub = \SI{150}{\meter}$, and dotted lines denote the rotor extent. }
    \label{fig:sbl_inflows}
\end{figure}

We begin by investigating rotor performance in the stably stratified concurrent-precursor ABL simulations. 
Increasing the surface cooling rate $C_r$ increases thermal stratification, limiting vertical motions in the ABL and inhibiting boundary layer growth. 
As a result, the thermal stratification affects the structure of the SBL, including the wind shear, wind veer, hub height freestream velocity $\Uhub$, Obukhov length, and ABL height. 
Profiles of wind speed, $|\vec{U}| = \UBmag$, wind direction, $\psi = \arctan(\vB/\uB)$, and potential temperature at the end of the concurrent-precursor simulation (i.e., after \SI{20}{\hour} for the SBL and after $\approx \SI{55}{\hour}$ in the TNBL) are shown in \cref{fig:sbl_inflows}. 
The hub height ($\zhub = \SI{150}{\meter}$) and top and bottom tip heights for the \iea reference wind turbine ($D = \SI{240}{\meter}$) are also shown in \cref{fig:sbl_inflows}. 

We observe several monotonic trends as stratification increases. 
First, the ABL height drops with increasing cooling rate due to the suppression of vertical motions. 
As a result, the wind shear in the rotor extent increases with increasing $C_r$. 
Additionally, the wind veer in the rotor extent also increases monotonically with increasing cooling rate. 

We can quantify the wind shear and wind veer in the ABL using several metrics, which are useful for different reasons. 
The canonical metric for describing the amount of wind shear is the power law exponent $\powerlaw$ \citep[c.f.][]{vansark_we_2019}, defined in \cref{eq:powerlaw}. 
However, despite its widespread use, the power law exponent $\powerlaw$ has an important limitation in the context of this study: the degree of wind speed variation across the rotor (i.e., $U(\zhub + R) - U(\zhub - R)$) depends not only on $\powerlaw$, but also on the reference height (here: $\zhub$). 
To illustrate this, we can relate the variation in wind speed across the rotor to the gradient of $\UBmag$ at hub height.
For a power law profile, this yields $\partial_z(\UBmag/\Uhub)\big|_{z=\zhub} = \powerlaw (z/z_h)$. 
%
%
Hence, the non-dimensional wind speed change across the rotor (linearized to $\powerlaw D/\zhub$) increases with decreasing $\zhub$ if $\powerlaw$ and $D$ are held fixed. 
Therefore, in conjunction with quantifying shear using a power law exponent, we will also describe shear using the slope of the non-dimensional wind speed profile at hub height: 
\begin{equation}
\label{eq:linear_shear}
    \shear \equiv \left. \pdv{(U/U_h)}{(z/D)} \right|_{z=z_h}. 
\end{equation}
The linearized relationship between the power law exponent $\powerlaw$ and the shear slope $\shear$ is $\shear = \powerlaw / (\zhub / D)$. 

Next, we quantify the wind veer. 
Because wind veer is empirically often approximately linear in the ABL \citep{kelly_shear_2023}, it is common to characterize wind veer as the rate of turning per unit length, i.e.,\ degrees per meter. 
In this study, we will non-dimensionalize this metric by the rotor diameter to yield total turning across the rotor: 
\begin{equation}
    \label{eq:linear_veer}
    \veer \equiv -\left[ \psi(z=\zhub+R) - \psi(z=\zhub-R) \right]. 
\end{equation}
We include a negative sign in our definition of \cref{eq:linear_veer}, following the typical convention in the northern hemisphere dictated by the Ekman spiral that winds turn rightward (increasingly negative $\psi$) with increasing height. 
A full summary of the ABL inflows studied using the concurrent-precursor LES setup is given \cref{tab:sbl_properties}, including the friction velocity $u_\star$, Obukhov length $L_\text{obu}$, cross-isobaric angle $\alpha_0$, and rotor-equivalent turbulence intensity $\ti_d$. 

\begin{table}[htb]
    \centering
    \caption{Properties of the ABLs studied using the concurrent-precursor LES method. All ABLs use a surface roughness of $z_0 = \SI{0.1}{\meter}$, geostrophic wind speed $G = \SI{12}{\meter\per\second}$, latitude of $\phi = \SI{45}{\degree}$, and are computed in a horizontal domain of $\SI{7.68}{\kilo\meter} \times \SI{3.2}{\kilo\meter}$.}
    \label{tab:sbl_properties}
    \begin{tabular}{llllllllll}
    \hline
    $C_r$ (\si{\kelvin\per\hour}) & $\Uhub$ (\si{\meter\per\second}) & $\powerlaw$ (-) & $\shear$ (-) & $\veer$ (deg) & $\psi_h$ (deg) & $u_\star$ (\si{\meter\per\second}) & $L_\text{obu}$ (\si{\meter}) & $\alpha_0$ (deg) & $\ti_d$ (\%) \\ \hline
    0 & 9.74 & 0.16 & 0.23 & 2.8 & -2.7 & 0.534 & $\infty$ & -15.2 & 6.8 \\
    0.1 & 9.51 & 0.30 & 0.54 & 12.7 & -1.1 & 0.443 & 489 & -30.1 & 4.2 \\
    0.2 & 9.77 & 0.34 & 0.62 & 16.6 & -1.4 & 0.423 & 261 & -32.7 & 3.8 \\
    0.3 & 10.05 & 0.38 & 0.69 & 21.1 & -1.6 & 0.405 & 178 & -34.6 & 3.5 \\
    0.4 & 10.39 & 0.41 & 0.76 & 26.6 & -2.0 & 0.390 & 133 & -36.0 & 3.4 \\
    0.5 & 10.79 & 0.44 & 0.82 & 32.5 & -2.6 & 0.376 & 106 & -37.2 & 3.3 \\ \hline
    \end{tabular}
\end{table}

\begin{figure}[htb]
    \centering
    \includegraphics[width=0.85\linewidth]{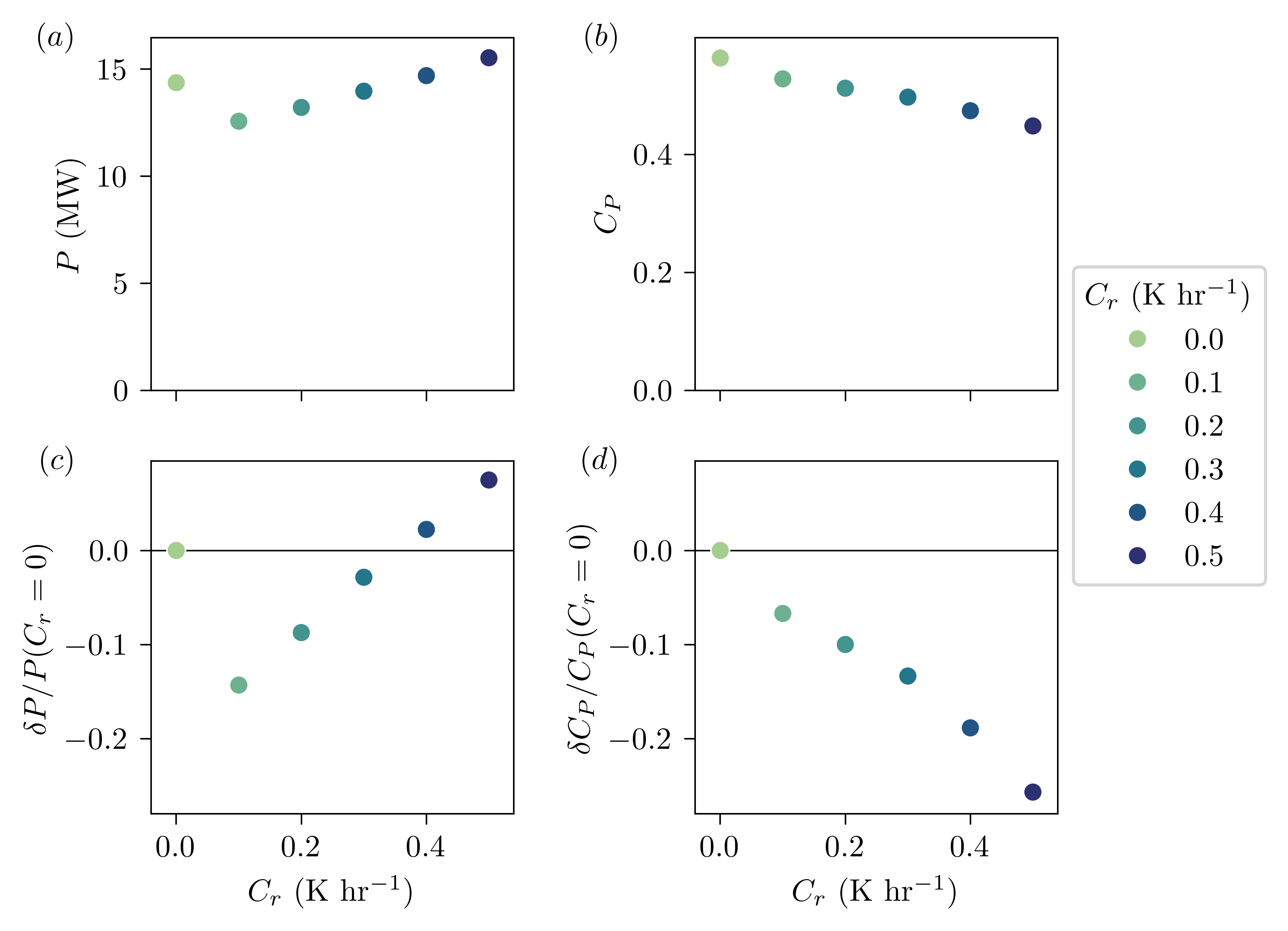}
    \caption{($a$) Turbine power $P$ and ($b$) power coefficient $C_P = 2P/(\rho A_d \Uhub^3)$ as a function of cooling rate $C_r$ for an actuator disk with dimensions of the \iea reference wind turbine. Relative deviations in turbine power and power coefficient from the $C_r=0$ case are shown in ($c$) and ($d$), respectively.}
    \label{fig:sbl_power}
\end{figure}

Next, we examine the power produced by an actuator disk-modeled wind turbine immersed in the ABL flow. 
Specifically, we are interested in how the multitude of ABL properties in \cref{tab:sbl_properties} affect the power production and efficiency of a turbine. 
We show the mean turbine power, defined as $P = -\vec{F}_T \cdot \vec{u}_d$, as a function of cooling rate in \cref{fig:sbl_power}($a$), assuming an air density of $\rho = \SI{1.225}{\kilo\gram\per\meter\cubed}$. 
Note that although the simulated actuator disk shares the dimensions ($\zhub$ and $D$) of the \iea reference turbine, the turbine power $P$ can exceed the rated power of $\SI{15}{\mega\watt}$ because the thrust coefficient, $C_T'$ is a prescribed parameter here, rather than depending on a turbine controller, and because the actuator disk does not have tip/root losses in these simulations. 

Generally, power production increases with increasing cooling rate. 
However, this trend is dominated by the cubic dependence of power on the wind speed magnitude. 
Normalizing the power by the power in the freestream flow at hub height $\tfrac 12 \rho A_d \Uhub^3$ yields a power coefficient, $C_P \equiv 2P/(\rho A_d \Uhub^3)$, which is a function of the inflow conditions beyond the hub height wind speed magnitude, and quantifies the aerodynamic efficiency of the turbine (\cref{fig:sbl_power}($b$)). 
Normalization reveals that the power coefficient has the opposite trend from the power itself: $C_P$ drops precipitously and monotonically with increasing cooling rate, which corresponds to increasing shear and veer magnitudes, as described earlier in \cref{tab:sbl_properties}. 
In the most stable ABL, where wind shear and veer are relatively strong, $C_P$ is more than $20\%$ lower than in the TNBL (\cref{fig:sbl_power}($d$)).
Recall that for an actuator disk, the thrust force $|\vec{F}_T|$ is proportional to $\ud^2$, and the turbine power $P$ is thus proportional to $\ud^3$. 
Therefore, changes in $C_P$ are directly related to changes in the disk velocity through the rotor-normal induction factor $a_n \equiv 1 - \ud/\Uhub$. 
In subsequent sections, we investigate the impacts of wind shear and wind veer on the efficiency (i.e., power coefficient, $C_P$), thrust coefficient ($C_T \equiv 2|\vec{F}_T|/(\rho A_d \Uhub^2)$), and induction factor ($a_n$) in sheared and veered inflow to understand the mechanisms that lead to high efficiency losses ($>20\%$) in stable ABLs.

\section{Wind shear effects in idealized inflows}
\label{sec:shear_idealized}
In this section, we parse the effects of wind shear separately from wind veer using a synthetic inflow method. 
Using idealized, parameterized inflows of the streamwise inflow profile $\uB(z)$, we investigate the impact of wind shear on power production and rotor aerodynamics in \cref{ssec:shear_rotor}, parsing the importance of wind shear by a geometric effect and an inductive effect. 
Then, in \cref{ssec:shear_nonlocal}, we study the inductive response of a rotor in sheared inflow conditions and how wall effects modulate the induction.

\subsection{Effect of shear on rotor quantities}
\label{ssec:shear_rotor}

We begin investigating wind shear effects on rotor performance using idealized, synthetic inflow simulations. 
%
We begin with a simple parameterization of wind shear, where the gradient of the streamwise wind speed is approximately linear across the rotor disk, and the rotor-equivalent wind speed is intentionally held constant. 
Although the simplest shear parameterization would be a linear profile ($\UBmag(z)/\Uhub = 1 + \shear(z-\zhub)/D$), far from the disk, linear profiles of $\UBmag(z)$ would cause high wind speeds and/or reverse flow (negative wind speed). 
Therefore, we use a $\tanh$ profile to transition from a quasi-linear profile in the rotor region to a constant value far from the rotor in high shear magnitudes \citep[c.f.][]{heck_unraveling_2025}:
\begin{equation}
\label{eq:tanh_shear}
    \UBmag/\Uhub = 1 + (1-\epsilon) \tanh \left(\frac{\shear (z - \zhub)}{D(1-\epsilon)}\right).
\end{equation}
In \cref{eq:tanh_shear}, profiles are approximately linear near the rotor ($\tanh(x) \approx x$ for small $x$) with the slope $\shear (z/D)$. 
The parameter $\epsilon = 0.2$ controls the threshold on the maximum and minimum velocities to prevent reverse-flow. 
Note that in \cref{eq:tanh_shear}, moving the turbine hub height shifts the entire wind profile such that the linear portion of the $\tanh$ profile is centered around the rotor. 
Furthermore, the rotor-equivalent wind speed, defined in \cref{eq:rews} (where $\theta$ is defined from the positive $y$-axis such that $\theta = \arctan((z-\zhub)/y)$), is constant for the $\tanh$ inflow shear profiles due to symmetry. 
We vary the strength of the $\tanh$ shear profiles between $\shear \in [-0.32, 0.8]$, which gives the same range of wind linearized wind shear at hub height as the range of power law exponents $\powerlaw$ observed in the ABL concurrent-precursor LES (recall that $\shear = \powerlaw D/\zhub$). 

\begin{figure}[htb]
    \centering
    \includegraphics[width=0.9\linewidth]{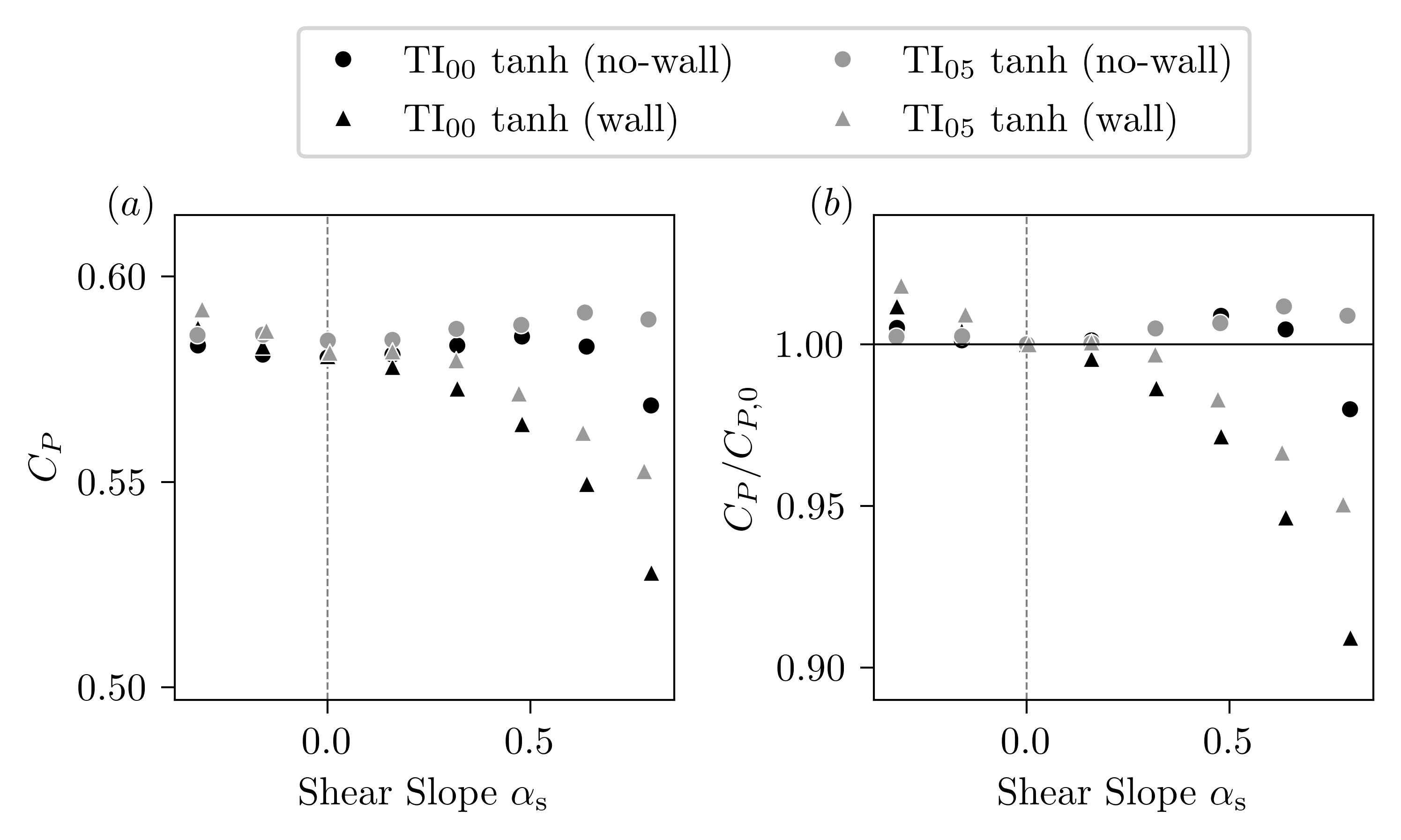}
    \caption{Actuator disk ($a$) power coefficient $C_P = 2P/(\rho A_d \Uhub^3$) and ($b$) power coefficient normalized to shear-free inflow conditions ($\shear=0$), as a function of wind shear slope at hub height ($\shear$). Four variations of $\tanh$ shear inflows (see \cref{eq:tanh_shear}) shown: $\ti = 0\%$ and $\ti = 5\%$ with wall proximity.}
    \label{fig:shear_cp}
\end{figure}

To study the effect of wind shear on rotor performance, we perform four sets of simulations using the two different parameterizations for shear.
In all sets of simulations, we collect time-averaged statistics for the rotor-averaged velocity $\ud$, rotor thrust $\vec{F}_T$, and power $P$. 
First, we place a turbine centered vertically in the domain ($\zhub = L_z/2$) such that it is far from the slip walls and impose a $\tanh$ shear inflow profile without inflow turbulence (labeled ``$\ti_{00}~\tanh$ (no-wall)''). 
Second, we duplicate the numerical setup but add $5\%$ turbulence intensity (labeled ``$\ti_{05}~\tanh$ (no-wall)'') using a synthetic turbulence method \citep{heck_unraveling_2025}. 
Finally, for both the $\ti_{00}$ and $\ti_{05}$ inflows, we move the turbine hub height to $\zhub/D = 0.625$, matching the dimensions of the \iea reference turbine (\SI{150}{\meter}/\SI{240}{\meter}), which we label ``$\ti_{00}~\tanh$ (wall)'' and ``$\ti_{05}~\tanh$ (wall)'', respectively. 
In \cref{fig:shear_cp}, we show the mean power coefficient as well as the power coefficient normalized by the efficiency in the shear-free inflow, $C_{P, 0} = C_P(\shear = 0)$. 

The extent to which turbine power production is affected by shear depends on the presence of the wall as well as the imposed profile ($\tanh$ versus power law). 
Beginning with the wall-free simulations, we observe that the changes in turbine power are marginal: only a $\pm 3\%$ change in power is observed even at very strong shear slopes ($\shear = 0.8$). 
Therefore, we conclude that the importance of freestream turbulence (up to $\ti = 5\%$) on the rotor dynamics is negligible, and that shear in the absence of wall effects only has a minor effect on rotor-averaged quantities. 
The $\ti=0\%$ and $\ti=5\%$ inflow simulations have nearly identical changes in power as a function of shear, except for the highest shear slope. 
By contrast, when the rotor is near the wall, wind shear can significantly modulate the power coefficient (triangles in \cref{fig:shear_cp}). 
Power coefficient decreases with increasing $\shear$, and efficiency losses are slightly higher for $\ti=0\%$ compared with $\ti=5\%$. 
At the highest value of shear, the combined effect of the wall and $\tanh$ shear leads to a $9\%$ drop in $C_P$, relative to the zero-shear baseline, despite the constant rotor-equivalent wind speed. 
Importantly, the deviations from rotor performance in uniform flow (e.g., $C_{P,0}$) for the $\tanh$ inflow profiles arise despite the rotor-equivalent wind speed being held constant. 
Therefore, we characterize these impacts of shear on the rotor as \textit{inductive} effects -- they modify the power coefficient through a change in the induction factor $\an = 1 - \ud / \Uhub$. 

\begin{figure}[htb]
    \centering
    \includegraphics[width=1\linewidth]{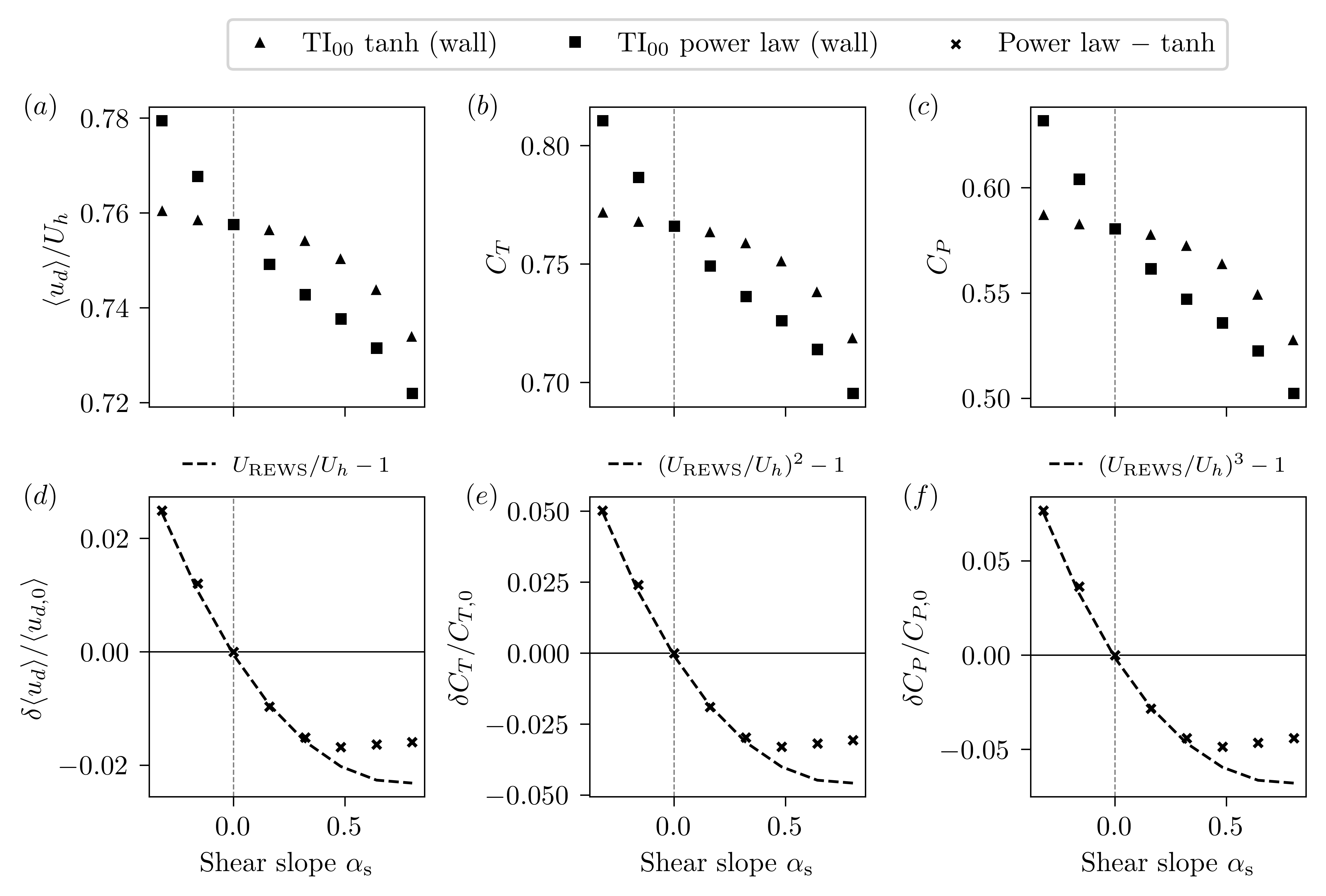}
    \caption{($a$) Disk-averaged velocity, ($b$) thrust coefficient $C_T$, and ($c$) power coefficient $C_P$ as a function of shear slope at hub height $\shear$ for $\tanh$ and power law shear profiles. ($d$-$f$) Normalized differences in disk velocity, thrust coefficient, and power coefficient between the power law and $\tanh$ inflow profiles, respectively. Dashed lines show the geometric effect from changing $\rews/\Uhub$.}
    \label{fig:shear_rews}
\end{figure}

However, shear is typically nonlinear in the ABL; rather, ABL shear is often approximated using a power law profile (see \cref{eq:powerlaw}). 
Therefore, we perform a fifth set of synthetic inflow LES using a power law profile, labeled ``$\ti_{00}$ power law (wall),'' where we vary $\powerlaw \in [-0.2, 0.5]$ to span the range of power law exponents fit to inflows of the ABL in \cref{tab:sbl_properties}. 
The inflow is laminar ($\ti = 0\%$). 
The actuator disk is placed at $\zhub/D = 0.625$, matching the concurrent-precursor ABL simulations, and the bottom wall is again a slip wall. 
The disk velocity, thrust coefficient, and power coefficient are shown in \cref{fig:shear_rews} ($a$-$c$, respectively). 

The power law profiles show an even greater drop in efficiency ($C_P$, \cref{fig:shear_rews}($c$)) than the $\tanh$ (wall) simulations for positive shear, and an increase in power for negative shear. 
For a given value of $\shear$, the primary difference between the power law and the $\tanh$ profiles is the curvature of the power law profile. 
Specifically, a consequence of the power law curvature---which is not present in the $\tanh$ profiles---is that the rotor-equivalent wind speed $\rews$ changes as a function of the power law exponent $\powerlaw$. 
At $\powerlaw = 0$, $\rews = \Uhub$, and increasing power law exponent $\powerlaw$ monotonically decreases $\rews$ (and vice versa, i.e., increasingly negative $\powerlaw$ monotonically increases $\rews$).
We call the impact of wind shear on the disk velocity $\ud$ (and therefore $C_T$ and $C_P$) due to a change in $\rews$ as the \textit{geometric} effect, which is separate from the \textit{inductive} effect.
If $\ud$ depends linearly on $\rews$, then the geometric and inductive effects of shear would linearly superimpose, and the difference in $\ud$ between the power law and $\tanh$ wind profiles would be exactly $\rews/\Uhub - 1$. 
In \cref{fig:shear_rews}($d$), we show that this is true for $\shear \lesssim 0.4$, after which there are small departures from the linear superposition result. 
At the same time, we observe the difference in $C_T$ and $C_P$ between the power law and $\tanh$ profiles scales almost exactly with $(\rews/\Uhub)^2$ and $(\rews/\Uhub)^3$ (\cref{fig:shear_rews}($e$) and ($f$)), respectively. 
This is because the thrust for an actuator disk is proportional to $\ud^2$, and the power is proportional to $\ud^3$. 
Henceforth, we will focus on the inductive effect of shear by showing only $\ud$, noting that changes in $\ud$ also translate directly to changes in power and thrust. 
The physical mechanisms that modulate the inductive response of the rotor as a function of wind shear will be explored next in \cref{ssec:shear_nonlocal}. 



\subsection{Induction response due to wind shear and wall effects}
\label{ssec:shear_nonlocal}

In this section, we investigate how the induction of an actuator disk is modulated by wind shear and wall proximity. 
In \cref{ssec:shear_rotor}, we observed that the disk-averaged velocity $\langle u \rangle_d$, and therefore the actuator disk thrust and power, have a dependence on wind shear that is modulated by the proximity of a wall. 
Without wall effects, a linear shear profile has a negligible effect on $\ud$, except at very high shear ($\shear \gtrsim 0.8$). 
To understand how shear and near-wall proximity affect the disk-averaged velocity $\ud$ and therefore disk-averaged induction factor $\an$, we investigate the spatial dependence of the disk velocity ($u_d(r, \theta) = u(x_T, r, \theta)$) on the inflow properties. 

\begin{figure}[htb]
    \centering
    \includegraphics[width=1\linewidth]{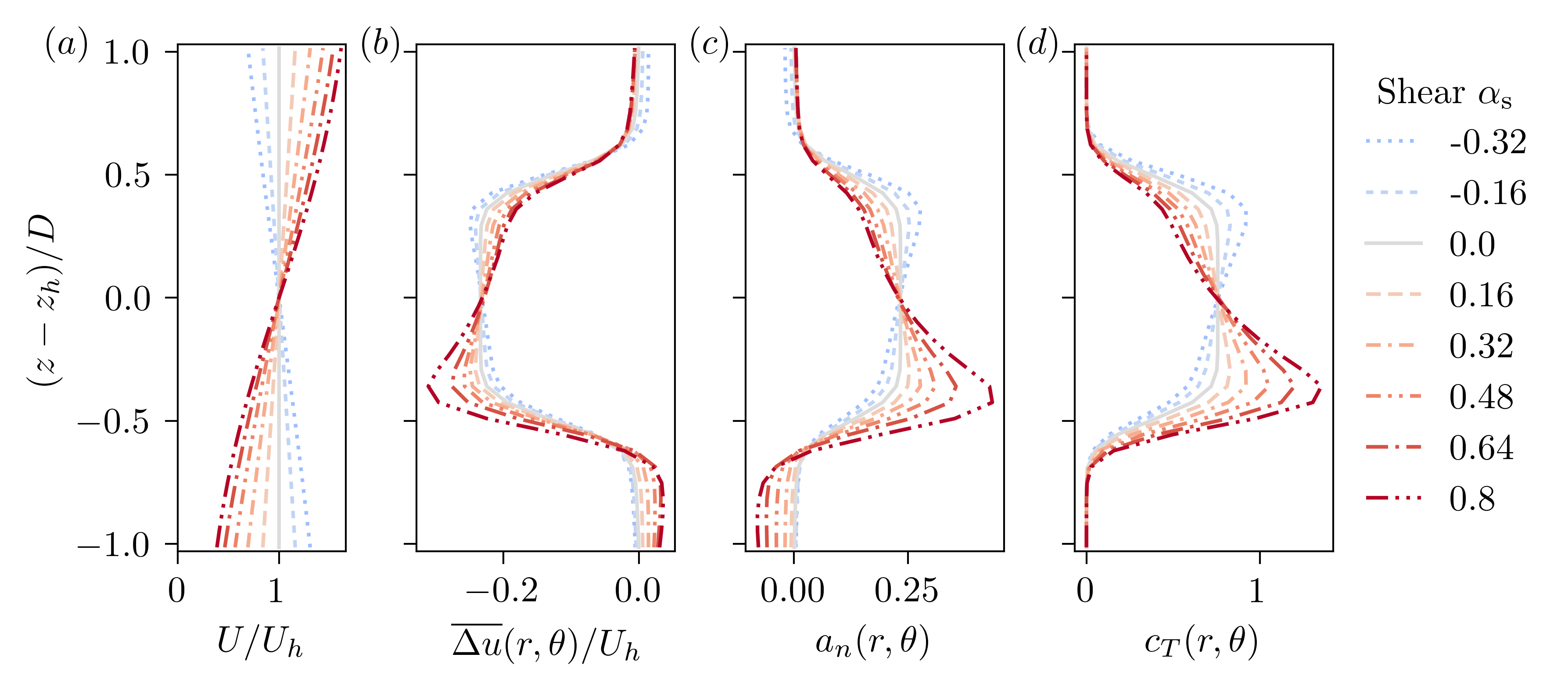}
    \caption{($a$) Sheared inflow profiles of wind speed, normalized by $\Uhub$. ($b$) Local induced velocity at the rotor $\overline{\Delta u}/\Uhub$. ($c$) Local axial induction factor $\an(r, \theta) = 1 - u_d(r, \theta)/\UBmag(z)$. ($d$) Local thrust coefficient $c_T(r, \theta) = -2f_x(r, \theta)/(\rho \UBmag^2(z))$. For rotor quantities ($b$-$d$), profiles are shown at the lateral centerline of the disk ($y=0$, i.e., $\theta = \pm \pi/2$). The disk is placed at $\zhub = L_z/2$.}
    \label{fig:shear_inflow_ct_a}
\end{figure}

First, we examine the no-wall case, where the turbine is centered vertically in the domain ($\zhub=L_z/2$). 
In \cref{fig:shear_inflow_ct_a}, we show vertical profiles of inflow velocity $\UBmag(z)$, induced velocity $\overline{\Delta u}(r, \theta)$, local induction factor $\an \equiv \overline{\Delta u}(r, \theta) / \UBmag(z) = 1 - u_d(r, \theta)/\UBmag(z)$, and local thrust coefficent $c_T(r, \theta)$. 
We define the local thrust coefficent $c_T(r, \theta)$ with respect to the local forcing per unit area $f_x$ and the freestream velocity magnitude $\UBmag(r, \theta)$ (note that $\UBmag(r, \theta)$ depends only on $z = r\sin(\theta)$ in this study): 
\begin{equation}
\label{eq:ct_local}
    c_T(r, \theta) \equiv -\frac{2 f_x(r, \theta)}{\rho \UBmag^2(r, \theta)}. 
\end{equation}
%
All vertical profiles of rotor quantities (\cref{fig:shear_inflow_ct_a}($b$-$d$)) are plotted along the lateral centerline of the disk, i.e., $\theta = \pm \pi/2$. 

In \cref{fig:shear_inflow_ct_a}, red hues (dashed and dash-dotted lines) denote positive shear while blue hues (dotted lines) denote negative shear. 
The uniform inflow case is shown in the solid gray line. 
As was noted previously, the parameterized $\tanh$ inflow profiles shown in \cref{fig:shear_inflow_ct_a}($a$) are linear near the rotor and smoothly asymptote to a constant positive value as $|z-\zhub| \gg D/\shear$ to avoid reverse-flow and numerical instabilities in the LES. 
In \cref{fig:shear_inflow_ct_a}($b$), we show the induced velocity at the disk non-dimensionalized by $\Uhub$.
Without the effect of the wall, the $\tanh$ inflow is rotationally symmetric.
Therefore, positive and negative shear values induce the same velocity $\overline{\Delta u}$, reflected over the vertical axis about $z=\zhub$. 
In uniform inflow ($\shear = 0$), the induced velocity across the disk is constant, except around the edges of the disk ($z-\zhub = \pm R$) due to the Gaussian-filtering applied to the actuator disk in LES (see \cref{eq:adm_kernel}). 
When the inflow is sheared, profiles of the induced velocity deviate from the uniform case in a couple of ways. 
First, the induced velocity $\Delta u$ is non-uniform, despite the uniform loading applied on the flow by the actuator disk. 
At low shear magnitude ($\shear \lesssim 0.32$), the induced velocity variation across the disk is nearly linear, but at high shear magnitude ($\shear = 0.8$), the induced velocity becomes increasingly nonlinear.  
Second, the induced velocity magnitude is not proportional to $\UBmag(z)$. 
That is, the local induction factor $\an(r, \theta) = \overline{\Delta u}(r,\theta)/\UBmag(r, \theta)$ is also not constant (\cref{fig:shear_inflow_ct_a}($c$)), even though the loading ($C_T'$) is constant over the disk. 
Instead, we observe that $|\overline{\Delta u}|$ is highest where the velocity is lowest, and vice versa.
Finally, we note that the shape of the local induction profile is qualitatively similar to the shape of the local thrust coefficient $c_T(r, \theta)$, shown in \cref{fig:shear_inflow_ct_a}($d$). 
However, the relationship between the two local variables can deviate considerably from conventional momentum models \citep[c.f.,][]{burton_wind_2011, madsen_implementation_2020, liew_unified_2024}, and we defer to future work to explore this local induction--thrust relationship. 
Furthermore, bladed rotors have non-uniform loading ($f_x$ varies as a function of the blade aerodynamics), and therefore should be explored in-depth in future work. 

\begin{figure}[htb]
    \centering
    \includegraphics[width=1\linewidth]{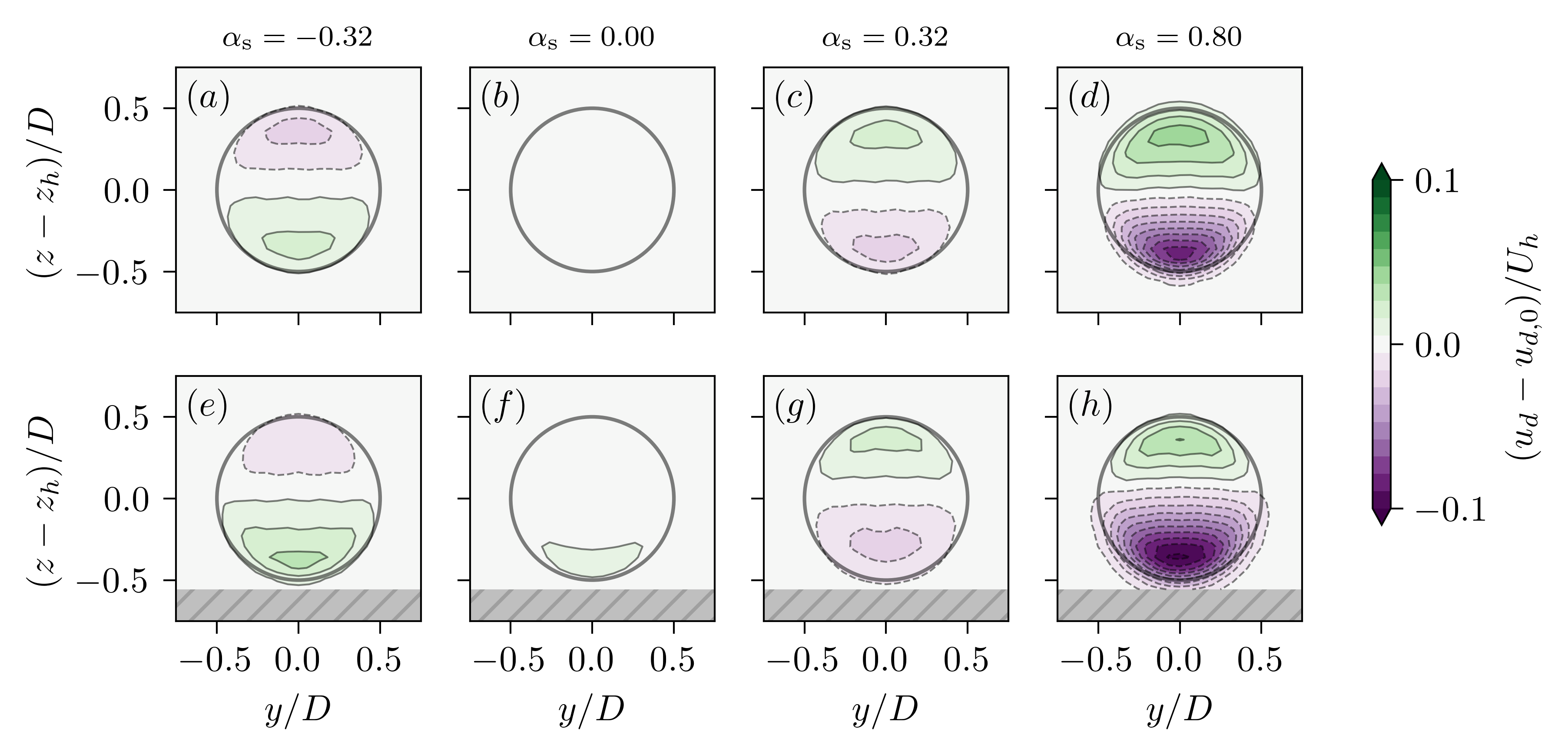}
    \caption{Contours of perturabations to the streamwise velocity $u$ at the rotor plane ($x=0$), relative to the rotor velocity $u_{d, 0}$ in the wall-free, uniform inflow case (subfigure $b$). Linear wind shear strength increases from left to right. ($a$-$d$) Rotor without wall effects, centered vertically in the domain. ($e$-$h$) Rotor near a slip wall, placed at the height of the \iea reference turbine ($\zhub/D=0.625$).}
    \label{fig:shear_ud_contour}
\end{figure}

The induced velocity profiles in \cref{fig:shear_inflow_ct_a}($b$) illustrate why disk-averaged velocity $\ud$ remains approximately constant in the $\tanh$ shear inflows without wall effects: perturbations of $\overline{\Delta u}$ relative to the uniform inflow case are approximately linear and thus integrate to zero. 
However, in \cref{ssec:shear_rotor}, we showed that the wall has a pronounced effect in modulating the disk velocity, thrust, and power in sheared inflow. 
In \cref{fig:shear_ud_contour}, we visualize the velocity at the disk, relative to the uniform inflow reference case where wall effects are negligible. 
We note that the disk velocity in the shear-free case without wall effects is axisymmetric and the radial dependence of $u_{d, 0} = \Uhub + \overline{\Delta u}$, due to the Gaussian-convolved turbine forcing function $I(\vec{x})$ (see \cref{eq:adm_kernel}), is shown in \cref{fig:shear_inflow_ct_a}($b$). 
The two-dimensional contours in \cref{fig:shear_ud_contour} show the anomaly of streamwise velocity at the rotor plane caused by shear and wall proximity. 
Contours are weighted by the actuator disk indicator function $I(0, r,\theta) / \max(I)$ such that they are non-zero only in the region of the (smoothed) actuator disk.
As a result, the integral of each contour plot in \cref{fig:shear_ud_contour} is equal to the deviation in $\ud$ from $\langle u_{d, 0}\rangle$, scaled by $A_d$. 

In the wall-free contours in \cref{fig:shear_ud_contour}($a$-$d$), the velocity perturbation at the disk displays a two-lobed dipole structure. 
The lobes are approximately anti-symmetric, reflecting the profiles in \cref{fig:shear_inflow_ct_a}($b$). 
This anti-symmetry holds for all but the strongest shear case ($\shear = 0.8$, \cref{fig:shear_ud_contour}($d$)), where the negative lobe is approximately twice as strong as the positive lobe in maximum magnitude. 
When the rotor is moved in the proximity of the slip wall (\cref{fig:shear_ud_contour}($e$-$h$)), which results in a no-penetration condition, the bottom lobe is amplified in magnitude relative to the no-wall case. 
The preferential amplification of the near-wall lobe occurs in both positive and negative shear. 
That is, the effect of the wall is to amplify the negative velocity perturbation for $\shear > 0$, and to amplify the positive velocity perturbation for $\shear < 0$. 
As a result, the presence of the wall increases $\ud$ in negative shear, which increases rotor thrust and power, and decreases $\ud$ in positive shear, which decreases thrust and power. 
In \cref{fig:shear_ud_contour}($h$), the negative lobe is amplified by the presence of the wall such that it is $\approx 175\%$ stronger than the positive lobe in peak magnitude. 
As the wall proximity clearly modulates the effects of wind shear, an additional sweep of hub height values ($\zhub/D$) is included in \cref{appx:wall_sweep}. 

\begin{figure}[htb]
    \centering
    \includegraphics[width=0.9\linewidth]{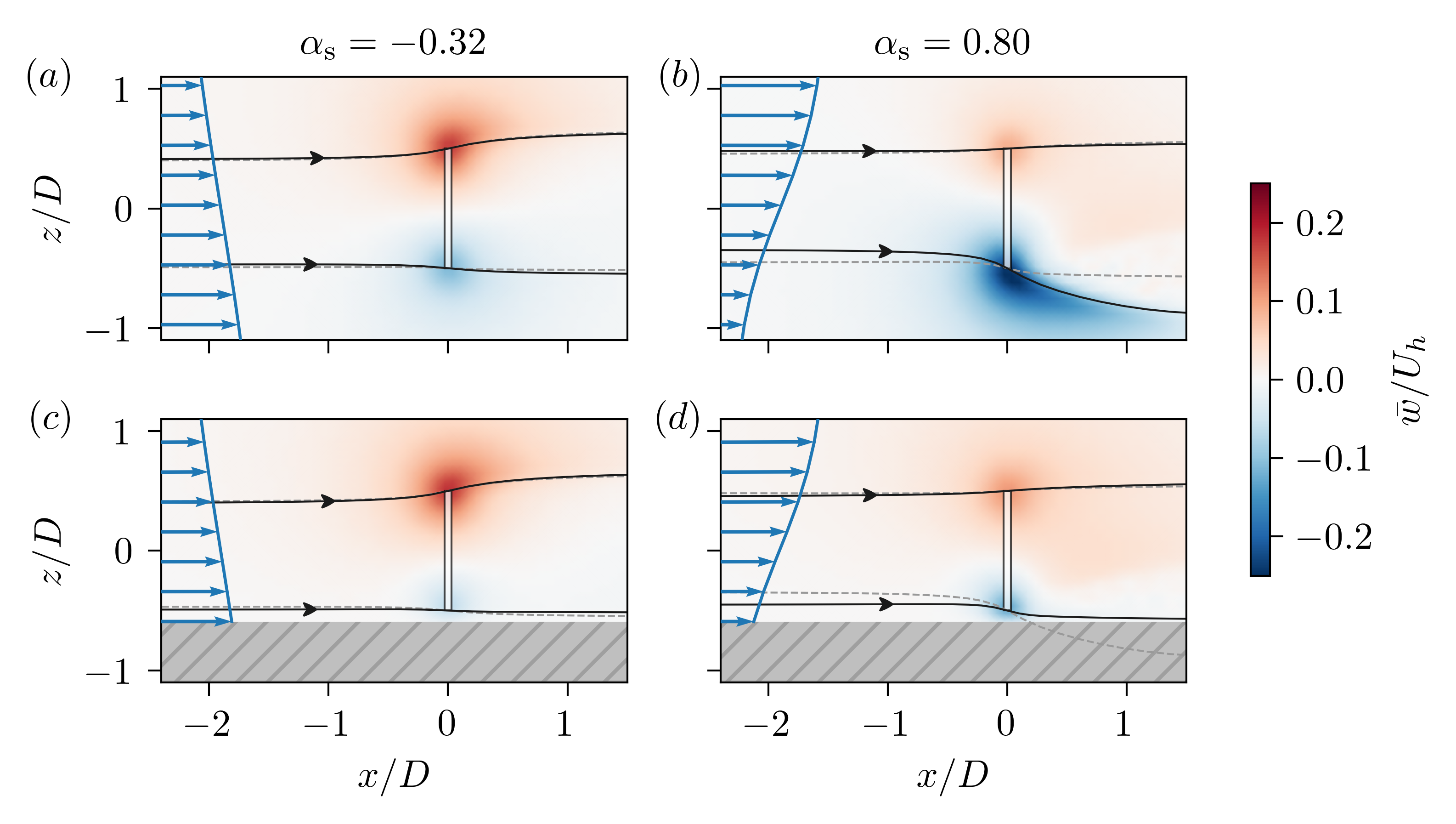}
    \caption{Slices through the lateral centerline of the rotor ($y=0$) showing contours of vertical velocity $\bar{w}$. Extremes of linear speed shear slope are shown: ($a$, $c$) moderate negative wind shear, ($b$, $d$) strong positive wind shear. ($a$, $b$) Rotor without wall effects, centered vertically in the domain. ($c$, $d$) Rotor with wall effects, placed at the height of the \iea reference turbine ($\zhub/D=0.625$). The boundary of an energy tube intersecting the rotor is shown with black solid lines; gray dashed lines denote an energy tube for the same shear strength but opposite wall condition. The inflow is shown in the blue quiver plot, left, scaled such that $x/D =  2\Uhub$.}
    \label{fig:shear_streamlines}
\end{figure}

To conclude this section, we examine the mechanism for the selective amplification of the lower lobe of the velocity perturbation due to the proximity of the wall. 
The wall in all of the synthetic inflow simulations is a slip wall, enforcing zero shear stress and zero vertical velocity ($w = 0$).
The presence of the wall modifies the bypass flow below the turbine. 
Therefore, in \cref{fig:shear_streamlines}, we visualize contours of the vertical velocity field to seek an explanation for a decrease in rotor power resulting from the interaction of positive shear and wall proximity (and vice versa for negative shear). 

\Cref{fig:shear_streamlines} shows that a key difference between the wall-free flows ($a$, $b$) and the near-wall flows ($c$, $d$) is the suppression of velocity flowing underneath the turbine ($\bar{w} < 0$). 
That is, regardless of $\shear$, the presence of the wall reduces bypass flow underneath the rotor bottom-tip due to the no-penetration condition. 
As a result, more of the near-wall flow that would otherwise be redirected underneath the rotor instead passes through the rotor and affects $\ud$ and therefore thrust and power. 
When the near-wall flow is slower than $\Uhub$ (in the case of positive shear), the effect on the rotor is a decrease in $\ud$ because low-velocity flow is forced through the rotor plane and displaces higher-velocity flow aloft, enhancing the negative lobe of the velocity perturbation (\cref{fig:shear_ud_contour}($g$, $h$)). 
By contrast, when the near-wall velocity is higher than $\Uhub$ (negative shear), the wall increases $u_d$ by forcing more high-velocity flow through the rotor that would have otherwise been redirected underneath the disk and displaces low-velocity fluid aloft. 

A complementary interpretation is revealed by analyzing from where the kinetic energy that is ultimately extracted by the actuator disk originated in the freestream flow. 
We answer this question by computing an energy tube \citep{meyers_flow_2013} seeded at the actuator disk plane at $R_s = R$. 
Energy lines for each flow in \cref{fig:shear_streamlines} are shown in solid black lines, and, for comparison, energy lines for the same shear strength $\alpha_s$ but opposite wall conditions are shown in the lighter dashed streamlines. 
The mechanism for shear to modify induction through a change in $\ud$ is most apparent in the $\shear = 0.8$ inflow (\cref{fig:shear_streamlines}($b, d$)). 
As shear increases, the local thrust $c_T$ at the bottom rotor tip is selectively amplified due to decreasing $\UBmag^2(z=\zhub-R)$. 
The amplified local thrust (which exceeds $c_T > 1$, see \cref{fig:shear_inflow_ct_a}($c$)) creates a high-induction region that strongly redirects the flow below the rotor downstream in the wall-free case. 
Looking upstream, the consequence of the high-induction region at the bottom tip is that the energy tube upstream of the turbine originates from slightly aloft of $\zhub$, where the flow contains more energy. 
When we impose a wall effect with a no-penetration condition, the low-velocity flow is forced through the actuator disk rather than redirected beneath it due to suppressed $\bar{w}$ below hub height. 
As a result, the energy tube in the near-wall case originates closer to the wall, where the flow contains lower energy. 
Downstream, the energy tube in the wall-bounded case presses against the wall, whereas the energy tube in the wall-free case would have clipped through the ground (dashed line in \cref{fig:shear_streamlines}($d$)).
These trends are all reversed in the negative shear case ($\alpha_s < 0$), although the effect is more subtle due to the weaker shear magnitude. 

To summarize, we use simulations of idealized sheared inflows with and without freestream turbulence to study the interaction between an actuator disk and inflow shear. 
We find that two effects, which linearly superimpose up to $\shear \lesssim 0.4$, cause a decrease in turbine efficiency ($C_P$) for positive power law shear exponents, and vice versa for negative power law shear exponents. 
The first effect is a geometric change in rotor-equivalent wind speed, $\rews$. 
For an actuator disk, rotor power is cubically dependent on $\rews$, and $\rews$ monotonically decreases with increasing $\powerlaw$ through a simple geometric relationship. 
The second effect of the wind shear is to create a non-uniform thrust coefficient and induced velocity across the rotor, which modifies the induction of the rotor and does not necessarily average to equal the uniform flow value. 
Furthermore, the presence of the wall modulates the non-uniformity in the induced velocity at the rotor by forcing the near-wall flow through the rotor extent. 
In total, the presence of the wall amplifies the inductive effect of shear, which may enhance (in negative shear) or degrade (in positive shear) the power production of the rotor.

\section{Wind veer effects in idealized inflows}
\label{sec:veer_ideal}
In this section, we will use LES data from idealized, synthetic inflow simulations to investigate the effects of wind veer on rotor aerodynamics. 
We explore the effect of veer on the disk velocity (induction), thrust, and power of an actuator disk in \cref{ssec:synthetic_veer_rotor}. 
Then, we analyze the mean kinetic energy (MKE) budget to explain the modulation of rotor induction and power in veered inflow conditions in \cref{ssec:synthetic_veer_dynamics}. 
Finally, the physical mechanism affecting rotor induction under veered inflows is presented in \cref{ssec:synthetic_veer_vorticity} using the streamwise vorticity dynamics.

\subsection{Effect of veer on rotor quantities}
\label{ssec:synthetic_veer_rotor}

We begin the investigation of the influence of wind veer in isolation on turbine power production by prescribing idealized, parameterized inflow profiles of wind speed magnitude $\UBmag(z)$ and wind direction $\wdB(z) = \arctan(\vB/\uB)$. 
The simplest profiles that vary the effects of veer would be a linear profile $\wdB(z)$. 
However, far from the disk, a linear wind direction profile $\wdB(z)$ could cause reverse flow (negative wind speed) if the wind direction exceeds $\pi/2$. 
Therefore, we use a $\tanh$ profile, similar to \cref{ssec:shear_rotor}, to prescribe a quasi-linear wind direction profile in the rotor region that smoothly transitions to a constant value far from the rotor in large veer magnitudes \citep{heck_unraveling_2025}:
\begin{equation}
        \label{eq:tanh_veer}
        \wdB(z) = (1-\epsilon) \tanh \left( -\frac{\veer (z - \zhub)}{D(1-\epsilon)}\right).
\end{equation}
In \cref{eq:tanh_veer}, wind direction profiles are approximately linear near the rotor with the slope $\veer (z/D)$ in radians per rotor diameter. 
The parameter $\epsilon = 0.2$ (same as in \cref{eq:tanh_shear}) controls the threshold on the maximum and minimum wind angle (in radians). 
Note that a negative sign is included in \cref{eq:tanh_veer}, such that clockwise wind turning with increasing $z$ corresponds with positive veer ($\veer > 0$), following the Ekman spiral in the northern hemisphere. 

\begin{figure}[htb]
    \centering
    \includegraphics[width=\linewidth]{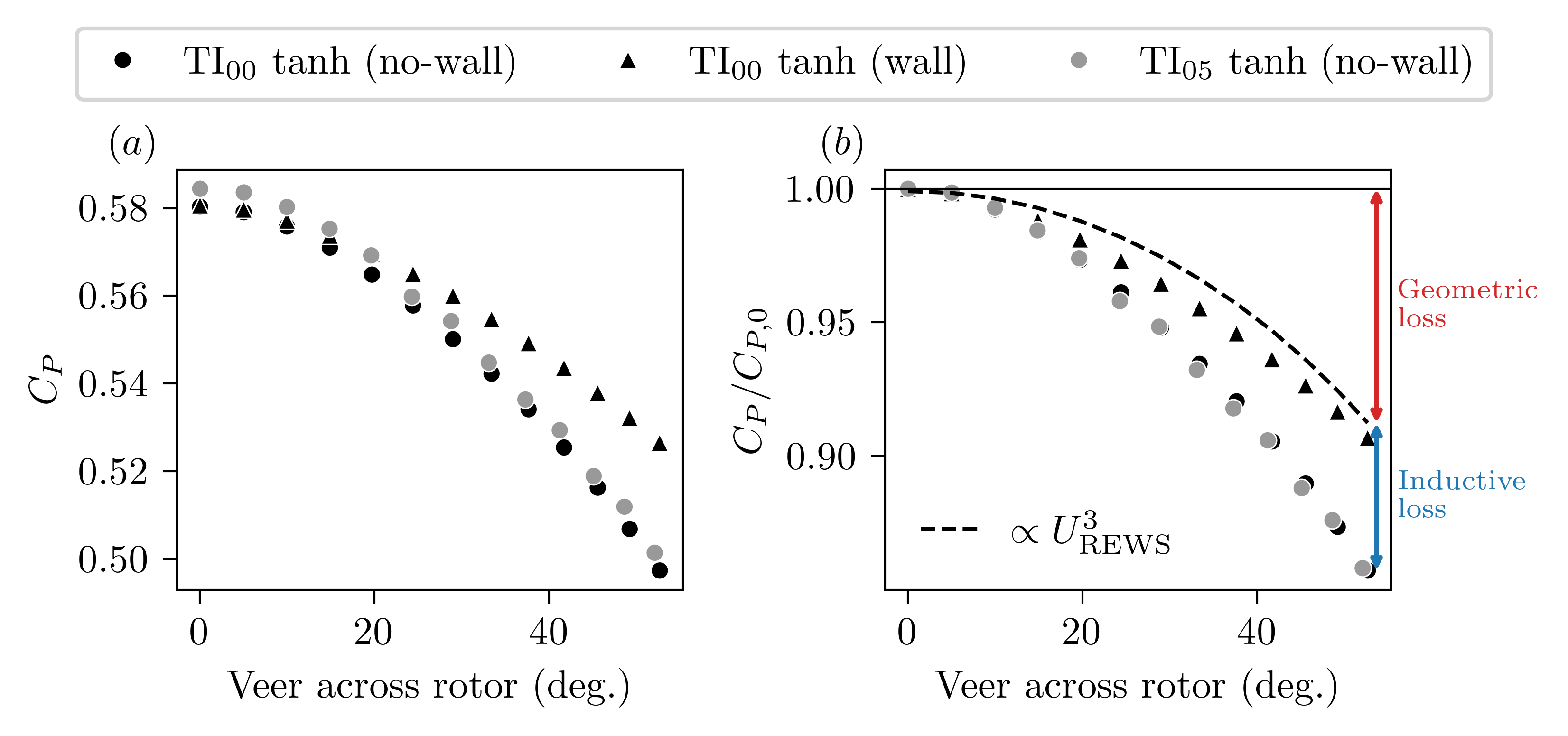}
    \caption{($a$) Actuator disk power coefficient $C_P = 2P/(\rho A_d \Uhub^3 )$ as a function of veer across the rotor ($\veer$). ($b$) Power coefficient, normalized to uniform inflow conditions without wind veer, as a function of veer across the rotor ($\veer$). Geometric losses $\propto \rews^3$ are shown in the dashed black line.}
    \label{fig:veer_cp_power}
\end{figure}

Similar to the investigation of wind shear (\cref{ssec:shear_rotor}), we perform three sets of idealized inflow simulations using the parameterization of wind veer in \cref{eq:tanh_veer}. 
All simulations sweep a range of wind veer $\veer \in [0, \SI{60}{\degree}]$ in increments of $\SI{5}{\degree}$.  
Because we simulate an irrotational actuator disk here (see \cref{ssec:padeops}), the effect of positive and negative veer is exactly symmetric. 
Hence, we only show $\veer \geq 0$. 
A rotational disk would break the symmetry of shear and veer and should be explored in future work. 
The first set of simulations uses laminar inflow, $\tanh$ wind direction profiles without wall effects. 
In these simulations, labeled ``$\ti_{00}$ $\tanh$ (no-wall),''  the actuator disk is centered vertically in the domain such that $\zhub = L_z / 2$. 
The second set of simulations, labeled ``$\ti_{05}$ $\tanh$ (no-wall),'' uses the same mean inflow profiles and turbine location as the first set, but superimposes turbulent fluctuations of $\ti = 5\%$ to the inflow using the synthetic inflow method described in \citet{heck_unraveling_2025} and \cref{sssec:setup_turbulent}. 
Finally, the third set of simulations, labeled ``$\ti_{00}$ $\tanh$ (wall),'' moves the turbine hub height to $\zhub/D = 0.625$ such that the proximity of the slip wall, which imposes a no-penetration condition, may influence the rotor dynamics. 
For each simulation, we compute the actuator disk power coefficient, which we show in \cref{fig:veer_cp_power}. 



In \cref{fig:veer_cp_power}($a$), we observe a significant dependence of $C_P$ on wind veer. 
In all three sets of simulations, $C_P$ decreases with increasing wind veer magnitude. 
The efficiency loss due to wind veer, relative to efficiency in uniform inflow ($C_{P,0} = C_P(\veer = 0)$), is nearly identical between the $\ti = 0\%$ and $\ti = 5\%$ inflows (\cref{fig:veer_cp_power}($b$)). 
Additionally, the turbines located far from the wall ($\zhub = L_z/2$) lose more power as the wind veer magnitude increases than when the turbine is located close to the wall ($\zhub/D = 0.625$). 

One mechanism through which wind veer lowers the power extracted by the actuator disk is by reducing the mass and kinetic energy flux through the rotor face due to local rotor-inflow misalignment. 
Using the rotor-equivalent wind speed definition (\cref{eq:rews}), the decrease in power due to the drop in mass flux (i.e., $(\rews / \Uhub)^3$) is shown by the dashed black line in \cref{fig:veer_cp_power}($b$). 
Similar to in sheared inflow, this source of power loss is due to geometry alone, which we call a \textit{geometric loss}. 
If the power loss due to increasing wind veer was explained solely by geometry (decreasing $\rews$), then $P(\veer)$ would decrease as $P_0 \cdot (\rews / \Uhub)^3$. 
However, from \cref{fig:veer_cp_power}, it is clear that the power drops faster than $(\rews / \Uhub)^3$, particularly when the turbine is located far from the slip wall. 
Therefore, in addition to a geometric power loss associated with wind veer, there must also be an additional \textit{inductive loss}: a physical mechanism in which wind veer affects the dynamics of the near-wake to further decrease the extractable energy at the rotor beyond the losses from geometry. 

\begin{figure}[htb]
    \centering
    \includegraphics[width=\linewidth]{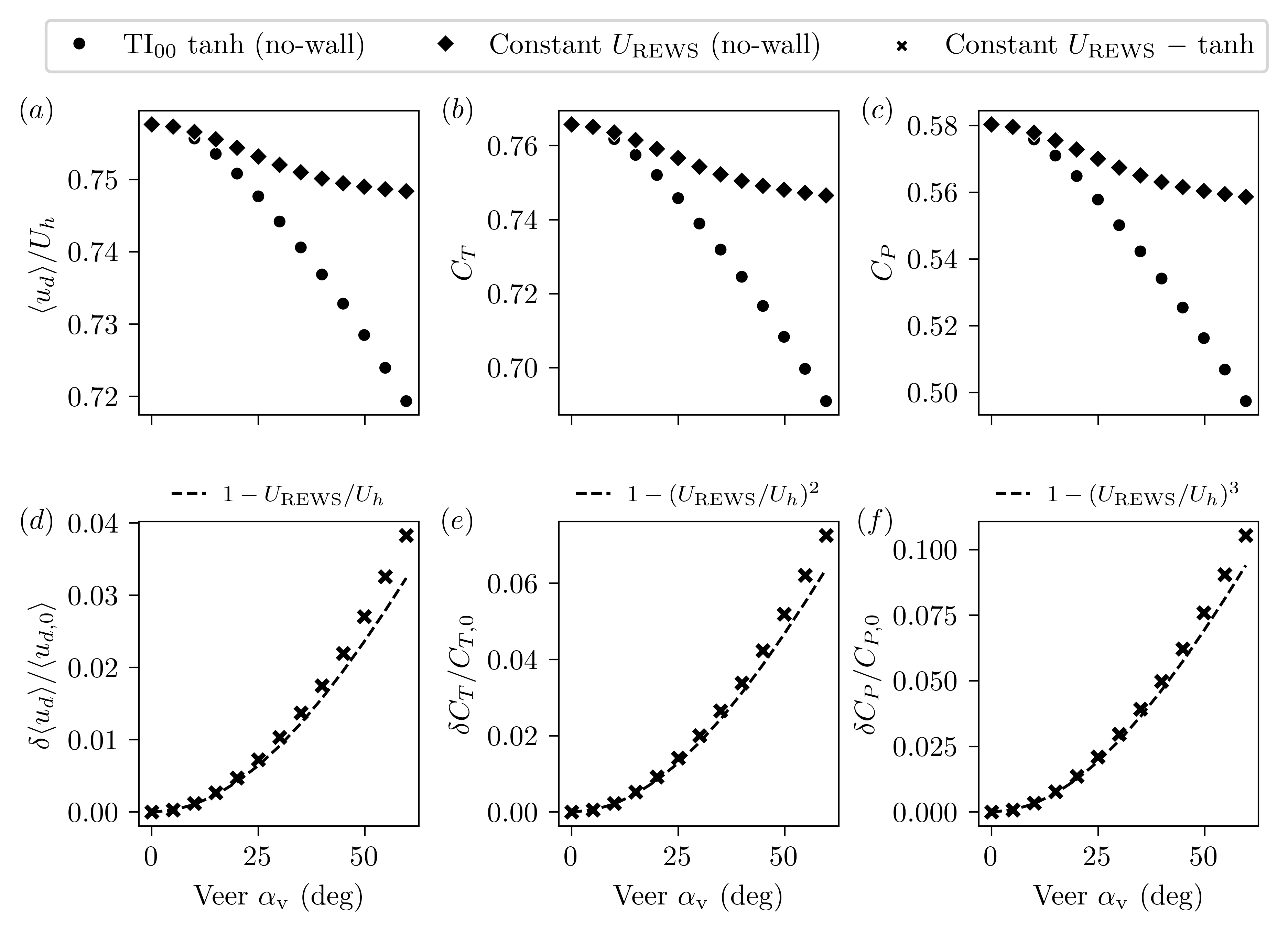}
    \caption{($a$) Disk-averaged velocity, ($b$) thrust coefficient, and ($c$) power coefficient for the constant $\UBmag$ $\tanh$ and constant $\rews$ inflow as a function of wind veer, $\veer$. ($d$-$f$) Normalized differences in disk velocity, thrust coefficient, and power coefficient between the $\tanh$ and constant $\rews$ inflows, respectively. The dashed black line in ($d$-$f$) represents the geometric effect of the changing rotor-equivalent wind speed $\rews$. }
    \label{fig:ud_veer_rews}
\end{figure}

To investigate the source of the inductive loss, we perform a fourth sweep of veer-only LES cases where $\rews$ is fixed as veer changes.
Here, we again set $\zhub = L_z/2$ to explore the effects of veer without wall effects present. 
Fixing $\rews$ isolates the effect of the inductive losses from geometric losses. 
We accomplish this by prescribing a custom inflow profile: 
\begin{subequations}
    \label{eq:const_rews_Uphi}
    \begin{align}
        \refstepcounter{equation}
        \uB(z) &= \Uhub;  &
        \vB(z) &= \UBmag(z) \cdot \sin(\wdB(z)),  \tag{\theequation, b}
    \end{align}  
\end{subequations}
where $\UBmag(z)$ and $\wdB(z)$ are given by \cref{eq:tanh_shear} and \cref{eq:tanh_veer}, respectively. 
Note that the lateral inflow velocity profile, $\vB(z)$, is the same in these new simulations as in \cref{fig:veer_cp_power}, and the streamwise flow is the same as the uniform inflow case. 
That is, the custom profile in \cref{eq:const_rews_Uphi} has the same mass flux through the rotor, but increased mechanical energy flux due to the addition of the lateral flow. 
If $\uB(z) = \UBmag(z) \cdot \cos(\wdB(z))$, then these profiles would exactly match \cref{eq:tanh_veer}, simply written in terms of $\uB$ and $\vB$ rather than $\UBmag$ and $\wdB$. 
Another consequence of the augmented inflow profiles is that the freestream streamwise vorticity $\wxB \equiv -\partial \vB/\partial z$ is kept constant, which will be useful later in \cref{ssec:synthetic_veer_vorticity} when analyzing vorticity dynamics. 
The disk-averaged velocity, thrust coefficient, and power coefficient for the ``$\ti_{00}$ $\tanh$ (no-wall)'' sweep from \cref{fig:veer_cp_power} and the augmented profiles (labeled ``Constant $\rews$ (no-wall)'') are shown in \cref{fig:ud_veer_rews} ($a$-$c$, respectively). 

In the Constant $\rews$ inflows, the kinetic energy flux in the freestream flow actually increases with increasing veer through energy in the lateral velocity, $(v^B)^2$. 
However, despite the increased kinetic energy flux available in the freestream flow, the Constant $\rews$ simulations still show a drop in disk velocity, thrust, and power as a function of increasing inflow veer. 
This is a rather counterintuitive result for which the physical mechanisms are investigated in \cref{ssec:synthetic_veer_dynamics}.
The magnitude of the drop in all variables is smaller than the $\tanh$ inflows because there is no geometric loss due to wind veer in the Constant $\rews$ inflows by construction. 
Furthermore, \cref{fig:ud_veer_rews}($d$) shows that the difference in $\ud$ between the $\tanh$ (no-wall) and Constant $\rews$ (no-wall) inflow profiles is almost exactly $\rews/\Uhub$, suggesting that the geometric and aerodynamic effects of veer can be modeled separately and their effects can be superimposed. 
Likewise, the difference in $C_T$ and $C_P$ between the constant $\rews$ and constant $\UBmag$ simulations is shown to track with $(\rews / \Uhub)^2$ and $(\rews / \Uhub)^3$, respectively (see \cref{fig:ud_veer_rews}($e, f$)). 
Finally, it is worth noting that both the geometric and inductive effects of wind veer introduce power losses of approximately equal orders of magnitude when wall effects are absent. 
That is, accounting only for the geometric loss through $\rews$ when estimating the power produced in veered atmospheric inflows may overestimate turbine power production by a significant margin. 
The degree to which the geometric loss or the inductive loss is larger in magnitude depends on other aspects, such as the thrust coefficient of the turbine, which should be investigated in future work.

\subsection{Aerodynamic losses due to wind veer}
\label{ssec:synthetic_veer_dynamics}
To investigate the source of the aerodynamic loss, we introduce the MKE budget. 
The mean kinetic energy $\bar{e} \equiv \tfrac 12 \bar{u}_i \bar{u}_i$ is defined based on the time-averaged velocity field $\bar{u}_i$ using a Reynolds decomposition, i.e., $u_i = \bar{u}_i + u_i'$, where $(\cdot)'$ denotes a zero-mean fluctuation. 
We derive a prognostic equation for the quasi-steady mean kinetic energy by multiplying the Reynolds-averaged Navier--Stokes equations by the mean velocity vector $\bar{u}_i$ and summing over $i$-components. 
The result is
\begin{equation}
\label{eq:mke_full}
\newcommand*{\vp}{ \vphantom{\frac{\overline{u_j'}}{x_j}} }
\underbrace{\vp \mathcal{R}}_\text{Residual} = 
\underbrace{\vp -\frac{\partial \bar{e}\bar{u}_j}{\partial x_j}}_{\substack{\text{Advection} \\ (\mathcal{A})}}
\underbrace{\vp + \overline{u_i' u_j'} \frac{\partial \bar{u}_i}{\partial x_j}}_{\substack{\text{Shear} \\ (\mathcal{S})}}
\underbrace{\vp -\frac{\partial \bar{u}_i\overline{u_i' u_j'}}{\partial x_j}}_{\substack{\text{Transport} \\ (\mathcal{T})}}
\underbrace{\vp -\frac{\partial \bar{p}\bar{u}_j}{\partial x_j}}_{\substack{\text{Pressure} \\ \text{work}~(\Pi)}}
\underbrace{\vp - \frac{2}{\Ro} \varepsilon_{ijk} \Omega_j G_k \bar{u}_i}_{\substack{\text{Geostrophic}~(\Pi_\infty)}}
\underbrace{\vp + \frac{\bar{w}}{Fr^2 \theta_0} (\bar{\theta} - \theta_0)}_{\text{Buoyancy}~(\mathcal{B})}
\underbrace{\vp -\bar{u}_i \frac{\partial \bar{\tau}_{ij}}{\partial x_j}}_{\substack{\text{Dissipation} \\ (\epsilon)}}
\underbrace{\vp + \bar{f}_i \bar{u}_i}_{\mathcal{P}_\text{AD}}.
\end{equation}
Note that summing over all $i, j$ components in the Coriolis term results in zero \citep[c.f.,][]{stull_introduction_1988}, therefore it does not appear \cref{eq:mke_full}. 
We set the left-hand side of \cref{eq:mke_full} to a residual term $\mathcal{R}$ and evaluate each right-hand side term using LES data. 

We are interested in the contribution of each term in the MKE budget, specifically in the near-wake region. 
Under the assumptions of traditional momentum theory \citep{burton_wind_2011}, the near-wake region is characterized by inviscid flow where turbulent transport, shear, buoyancy, and dissipation are small.
Furthermore, in momentum theory, the pressure field $\bar{p}$ fully recovers to freestream conditions, and therefore the net pressure work is zero. 
Therefore, control volume analysis of the MKE budget for flow around an actuator disk under traditional assumptions yields a balance of advection (energy fluxes) with a sink of power due to the actuator disk. 

In ABL inflow conditions, several assumptions of classical momentum theory are violated: inflow is non-uniform (both in speed and direction), freestream turbulence is present, and the wake pressure may not recover fully to freestream conditions. 
Recent studies have challenged and relaxed these assumptions of traditional momentum theory in quasi-steady, turbulent inflows. 
For example, heavily loaded actuator disks ($\an \gtrsim 0.37$, \citep{wilson_applied_1974}) create a persistent pressure drop in the wake called base suction \citep{steiros_drag_2018}.
The extraction of momentum and energy from the wake pressure field by the actuator disk increases the thrust and power coefficients above predictions from classical momentum theory at high induction factors $\an$ \citep{steiros_drag_2018, liew_unified_2024}. 
Furthermore, several studies have challenged inviscid assumptions of classical momentum theory, arguing that increased turbulent fluxes modify wake velocities and rotor performance \citep{revaz_effect_2025, bastankhah_generalised_2026}. 
Analyzing the MKE budget reveals dynamically important departures from traditional assumptions that may cause classical theory to break down. 

\begin{figure}
    \centering
    \includegraphics[width=1\linewidth]{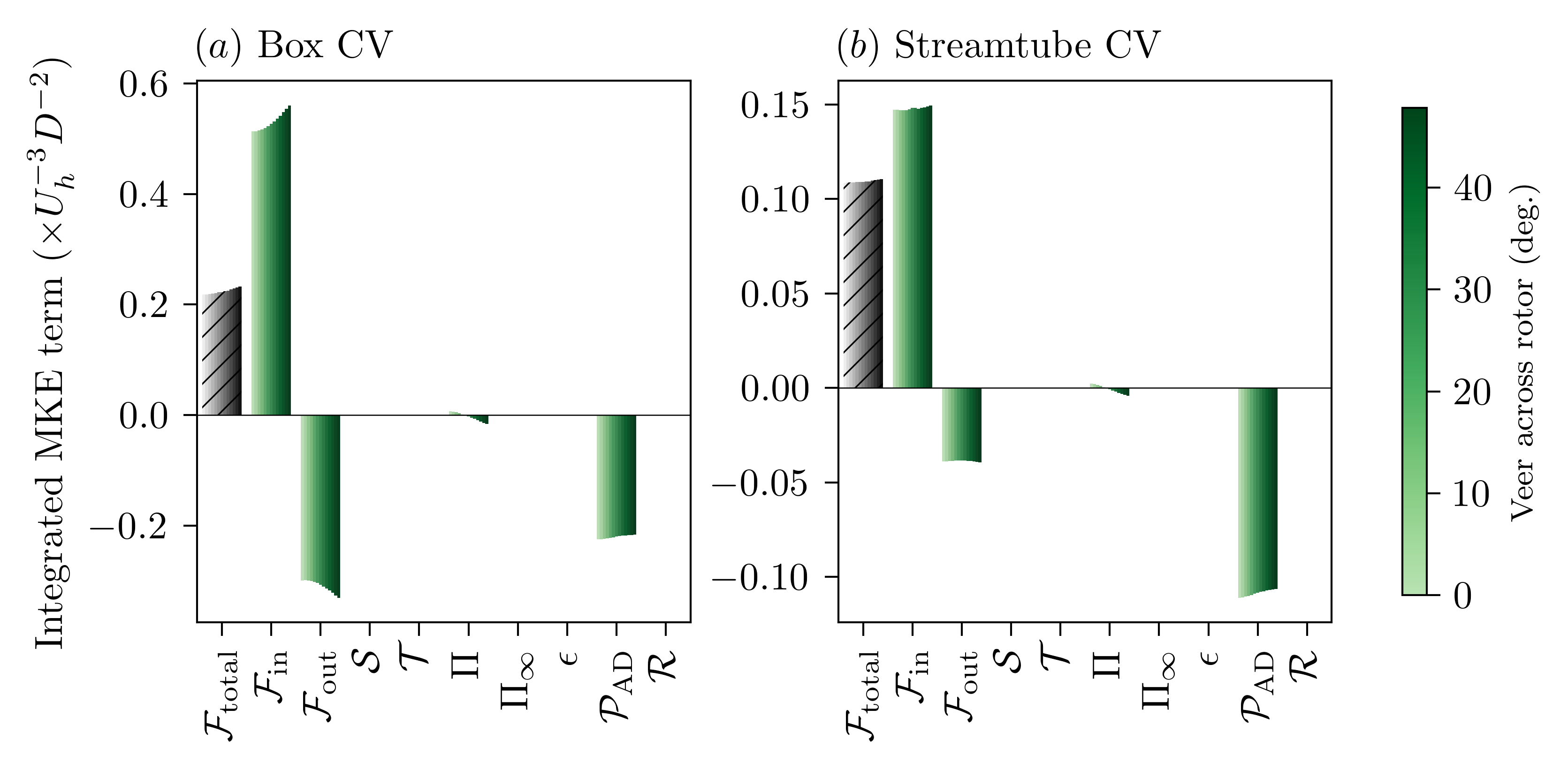}
    \caption{Volume-integrated mean kinetic energy budget around the constant $\rews$ flow for increasing wind veer $\veer$ from $0$ to $60^\circ$. ($a$) Box control volume spanning $y, (z-\zhub) \in [-R, R]$. ($b$) Streamtube control volume with $R_s/R = 0.7$. Control volumes extend from the domain inlet ($x/D = -5$) to the end of the near-wake, $x=x_0$. For both control volumes, $\mathcal{A}_\text{total} = \mathcal{A}_\text{in} + \mathcal{A}_\text{out}$.}
    \label{fig:mke_budget_rews}
\end{figure}

Using LES data integrated in a control volume around the turbine, we evaluate the volume-integrated contribution of each term in the MKE budget. 
To isolate the source of the inductive loss, we perform the MKE budget analysis using the Constant $\rews$ (no-wall) simulation set. 
The MKE budget associated with two control volumes is shown in \cref{fig:mke_budget_rews}. 
First, in \cref{fig:mke_budget_rews}($a$), a streamtube control volume (CV) is used, where the streamtube is seeded at $R_s / R = 0.7$. 
Next, in \cref{fig:mke_budget_rews}($b$), MKE budget terms are integrated within a box CV for $y \in [-R, R]$ and $z-\zhub \in [-R, R]$. 
In both CVs, streamwise extent begins at the inlet of the domain ($x/D = -5$) and ends at the near-wake length $x_0$, defined here as the $x$-location where the streamtube-averaged velocity $\langle u \rangle$ is minimum \citep{shapiro_modelling_2018, liew_unified_2024}. 
The qualitative trends in the MKE budget are not sensitive to $R_s < R$, nor do the qualitative trends change if the streamwise extent of the control volume is fixed (e.g., $x/D \in [-5, 5]$). 
We also parse the net advection of MKE ($\mathcal{A}_\text{total}$) into the MKE flux through the inlet of the CV ($\mathcal{A}_\text{in}$) and the MKE flux through all other faces of the CV (sides and outlet, collectively $\mathcal{A}_\text{out}$). 

Using the Constant $\rews$ (no-wall) inflow, the MKE available to the wind turbine increases as wind veer magnitude increases ($\mathcal{A}_\text{in}$ increases with increasing veer magnitude). 
This is because wind veer contains additional energy in the lateral flow, $(\vB)^2$, perpendicular to the rotor disk, while the velocity normal to the disk is intentionally held constant. 
The increase in $\mathcal{A}_\text{in}$ is best shown in the box CV (\cref{fig:mke_budget_rews}($a$)) because the inlet area $A_\text{in} = D^2$ is held fixed. 
The increase in $\mathcal{A}_\text{in}$ is less noticeable for the streamtube CV (\cref{fig:mke_budget_rews}($b$)) because the inlet area $A_\text{in}$ changes in size as a function of veer due to changes in $\ud$ to conserve mass flux. 
Nonetheless, in both CVs, the actuator disk sink $\mathcal{P}_\text{AD}$ (aerodynamic turbine power) decreases in magnitude with increasing wind veer despite the increase in available MKE.

As the veer magnitude increases, the extracted energy $\mathcal{P}_\text{AD}$ decreases despite the increase in available MKE because the pressure-work term ($\Pi$) takes energy from the actuator disk. 
The dynamic contribution of the pressure-work term is a clear departure from classical momentum theory, which assumes that pressure recovers to freestream conditions in the wake and therefore does no work on the CV.
Applying the divergence theorem and continuity to the streamtube-integrated MKE budget, the pressure work term can be written as $\Pi = A_d \ud  (\langle p_\text{in} \rangle - \langle p_\text{out} \rangle)$, where $A_d \ud$ is the volumetric flow through the rotor and $\langle p_\text{in} \rangle$ and $\langle p_\text{out}\rangle$ are the streamtube-averaged pressures at the inlet and outlet of the CV, respectively. 
As $A_d \ud > 0$, the pressure work term decreases with increasing $p_\text{out}$. 
One can interpret the increase in $p_\text{out}$ as the formation of an adverse pressure gradient, which develops due to the interaction between the wind turbine wake and the freestream veer. 
As a result, kinetic energy is stored in $p_\text{out}$ that cannot be harnessed by the turbine, sapping away energy available to the actuator disk ($\mathcal{P}_\text{AD}$) \citep[c.f.][]{dar_experimental_2023}. 

\begin{figure}
    \centering
    \includegraphics[width=1\linewidth]{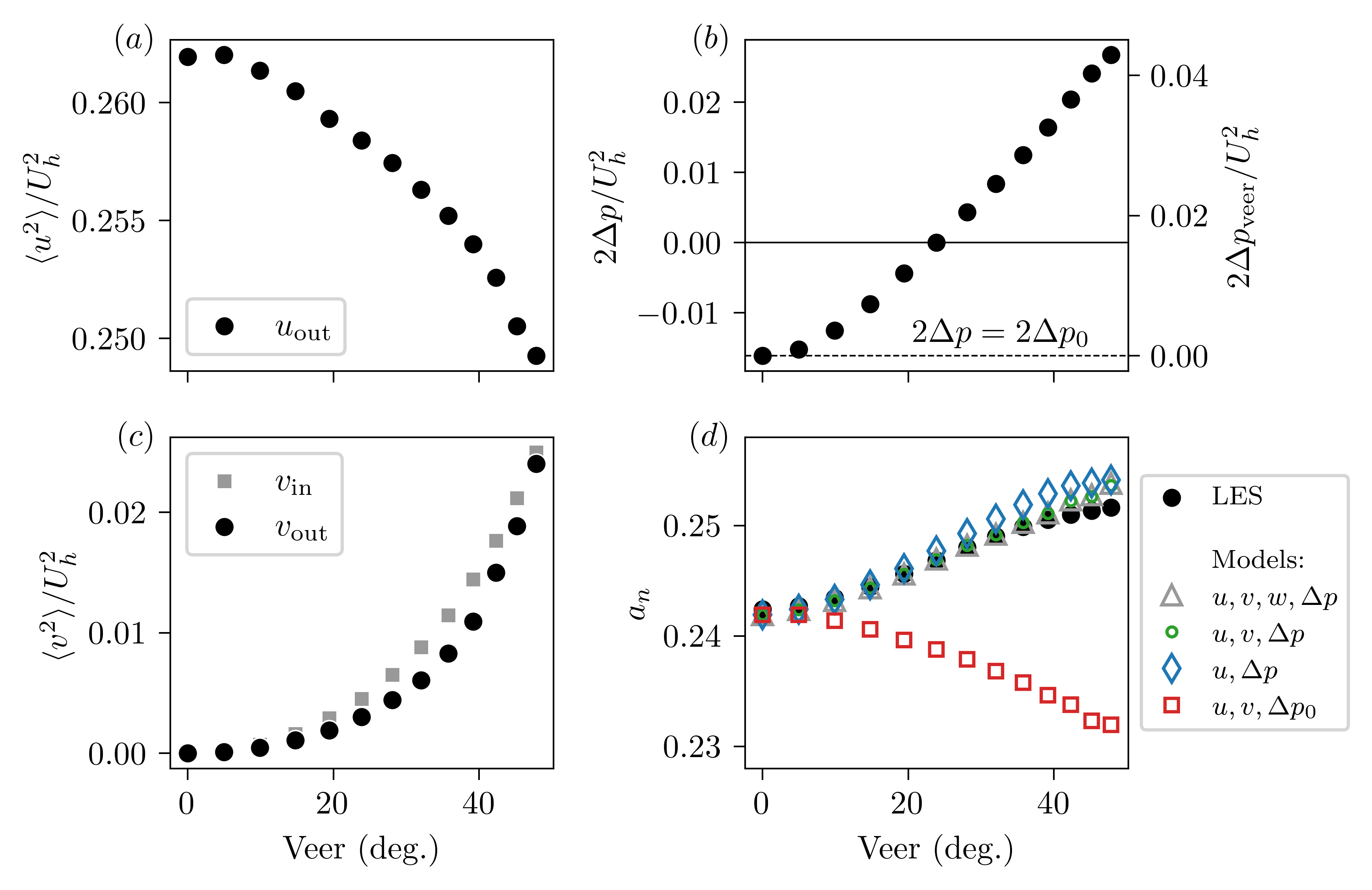}
    \caption{Streamtube-averaged near-wake quantities and rotor induction, measured in LES, as a function of inflow veer, showing ($a$) outlet velocity $\langle u_\text{out} \rangle$, ($b$) pressure drop $\Delta p$, ($c$) lateral velocity variance $\langle v^2 \rangle$, and ($d$) rotor-averaged induction factor $\an = 1 - \ud / \Uhub$. Closures to \cref{eq:an_veer_final} using LES data from ($a$-$c$) labeled as induction `Models'.}
    \label{fig:an_model_les}
\end{figure}

To quantify the importance of the pressure work ($\Pi$) on the induction $\an$, we can rearrange the MKE budget equation, keeping only the advection, pressure-work, and turbine forcing terms to solve for the disk velocity $\ud$.
Choosing a streamtube CV results in 
\begin{equation}
\label{eq:mke_integrated}
\iiint_\mathcal{V} -\bar{f}_i \bar{u}_i \, d\mathcal{V}
=
\iiint_\mathcal{V}
\left(
- \frac{\partial (\bar{e} + \bar{p})\bar{u}_j}{\partial x_j}
\right) \, d\mathcal{V}
\end{equation}
The definition of the actuator disk sink term is $-\vec{F}_T \cdot \vec{u}_d = +\tfrac 12 A_d \ud^3 C_T'$. 
Furthermore, we write the volume integral of the divergence of the flux of total mechanical energy $\bar{e} + \bar{p}$ using the divergence theorem. 
Because the CV is a streamtube, there is no mass flux out of the mantle (sides) of the CV by definition; the only flux of $\bar{e} + \bar{p}$ is through the inlet and outlet area. 
Furthermore, the flow variables $\bar{e}$ and $\bar{p}$ are approximately constant across the streamtube cross-section. 
As a result, we can write: 
\begin{equation}
\label{eq:an_veer_derivation_1}
\frac 12 A_d \ud^3 C_T' \eta_P
= 
u_\text{in} A_\text{in} \langle e_\text{in} + p_\text{in}\rangle - u_\text{out} A_\text{out} \langle e_\text{out} + p_\text{out}\rangle,
\end{equation}
where $\eta_P$ is the fraction of the actuator disk power sink encapsulated by the streamtube. 
Finally, we apply conservation of mass to cancel $u_\text{in} A_\text{in} = u_\text{out} A_\text{out} = \ud A_d \eta_u$, where $\eta_u$ is the fraction of mass flux through the rotor encapsulated by the streamtube. 
For a sharp, ideal actuator disk, $\eta_P/\eta_u = 1$, but for a Gaussian-regularized actuator disk (see \cref{eq:adm_kernel}), this ratio depends on the streamtube radius $R_s$ and the Gaussian kernel width $\Delta$ (for our analysis, $R_s = 0.35$ and $\Delta = 2.5h$, and $\eta_P/\eta_u = 0.98$). 
Then, we rarrange \cref{eq:an_veer_derivation_1} as a function of $\an = 1 - \ud/U_h$ and substitute $\bar{e} = \tfrac 12 (\bar{u}^2 + \bar{v}^2 + \bar{w}^2)$, assuming zero vertical velocity in the inflow ($\wB(z) = 0$).
The final result, assuming an ideal actuator disk ($\eta_P/\eta_u = 1$) is
\begin{equation}
\label{eq:an_veer_final}
a_n = 1 - \sqrt{\frac{\langle u_\text{in}^2 + v_\text{in}^2 - u_\text{out}^2 - v_\text{out}^2 - w_\text{out}^2\rangle - 2\Delta p}{\Uhub^2 C_T'}},
\end{equation}
where we define $\Delta p \equiv \langle p_\text{out}\rangle - \langle p_\text{in} \rangle$. 
The streamtube-averaged velocities $\langle u^2 \rangle$ and $\langle v^2 \rangle$ are shown in \cref{fig:an_model_les}, alongside the wake pressure difference $\Delta p$ and the induction factor $\an$.
Furthermore, predictions of \cref{eq:an_veer_final} are shown in \cref{fig:an_model_les}($d$) using LES data as an \textit{a~priori} evaluation of the model equation. 

Beginning with \cref{fig:an_model_les}($a$), we see that there is a slight decrease in the streamwise velocity $\langle u_\text{out}^2 \rangle$ as a function of wind veer. 
However, the primary effect of wind veer is an increase in the wake pressure ($p_\text{out}$), which increases $\Delta p$, as shown in \cref{fig:an_model_les}($b$) (multiplied by two such that it is directly comparable to $\langle u_i^2 \rangle$ in \cref{eq:an_veer_final}).
Note that in zero veer conditions, $\Delta p(\veer=0) \equiv \Delta p_0 <0$ due to the incomplete pressure recovery in the near-wake known as base-suction \citep{steiros_drag_2018, liew_unified_2024}. 
Relative to the uniform inflow case, the wake pressure increases with increasing veer magnitude. 
We define the additional wake pressure contribution from wind veer as $\Delta p_\text{veer} = \Delta p - \Delta p_0$ in \cref{fig:an_model_les}($b$), right $y$-axis, to illustrate the veer-induced pressure effect. 
In \cref{fig:an_model_les}($c$), we can see that the energy contained in the lateral flow increases with increasing wind veer, but that the energy in the lateral flow is approximately conserved within the streamtube ($\langle v_\text{in}^2 \rangle \approx \langle v_\text{out}^2\rangle$), meaning the actuator disk extracts a very limited quantity of the lateral velocity (tangent to the disk), as expected.

Using the flow at the inlet and outlet of the streamtube, we can compute the turbine induction factor. 
In \cref{fig:an_model_les}($d$), we perform an \textit{a~priori} evaluation of \cref{eq:an_veer_final} using the LES quantities shown in \cref{fig:an_model_les}($a$-$c$) with various simplifications. 
For all inflows in \cref{eq:an_veer_final}, $u_\text{in}^2 = \Uhub^2$ by construction. 
The gray triangles include the energy from all components of the velocity field, while the green circles neglect the energy contained in the vertical velocities $\langle\bar{w}\rangle^2$, and the blue diamonds further neglect the energy in the lateral velocity $\langle \bar{v}^2 \rangle$. 
Additionally, we close \cref{eq:an_veer_final} with the pressure drop $\Delta p$ from the zero veer case in the red squares, i.e., $\Delta p = \Delta p_0$ is assumed not to depend on veer. 
Notably, without consideration of the veer-induced pressure gradient $\Delta p_\text{veer}$ (red squares), the trend in the induced velocity is opposite to the trend observed in LES (black circles). 
This indicates that the inclusion of $\Delta p_\text{veer}$ is critical to predicting the induction---and, by extension, the thrust and power---of an actuator disk in veered inflow. 
By contrast, closing \cref{eq:an_veer_final} with only $u^2$ (i.e., assuming $\langle v_\text{in}^2 \rangle \approx \langle v_\text{out}^2\rangle$) is nearly the same result as including the energy in the lateral velocity. 
Furthermore, the inclusion of $\langle w^2 \rangle$ does not noticeably change the model predictions. 
In total, \cref{fig:an_model_les} shows that the veer-induced wake pressure in veered inflow is critical to accurately modeling the induction, thrust, and power of an actuator disk. 
This increase in induction factor is associated with a reduction in the actuator disk power efficiency (coefficient of power).
In \cref{ssec:synthetic_veer_vorticity}, we explore the source of the adverse pressure gradient through the lens of vorticity dynamics. 


\subsection{Streamwise vorticity dynamics and wall effects}
\label{ssec:synthetic_veer_vorticity}

Here, we investigate the source of the veer-induced adverse pressure gradient through an analysis of the streamwise vorticity dynamics.
It is useful to reframe the additional lateral flow in terms of freestream vorticity in the streamwise direction: 
\begin{equation}
    \label{eq:wx_base}
    \omega_x = \pdv{w}{y} - \pdv{v}{z}.
\end{equation}
In the absence of vertical velocities in the inflow, the role of inflow wind veer is to introduce streamwise vorticity into the inflow: $\wxB = -\partial \vB/\partial z$.
In the northern hemisphere, positive streamwise vorticity indicates positive wind veer (rightward turning with increasing height). 

\begin{figure}[htb]
    \centering
    \includegraphics[width=\linewidth]{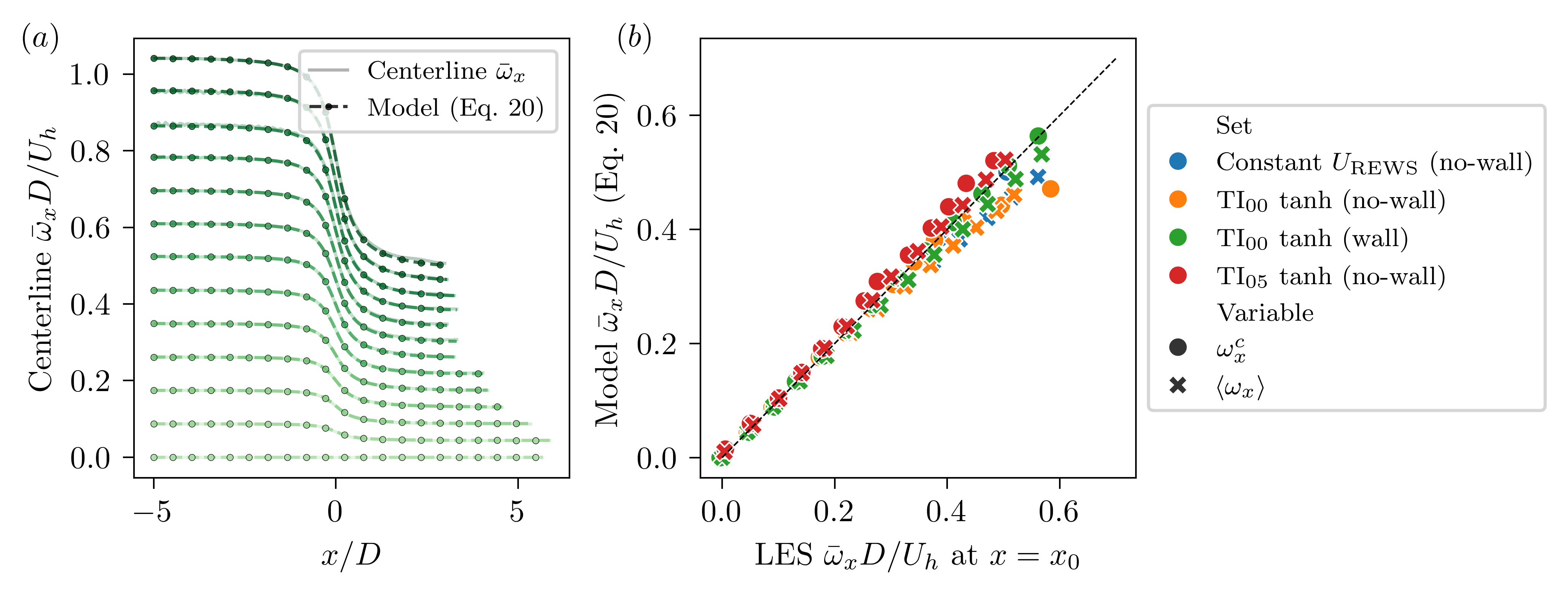}
    \caption{($a$) Streamwise evolution of centerline vorticity $\bar{\omega}_x^c$ as computed in LES (solid lines) and compared against the vorticity model \cref{eq:wx_model} (dashed lines). ($b$) Predictions of $\bar{\omega}_x^c$ and $\langle \bar{\omega}_x \rangle$ using \cref{eq:wx_model} at the end of the near-wake region ($x=x_0$) for idealized inflows.}
    \label{fig:wx_evolution}
\end{figure}

We investigate the connection between vorticity and wake pressure using the time-averaged streamwise vorticity budget, which can be derived by taking the curl of the Navier--Stokes equations (\cref{eq:NSE_filtered}) and Reynolds-averaging \citep[c.f.][]{heck_coriolis_2025}. 
%
%
Integrating the mean vorticity budget within a streamtube control volume from $x\in [-5D, x_0]$ (not shown) indicates that the dominant terms in the streamwise vorticity budget are advection balanced by vortex stretching in the streamwise direction: 
\begin{equation}
    \label{eq:wx_simplified}
    \bar{u} \pdv{\bar{\omega}_x}{x}
    = 
    \bar{\omega}_x \pdv{\bar{u}}{x}.
\end{equation}
The simplified vorticity budget equation (\cref{eq:wx_simplified}) has the following solution: 
\begin{equation}
    \label{eq:wx_model}
    \bar{\omega}_x(x) = \wxB \frac{\bar{u}(x)}{\uB}. 
\end{equation}
That is, any change in vorticity is linearly proportional to a change in streamwise velocity (induced by the rotor and wake).
An alternative interpretation of \cref{eq:wx_model} is that angular momentum within the streamtube must be conserved in the absence of external torques.
Therefore, the expansion of the streamtube cross-sectional area (due to the deceleration of streamwise velocity in the induction region and wake) must induce a proportional decrease in streamwise vorticity. 
In \cref{fig:wx_evolution}, we evaluate the wake vorticity equation (\cref{eq:wx_model}) for $\bar{\omega}_x$ \textit{a~priori} using $\bar{u}$ and $\wxB$ from LES, and compare the model equation to the LES value of $\bar{\omega}_x(x)$. 

As shown in \cref{fig:wx_evolution}($a$), the simple model equation is an excellent predictor of the evolution of centerline wake vorticity as a function of $x$. 
After the near-wake region ends ($x\geq x_0$), the model equation is no longer valid as turbulence terms are significant in the vorticity budget. 
Because \cref{eq:wx_model} is analytic, we can predict $\bar{\omega}_x$ at an arbitrary point $x \leq x_0$ downstream, given $\bar{u}(x)$.
A particular point of interest is the ending of the near-wake region, $x=x_0$. 
This is because if all of the wake variables are known at $x=x_0$ (e.g., $u_\text{out}$, $p_\text{out}$), we can compute the rotor-averaged induction factor (\cref{eq:an_veer_final} in \cref{ssec:synthetic_veer_dynamics}). 
Given the freestream vorticity $\wxB$ and centerline velocity $\bar{u}^c(x=x_0)$ or the streamtube-averaged velocity $\langle u_\text{out} \rangle$, \cref{eq:wx_model} accurately predicts the vorticity at the end of the near-wake, shown in \cref{fig:wx_evolution}($b$). 

\begin{figure}[htb]
    \centering
    \includegraphics[width=\linewidth]{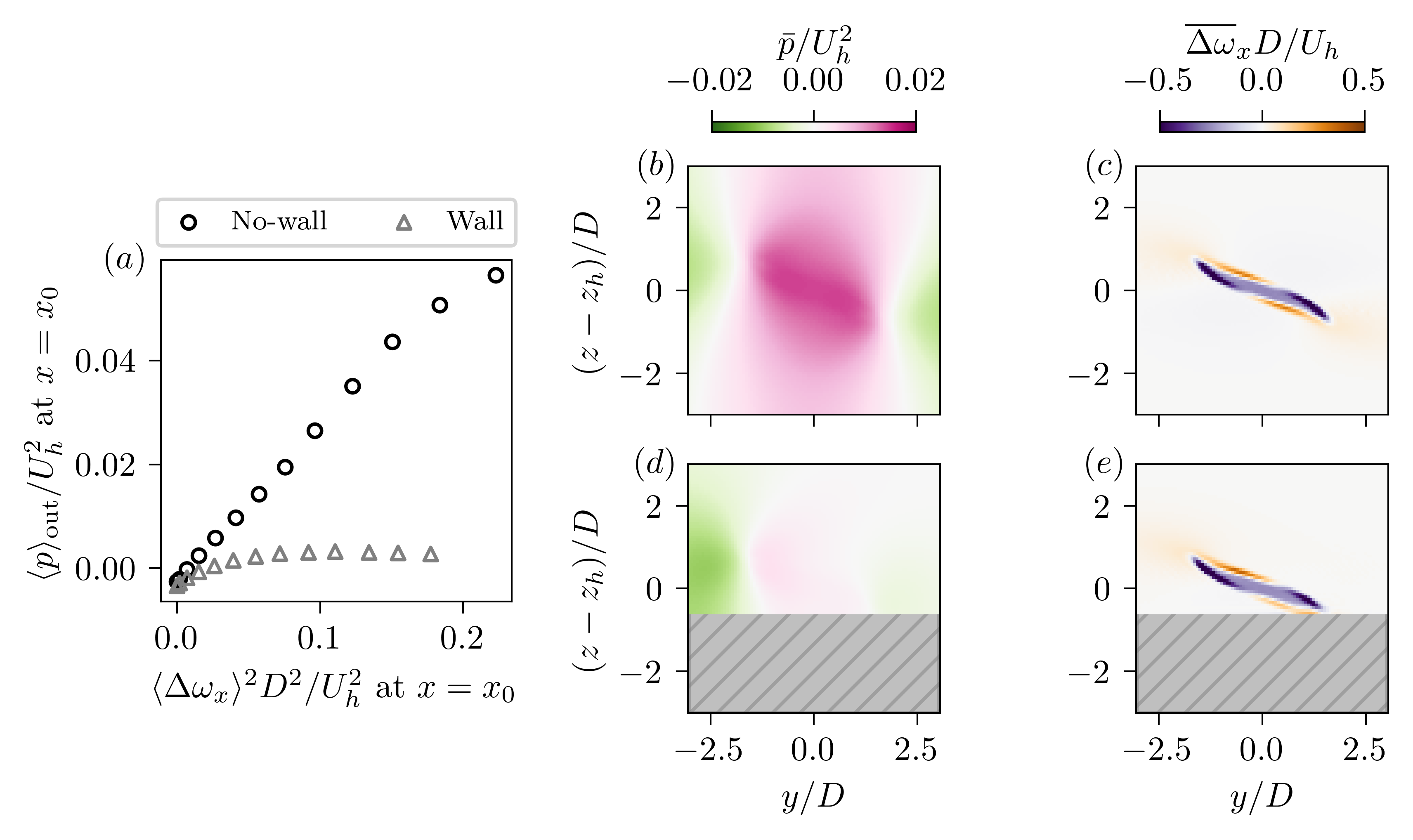}
    \caption{($a$) evaluation of the scaling between wake vorticity and pressure for $m=2$. ($b$, $c$): flow fields of pressure and wake vorticity, respectively, for $\zhub = L_z/2$ (no-wall case). ($d$, $e$) same as ($b$, $c$) for the near-wall case, $\zhub/D = 0.625$.}
    \label{fig:veer_pressure_wall}
\end{figure}

We now link the perturbation in vorticity to a perturbation in pressure. 
Here, we offer a simple approach to the pressure approximation through scaling assumptions between the veer-induced pressure and the veer-induced vorticity.
An alternative, more rigorous approach is described in \cref{appx:prss_veer} through a derivation of a pressure Poisson equation. 
The underlying assumption that we invoke is that a vortex core (i.e., a region of high vorticity relative to the background flow) is demarcated by a low-pressure center \citep[c.f.,][]{jeong_identification_1995}.
In the core of a wake in veered inflow, the vorticity drops relative to the background veer (\cref{fig:wx_evolution}($a$)).
Therefore, the wake is a region of low vorticity, relative to the background flow, and induces a high-pressure core relative to the freestream pressure implied by the freestream vorticity. 
Balancing the radial pressure gradient $\partial p/\partial r$ with the centrifugal velocity $u_\theta$ yields a Bernoulli-like scaling relation
\begin{equation}
    \Delta p_\text{veer} / \Uhub^2 \sim (\overline{\Delta \omega}_x D/\Uhub)^m, 
\end{equation}
where $\Delta \omega_x \equiv \omega_x - \wxB$ is the wake deficit vorticity. 
The empirical scaling exponent $m$ depends on the evolution of the vortex, such as core size, and also on non-local factors such as the strain of the velocity field. 
In \cref{fig:veer_pressure_wall}($a$), we find that $m = 2$ yields approximately linear scaling between vorticity and pressure when wall effects are absent. 

Contours of pressure and wake vorticity at $x=x_0$ are shown in \cref{fig:veer_pressure_wall}($b$, $c$) for the no-wall case at $\veer = \SI{30}{\degree}$ of veering across the rotor. 
A high-pressure core in the vicinity of the wake region is accompanied by a decrease in vorticity magnitude. 
However, because the pressure response from the vorticity field is non-local, the effect of the wall significantly modulates the wake pressure response in veered conditions, shown in \cref{fig:veer_pressure_wall}($d$, $e$). 
As a result, the wake pressure perturbation is severely suppressed, shown in \cref{fig:veer_pressure_wall}($d$), despite the streamwise wake vorticity field $\Delta \omega_x$ remaining mostly unchanged. 
The suppressed wake pressure response (and suppression of the veer-induced adverse pressure gradient) decreases the inductive losses of veer when the rotor is near the wall, explaining the driving flow mechanisms of the influence of wall proximity on the veer effects observed in \cref{fig:veer_cp_power}. 
While the influence of the wall may limit the induction loss in veered inflow, which we explore further in \cref{appx:wall_sweep}, we emphasize that in general, there is a non-negligible inductive effect induced by wind veer triggered by an adverse pressure gradient in the wake. 
The inductive response decreases the rotor power production and depends on factors including wall proximity and the thrust coefficient of the turbine. 

In summary, we have shown that there are two effects of wind veer on the induced velocity, thrust, and power of an actuator disk. 
The first effect is purely geometric: a lower velocity directed normal to the disk face decreases the flux of mass, momentum, and energy. 
We show that superimposed on the geometric effect is a second inductive effect which depends on the wall proximity, but does not depend on freestream turbulence. 
In the no-wall case, the inductive effect and geometric effect are approximately equal in magnitude for the thrust coefficient investigated here ($C_T' = 4/3$). 
This aerodynamic effect is driven by an adverse pressure gradient that forms in the wake due to the expansion and deceleration of the wake, which decreases the streamwise vorticity to conserve angular momentum. 
In idealized inflows of wind veer, the disk velocity, thrust, and power are a product of both the geometric and aerodynamic effects of wind veer.

\section{Consequences of shear and veer for turbine modeling and performance}
\label{sec:results_2}

Sections~\ref{sec:shear_idealized} and \ref{sec:veer_ideal} illustrate how wind shear, wind veer, wall proximity, and other properties in the atmospheric boundary layer affect the rotor performance of an actuator disk. 
Shear and veer both modify rotor performance by changing the rotor at the disk through two effects: (1) a geometric change in the rotor-equivalent wind speed $\rews$, which is straightforward to compute, and (2) an inductive (aerodynamic) response that modulates the relationship between rotor thrust and induced velocity. 
In this section, we first show how these combined effects lead to changes in turbine efficiency in ABL inflows in \cref{ssec:results_sbl_cumulative}. 
Then, in \cref{ssec:results_momentum_theory}, shear and veer-induced deviations from momentum theory, ubiquitously used in predictive rotor modeling, are investigated.

\subsection{Cumulative effects of shear and veer in atmospheric boundary layers}
\label{ssec:results_sbl_cumulative}

Throughout \cref{sec:shear_idealized} and \ref{sec:veer_ideal}, we have intentionally studied the inductive and geometric effects of shear and veer in isolation to more clearly understand the mechanisms independently. 
As a final step to the investigation of wind shear and veer effects on the power produced by an actuator disk, we study the superposition of wind shear (through the streamwise velocity, $\uB(z)$) and the wind veer (through the lateral velocity, $\vB(z)$). 
Specifically, we will investigate the extent to which the independent effects of shear and veer can be superimposed, using the ABL simulations presented in \cref{sec:results_sbl_power} for comparison. 

We can systematically deconstruct the interaction between the rotor and ABL inflow, simulated using the concurrent-precursor method in \cref{sec:results_sbl_power}, using the synthetic inflow method. 
These new simulations use the exact inflow profiles from the ABL (see \cref{fig:sbl_inflows}) as the mean inflow targets. 
We break down the flow into four sets of simulations of varying complexity, shown in \cref{tab:sbl_synthetic}. 
All four sets of synthetic inflow simulations impose properties of the ABL inflow from the concurrent-precursor simulations. 
However, by only imposing profiles of $\uB(z)$ or $\vB(z)$ (for the ``$\ti_{00}$ shear'' and ``$\ti_{00}$ veer'' sets, respectively), we can create engineered inflows that explicitly separate the effects of shear and veer within the ABL profiles.  
In all sets of simulations, we perform a synthetic inflow simulation for each cooling rate $C_r \in [0, 0.1, \ldots 0.5]$ from the concurrent-precursor simulations, totaling 24 cases. 
For all simulations, the turbine is placed at $\zhub/D = 0.625$, matching the concurrent-precursor simulations, and the bottom boundary is a slip wall. 
An additional simulation is performed using uniform, zero freestream turbulence inflow conditions to serve as a baseline case. 

\begin{table}[htb]
\renewcommand{\arraystretch}{1.5} 
\caption{Synthetic inflow simulation sets used to reproduce and parse the effects of ABL shear and veer on rotor power coefficient $C_P$. Inflow properties in the ABL are shown in \cref{sec:results_sbl_power}. The inclusion of the geometric and inductive effects of shear and veer on each set of simulations is denoted with the checkmark $\checkmark$. ``Other'' comprises stratification and turbulence effects. }
\label{tab:sbl_synthetic}
\begin{tabular}{|l|cccc|cc|cc|c|}
\hline
\multirow{2}{*}{\textbf{Inflow name}} & \multicolumn{1}{l}{\multirow{2}{*}{\textbf{$\uB(z)/\Uhub$}}} & \multicolumn{1}{l}{\multirow{2}{*}{\textbf{$\vB(z)/\Uhub$}}} & \multicolumn{1}{l}{\multirow{2}{*}{\textbf{$\theta^B(z)$}}} & \multicolumn{1}{l|}{\multirow{2}{*}{\textbf{$\ti$}}} & \multicolumn{2}{c|}{\textbf{Geometric}} & \multicolumn{2}{c|}{\textbf{Inductive}} & \textbf{Other} \\
 & \multicolumn{1}{l}{} & \multicolumn{1}{l}{} & \multicolumn{1}{l}{} & \multicolumn{1}{l|}{} & Shear & Veer & Shear & Veer &  \\ \hline
Turbulent synthetic & ABL & ABL & ABL & ABL & $\checkmark$ & $\checkmark$ & $\checkmark$ & $\checkmark$ & $\checkmark$ \\
$\ti_{00}$ synthetic & ABL & ABL & - & 0 & $\checkmark$ & $\checkmark$ & $\checkmark$ & $\checkmark$ &  \\
$\ti_{00}$ shear & ABL & 0 & - & 0 & $\checkmark$ & $\checkmark$ & $\checkmark$ &  &  \\
$\ti_{00}$ veer & 1 & ABL & - & 0 &  &  &  & $\checkmark$ &  \\ \hline
\end{tabular}
\end{table}

The deconstruction of the SBL inflow to the four sets of simplified flows, shown in \cref{tab:sbl_synthetic}, parses the effects of the inflow conditions on the actuator disk power generation in the following ways. 
The ``Turbulent synthetic'' simulations are the closest comparison to the concurrent-precursor simulations achievable with this synthetic inflow setup. 
The ABL mean velocity profiles are imposed, and the rotor-averaged turbulence intensity ($\ti_d$) is matched to ABL simulations with values given in \cref{tab:sbl_properties}. 
In the ``Turbulent synthetic'' simulations, the potential temperature equation is also solved, and the mean potential temperature profile is imposed in the inflow to mediate the production of turbulence with buoyant destruction of turbulence kinetic energy. 
Next, in the ``$\ti_{00}$ synthetic'' simulations, we parse the effects of stratification and freestream turbulence on the rotor performance from effects of shear and veer by imposing the exact mean inflow velocity profiles, but with zero freestream (laminar) inflow. 
Moving next to the engineered sets of inflow profiles, we isolate the geometric and inductive effects of wind shear by imposing $\uB(z)$ from the ABL profiles, but explicitly setting $\vB(z)=0$ in the ``$\ti_{00}$ shear'' simulations. 
Note that by imposing $\uB(z)$ from the ABL, the ``$\ti_{00}$ shear'' simulations have the same rotor-equivalent wind speed $\rews$ as the ABL simulations. 
That is, the ``$\ti_{00}$ shear'' simulations include the geometric effects of both shear \textit{and} veer.
Finally, in the ``$\ti_{00}$ veer'' simulations, we isolate the inductive effects of wind veer in the ABL using the ABL $\vB(z)$ profiles, but setting $\uB(z) = \Uhub$. 

\begin{figure}[htb]
    \centering
    \includegraphics[width=0.9\linewidth]{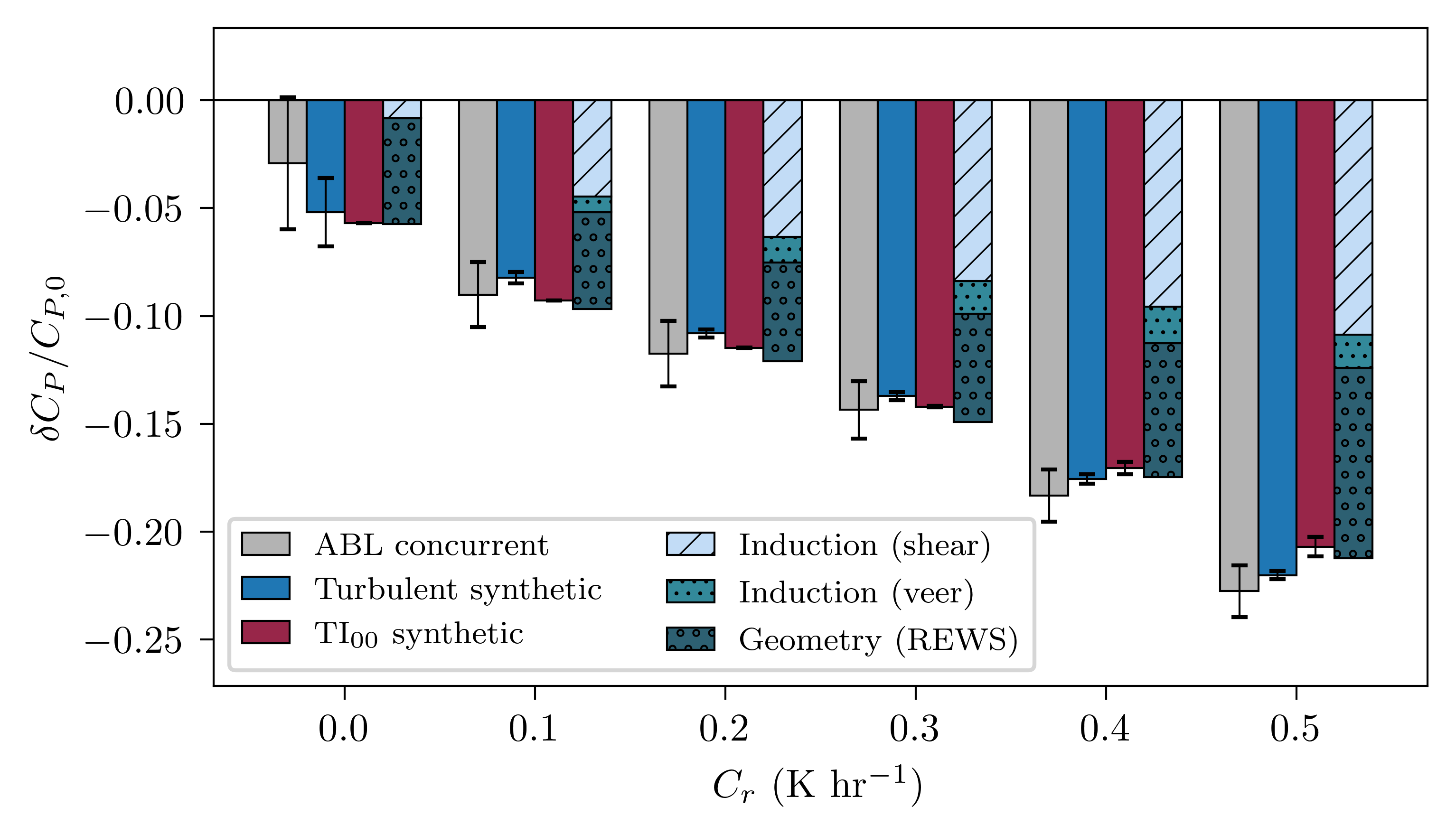}
    \caption{Gain/loss in power coefficient, relative to a uniform inflow case with a wall ($C_{P,0}$), as a function of inflow controlled by the surface cooling rate, $C_r$. The stacked bar represents individual effects of shear and veer, parsed from zero freestream turbulence ($\ti_{00}$) synthetic inflow simulations (see \cref{tab:sbl_synthetic}). Error bars show $\pm 2$ standard errors around the mean power coefficient.}
    \label{fig:sbl_cp_loss_barplot}
\end{figure}

For each simulation, including the ABL simulations in \cref{sec:results_sbl_power}, we extract the power coefficient from the time-averaged actuator disk power. 
Then, we compute the difference in $C_P$ between each simulation and the reference, uniform-inflow simulation: $\delta C_P = C_P - C_{P, 0}$. 
Finally, for the ``$\ti_{00}$ shear'' simulations, we additionally parse the change in power $\delta C_P$ into an inductive effect and a rotor-equivalent wind speed effect. 
The rotor-equivalent wind speed effect, labeled ``Geometry (REWS),'' is computed as $\delta C_P = \delta C_{P, 0} (1 - (\rews / \Uhub)^3)$, which assumes that the geometric and inductive effects of shear can be linearly decomposed (see \cref{ssec:shear_rotor}). 
Error bars show two standard errors of $C_P$ around the mean value for all simulations ($\approx 95\%$ confidence interval), where power measurements have been binned into 10-minute windows to avoid autocorrelations due to turbulence fluctuations. 
The various losses ($\delta C_P$), normalized by the uniform inflow performance $C_{P, 0}$, are shown in \cref{fig:sbl_cp_loss_barplot} as a function of cooling rate. 

Across the concurrent-precursor ABL simulations (gray), Turbulent synthetic simulations (blue), and $\ti_{00}$ synthetic simulations (red), the power losses as a function of ABL inflow follow the same trend: increasingly complex shear and veer profiles at increasing $C_r$ cause greater power losses. 
The power coefficient measured in Turbulent synthetic LES power estimates and the concurrent-precursor LES match within the $95\%$ confidence interval, and the $\ti_{00}$ synthetic simulations also lie within the $95\%$ confidence interval, except at the highest $C_r$. 
Therefore, the synthetic inflow method is able to reproduce the interactions between the rotor and the inflow. 
Furthermore, we only observe minor differences between the $\ti_{00}$ synthetic and Turbulent synthetic sets of simulations, indicating that freestream turbulence (up to $\ti \approx 7\%$ studied here in the $C_r = 0$ simulation, see \cref{tab:sbl_properties}) has only a minor effect on the power production (and therefore the thrust coefficient, power coefficient, and induction factor) of the rotor, relative to the effects of shear and veer.  
Finally, it is worth noting that in the TNBL, the concurrent-precursor simulation has smaller losses (lower $|\delta C_P|$) than the synthetic inflow simulations. 
This is due to a slight increase in power from turbulent fluctuations from long streamwise structures in the ABL, which are not captured by the synthetic inflow simulations. 
Normalizing by $\overline{\Uhub(t)^3}$ rather than $\Uhub^3 \equiv \overline{\Uhub(t)}^3$ (c.f., \citet{revaz_effect_2025}) closes the discrepancy between the concurrent-precursor and synthetic inflow simulations in the TNBL inflow (not shown). 
However, we opt to compute the power coefficient using the standard definition (i.e., normalized by the cube of the mean wind speed) for consistency throughout this study.  

The stacked bars in \cref{fig:sbl_cp_loss_barplot} parse the individual contributions to changes in power coefficient $\delta C_P$ due to the geometric and inductive effects from shear and veer. 
These individual contributions are computed using the engineered inflow simulations.
As a general observation, the effects of shear and veer seem to approximately superimpose for every ABL inflow condition (individual contributions sum to the ``$\ti_{00}$ synthetic'' value of $\delta C_P$ within $0.01 C_{P,0}$ for all cases). 
While there is no guarantee that shear and veer effects should linearly superimpose, it is worth noting that the mechanisms for inductive losses in shear and veer differ significantly; as such, it is reasonable that the effects of both mechanisms can be observed in the superimposed flow when both shear and veer are present. 
For the \iea reference turbine in the given ABL inflows, we observe that wind shear and geometric effects (which contain both shear and veer) comprise a significant fraction of the loss in efficiency ($C_P$). 
Because the \iea reference turbine is relatively close to the ground ($\zhub/D = 0.625$), the inductive effects of veer are suppressed (see \cref{ssec:synthetic_veer_vorticity}), while the inductive effects of shear are amplified (see \cref{ssec:shear_nonlocal} and also \cref{appx:wall_sweep}). 
Critically, while wind shear is the primary loss mechanism in \cref{fig:sbl_cp_loss_barplot}, we stress that the relative importance of the inductive effects of shear and veer and the geometric effect (changing $\rews$) will, in general, depend on properties of the inflow, as well as properties of the rotor and turbine (such as through $\zhub/D$). 

\begin{figure}[htb]
    \centering
    \includegraphics[width=0.9\linewidth]{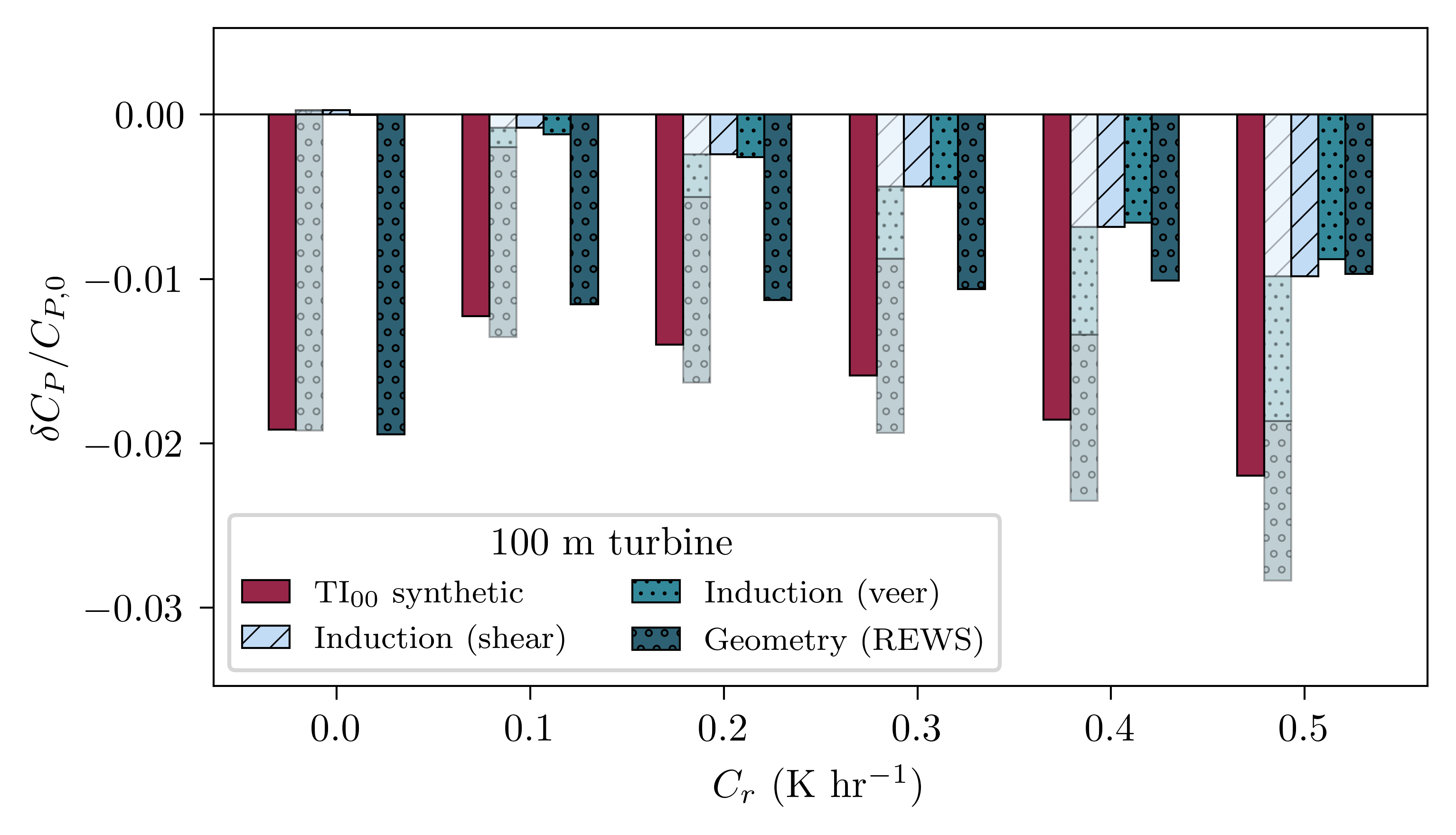}
    \caption{Gain/loss in power coefficient, relative to a uniform inflow case with a wall, varying the cooling rate $C_r$ for a turbine of $D = \SI{100}{\meter}$ and $\zhub/D = 1$ rather than the dimensions of the \iea ($D=\SI{240}{\meter}$, $\zhub/D=0.625$) shown in \cref{fig:sbl_cp_loss_barplot}. The stacked bar plot is the sum of the individual contributions from Induction~(shear), Induction~(veer), and Geometry~(REWS). }
    \label{fig:sbl_cp_loss_100m}
\end{figure}

To illustrate how the losses/changes in $C_P$ depend on the inflow conditions and turbine geometry, we use the synthetic inflow method and ABL profiles from \cref{sec:results_sbl_power} to run additional zero freestream turbulence inflow simulations of a \SI{100}{\meter}-diameter turbine with hub height $\zhub = \SI{100}{\meter}$ (i.e., ($\zhub/D =1$). 
In these simulations, the number of grid points is kept constant (\cref{sssec:setup_laminar}), and the domain size is $10\pi D \times 4\pi D \times 2\pi D$ using the smaller $\SI{100}{\meter}$ turbine size. 
We perform the same analysis as in \cref{fig:sbl_cp_loss_barplot} to isolate $\delta C_P$ for each flow and parse the total effect of shear and veer into separate inductive effects from shear and from veer, as well as from the geometric effect in varying $\rews$. 
The resulting $\delta C_P$ factors for the $\SI{100}{\meter}$ turbine are shown in \cref{fig:sbl_cp_loss_100m}. 

Compared with the \SI{240}{\meter}-diameter \iea reference turbine, the \SI{100}{\meter}-diameter turbine spans a smaller vertical extent in the atmosphere. 
As a result, the non-dimensional shear ($\shear$ and $\powerlaw$) and veer ($\veer$), which control the aerodynamic response of the rotors, differ, even though the ABL profiles are unchanged. 
For example, the background veer in the $C_r = \SI{0.1}{\kelvin\per\hour}$ case is approximately linear, with a veering rate of $\SI{0.053}{\degree\per\meter}$. 
The total turning across the \SI{240}{\meter}-diameter rotor is thus $\SI{12.7}{\degree}$, but the \SI{100}{\meter}-diameter rotor only spans $\SI{5.3}{\degree}$ of turning. 
In addition to the change in relevant non-dimensional inflow properties, the relative wall proximity ($\zhub/D$) also changes. 
In sum, all of these effects result in differing amounts of power loss ($\delta C_P$) for the $\SI{100}{\meter}$ rotor, as well as a different partitioning of the contributions to the total power loss from the various inductive and geometric effects. 
For example, the fraction of losses due to wind veer is relatively much greater for the $\SI{100}{\meter}$ turbine because of the increased $\zhub/D$, compared to the \iea reference turbine. 
Additionally, the magnitude of losses is approximately one order of magnitude greater for the \SI{240}{\meter} turbine in \cref{fig:sbl_cp_loss_barplot} compared with the \SI{100}{\meter} turbine for the same inflow profiles due to larger values of $\shear$ and $\veer$ for the larger rotor extent. 
This suggests that contemporary-scale wind turbines, with rotor diameters extending beyond $\SI{200}{\meter}$, will operate more frequently in conditions where wind shear and wind veer have a significant effect (more than $20\%$ efficiency loss, in the cases shown here) on rotor performance and power extraction, compared with smaller turbines of the past. 
Continued investigation of contemporary-scale rotors is critical to understand and model how wind turbine aerodynamics may be affected by properties of the atmospheric inflow.

\subsection{Shear and veer-induced deviations from momentum theory}
\label{ssec:results_momentum_theory}

\begin{figure}[htb]
    \centering
    \includegraphics[width=\linewidth]{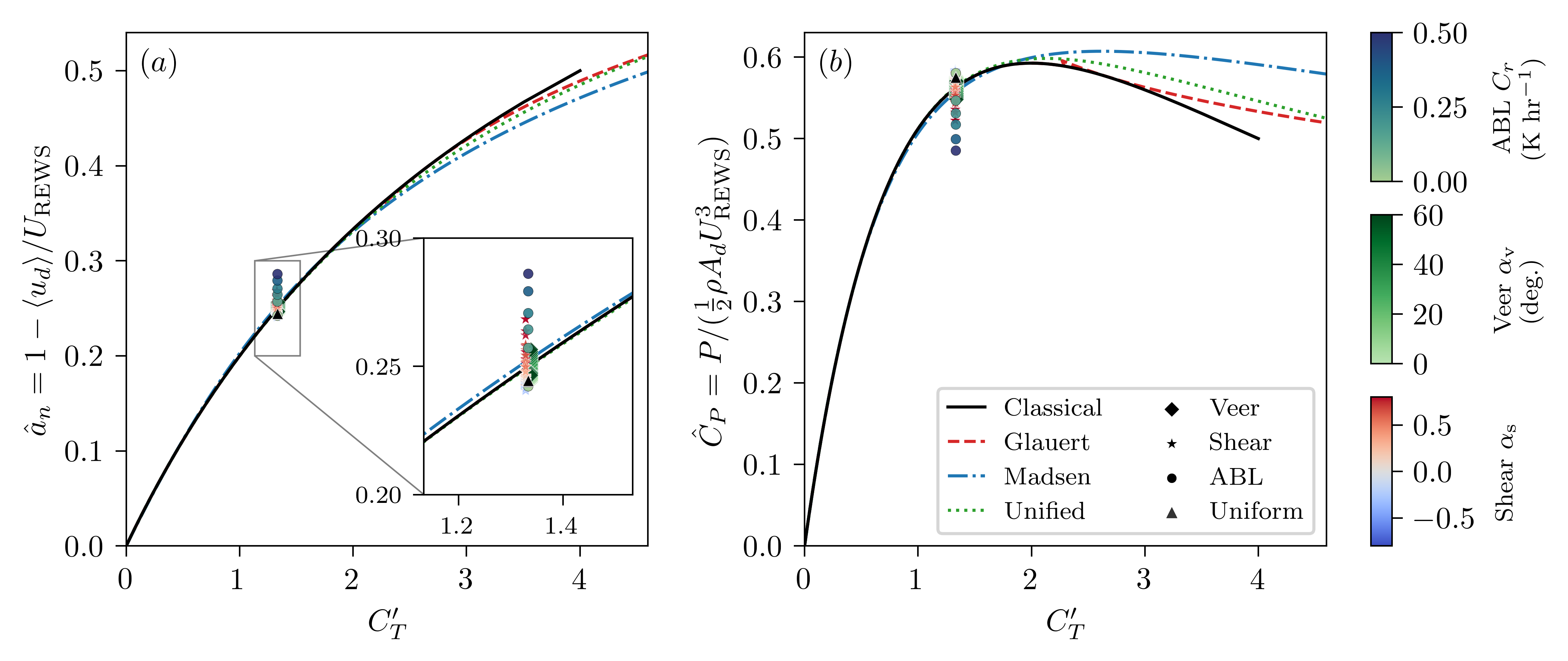}
    \caption{Momentum theory predictions of the relationship between ($a$) thrust coefficient $C_T'$ and induction factor $\hat{a}_n \equiv 1 - \ud / \rews$ and ($b$) $C_T'$ and power coefficient $\hat{C}_P \equiv P / (\tfrac 12 \rho A_d \rews^3)$, compared with LES cases of ABL concurrent-precursor flows ($\circ$), idealized shear ($\star$), and idealized veer ($\diamond$). Some LES points have been slightly offset horizontally for improved legibility. Uniform inflow is shown with a black triangle ($\blacktriangle$). }
    \label{fig:theory_ctprime_an}
\end{figure}

Shear and veer both have an inductive effect that modulates the relationship between the rotor thrust and its induced velocity. 
Because turbines produce power and thrust based on velocities at the rotor (i.e., $\ud$), which is modified by induction, accurate predictive models of induction are critical to engineering rotor modeling for turbine design \citep[c.f.,][]{burton_wind_2011}. 
Engineering modeling of the turbine induction, also called momentum theory, makes simplifying assumptions to reduce computational cost.
One assumption made in commonly used induction closure models is uniform inflow (zero wind shear and zero wind veer). 
As a result, existing induction models, such as classical theory \citep{burton_wind_2011}, high-thrust correction from \citet{glauert_general_1926}, empirical model from \citet{madsen_implementation_2020}, or the Unified momentum model by \citet{liew_unified_2024} (among others), do not capture deviations in the induction--thrust relationship of a rotor induced by wind shear or wind veer. 
To highlight deviations from classical theory induced by wind shear and wind veer effects, the relationship between the modified induction factor $\hat{a}_n \equiv 1 - \ud / \rews$ and modified power coefficient $\hat{C}_P = P/(\tfrac 12 \rho A_d \rews^3)$ to the prescribed turbine thrust coefficient $C_T'$ from LES simulations in Sections~\ref{sec:results_sbl_power}--\ref{sec:veer_ideal} and \cref{appx:wall_sweep} is shown in \cref{fig:theory_ctprime_an}. 
Momentum theory predictions from four induction closure models are also shown. 

Momentum theory predicts the relationship between turbine thrust ($C_T'$) and induced velocity.
We normalize by $\rews$, rather than $\Uhub$, to compute induction ($\hat{a}_n$) in \cref{fig:theory_ctprime_an}($a$) to explicitly isolate the inductive effects of wind shear and wind veer from any geometric effect; an even greater deviation from classical theory due to shear and veer is present if normalizing by $\Uhub$. 
The uniform inflow case ($\blacktriangle$) falls very close to the classical theory prediction, $\an = C_T'/(4 + C_T')$ (with a slight deviation due to the LES regularization, c.f., \citet{shapiro_filtered_2019}).
However, as wind shear and wind veer increase in magnitude, the LES points deviate further from the momentum theory prediction, reaching up to $\hat{a}_n = 0.286$ in the ABL $C_r = \SI{0.5}{\kelvin\per\hour}$ case where both strong shear and veer are present (a $+15\%$ deviation from classical theory, $\hat{a}_n = 0.25$). 
As the inductive effects of shear and veer generally reduce the disk velocity, thrust coefficient, and power coefficient, we observe that deviations from classical theory predominantly increase $\hat{a}_n$ across the range of parameters simulated in this study ($\shear \in [-0.4, 0.8]$; $\veer \in [0, \SI{60}{\degree}]$; $\zhub/D \in [0.625, \pi]$, $\ti \in [0, 7\%]$).
The corresponding change in power coefficient $\hat{C}_P$ is shown in \cref{fig:theory_ctprime_an}($b$), where $\hat{C}_P$ is also non-dimensionalized by $\rews$ to isolate the inductive effects of wind shear and wind veer.
For $C_T' = 4/3$, the most extreme inductive effects of shear and veer reduces $\hat{C}_P$ to $\hat{C}_P=0.485$ (a $-14\%$ deviation from classical theory of $\hat{C}_P = 0.5625$). 
Improved induction modeling in atmospheric inflow conditions is therefore critical to the power and loads predictions of all wind turbines. 
We further emphasize that modeling these inductive effects of shear and veer is particularly important for emerging wind turbine designs such as $200$+ \si{\meter}-diameter rotors that will operate in a wide range of sheared and veered conditions.


\section{Conclusions}
\label{sec:conclusions}

In this study, we investigate the impacts of non-uniform inflow, which is ubiquitous in the atmospheric boundary layer (ABL), on the performance of an actuator disk-modeled wind turbine. 
In large-eddy simulations of the stratified ABL, varying the surface cooling rate $C_r$ affects the ABL turbulence intensity, stratification, hub height wind speed, wind shear (vertical gradients of wind speed), and wind veer (vertical gradients of wind direction). 
We simulate the interaction between the ABL and an actuator disk with the dimensions of an \iea reference turbine ($D=\SI{240}{\meter}$, $\zhub/D = 0.625$) using a concurrent-precursor LES approach. 
As the cooling rate increases, the ABL height drops, and the nose of a low-level jet increases the wind speed at hub height. 
As a result, we observe that the power extracted by the actuator disk generally increases with increasing stability. 
However, normalizing by the cube of the hub height wind speed $\Uhub$ reveals that there is a monotonic decrease in the power coefficient $C_P$ (efficiency) with increasing cooling rate. 
At the highest cooling rate simulated ($C_r = \SI{0.5}{\kelvin\per\hour}$), the turbine efficiency is more than $20\%$ lower than an equivalent turbine operating in freestream conditions. 

To understand why there is a significant drop in $C_P$ as the cooling rate increases, we perform synthetic laminar and turbulent inflow simulations using idealized parameterizations of wind shear and wind veer. 
We find that both wind shear and wind veer can significantly affect rotor performance (induction, thrust coefficient, and power coefficient), but adding $\ti = 5\%$ inflow turbulence has a relatively small effect by comparison, particularly in veered inflow. 
First, shear and veer may change the rotor-equivalent inflow wind speed $\rews$, which results in an efficiency loss of approximately $(\rews/\Uhub)^3$. 
We call this the \textit{geometric} effect, because this effect is simply due to the inflow geometry affecting the mean flux of mass and energy into the rotor. 
Second, both shear and veer can change the rotor-averaged disk velocity $\ud$ through distinctive physical mechanisms which we describe as \textit{inductive} effects. 
In both shear and veer, the geometric and inductive effects superimpose approximately linearly, and we investigate the inductive effects through a control volume analysis of the flow physics. 
Wind shear creates non-uniformity in the local thrust coefficient (defined based on local $\UBmag(z)$ rather than $\Uhub$), which induces local variations in the disk velocity $u_d(r, \theta)$.
However, the rotor-averaged disk velocity (and, by extension, thrust and power) is only affected in sheared inflow when the rotor is close to the ground ($\zhub/D \lesssim 1$), due to a selective amplification of the induction effect by the wall. 
For positive power law wind shear (which is most commonly observed in the ABL), both the geometric and inductive shear effects decrease turbine power, with losses of almost $15\%$ at $\alpha = 0.5$ for a $\SI{240}{\meter}$-diameter wind turbine. 
Wind veer, by contrast, modifies the turbine induction by inducing a large-scale pressure response in the wake, creating an adverse pressure gradient. 
As a result of the adverse pressure gradient, the rotor induction increases, lowering the turbine thrust and power, but the adverse pressure gradient is mitigated as near-wall proximity increases. 
Between the geometric and inductive effects of wind veer at $\SI{40}{\degree}$ across the rotor, the power coefficient $C_P$ drops by approximately $10\%$ for a $\SI{240}{\meter}$-diameter turbine. 

Using a synthetic inflow method, we simulate the mean inflow profiles from the ABL to parse the effects of geometry, shear, and veer. 
The laminar inflow wind profiles in synthetic inflow simulations reproduce the power losses observed in the ABL. 
Furthermore, we perform separate simulations of isolated inflow shear (where the lateral inflow is explicitly set to zero) and isolated inflow veer (where the streamwise inflow is set to $\Uhub$) to parse the losses in the ABL into the geometric and inductive effects from both shear and veer. 
We find that the rotor diameter, hub height ($\zhub/D$), shear, and veer all affect the magnitude of efficiency ($C_P$) loss in the ABL. 
Larger rotor diameters have a greater variation in wind speed and direction across the rotor extent, given the same ABL inflows, leading to higher losses. 
Additionally, near-wall effects are generally amplified for large rotors due to decreasing $\zhub/D$ (i.e., for fixed bottom-tip clearance height, $\zhub/D$ decreases as $D$ increases). 
As a result, the magnitude of power losses and the relative contribution from the individual effects (geometric and inductive) from shear and veer depend on both turbine geometry and the inflow. 
For the same set of ABL inflows, we find that the power of the \SI{240}{\meter}-diameter \iea reference turbine is primarily affected by wind shear, while the power of a smaller \SI{100}{\meter}-diameter turbine (with $\zhub/D=1$) is affected approximately equally by shear and veer effects. 
It is further important to emphasize that there is not yet a model of the induction factor (i.e., the relationship between rotor thrust and rotor induced velocities) that captures the effects of wind speed shear and wind direction shear (veer) presented in this study.

This work models turbines as irrotational actuator disks, in which power and thrust depend only on the rotor-averaged induction factor (i.e., disk-averaged velocity). 
However, for a bladed turbine rotor, power and thrust are generated locally along the airfoils of each blade. 
Therefore, we expect the rotor response to inflow shear and veer may differ when rotational effects and non-uniform rotor loading are included, which should be explored in future work. 
Furthermore, the inductive effects of shear and veer uncovered in this work have a significant impact on power production, but existing models (such as the rotor-equivalent wind speed model, which captures the geometric effects only) neglect these interactions, even though a significant contribution to power losses from shear and veer is due to these inductive effects rather than inflow geometry. 
The combined effects of wind shear and wind veer induce deviations in the thrust--induction relationship from classical momentum theory models upwards of $+15\%$, and corresponding deviations in efficiency (power coefficient) upwards of $-15\%$. 
Therefore, future work is needed to develop models of induction under shear and veer that are sufficiently computationally efficient for wind energy engineering applications.
Finally, future work should investigate the geometric and inductive effects of non-uniform inflows using field measurements and SCADA power. 
Uncovering the physical mechanisms by which ABL inflows affect rotor power production will lead to improved rotor models, enabling the design and deployment of future wind turbines.

\paragraph{Acknowledgments}
Simulations were performed on the Stampede3 supercomputer under the NSF ACCESS project ATM170028.

\paragraph{Funding Statement}
K.S.H. and M.F.H. gratefully acknowledge funding from the National Science Foundation (Fluid Dynamics program, grant number FD-2542240, Program Manager: Dr. Ronald D. Joslin).
K.S.H. acknowledges additional funding through a National Science Foundation Graduate Research Fellowship under grant number DGE-2141064. 
S.A.M.~gratefully acknowledges support from the Gates Millennium Scholars Program and the United States Department of Energy Computational Science Graduate Fellowship under Award Number DE-SC0023112.

\paragraph{Data availability}
The code and data in this manuscript will be published open-access along with the publication of the paper.

\begin{appendix}
\crefalias{section}{appendix}

\section{Wall proximity effects of shear and veer}
\label{appx:wall_sweep}

Simplified inflow large-eddy simulations of wind shear and wind veer in isolation from \cref{sec:shear_idealized} and \cref{sec:veer_ideal}, respectively, revealed that the wall has a significant impact on the induction, thrust, and power of an actuator disk in sheared or veered inflow. 
The no-penetration wall condition ($w=0$ at $z=0$) modulates the inductive effects of shear and veer in separate ways. 
Furthermore, it was observed in \cref{ssec:shear_nonlocal} that the presence of the wall has an insignificant impact on the rotor performance in the uniform inflow case. 
To show intermediate effects of the wall between the two hub height extremes presented throughout the study (the `no-wall' case, $z_h=L_z/2 = \pi D$, and $\zhub/D = 0.625$, following the \iea reference turbine dimensions), we perform two new sets of simulations with six values of $\zhub/D$ between $0.625$ and $1.5$.
The `no-wall' cases are also shown for reference. 

\begin{figure}[htb]
    \centering
    \includegraphics[width=0.9\linewidth]{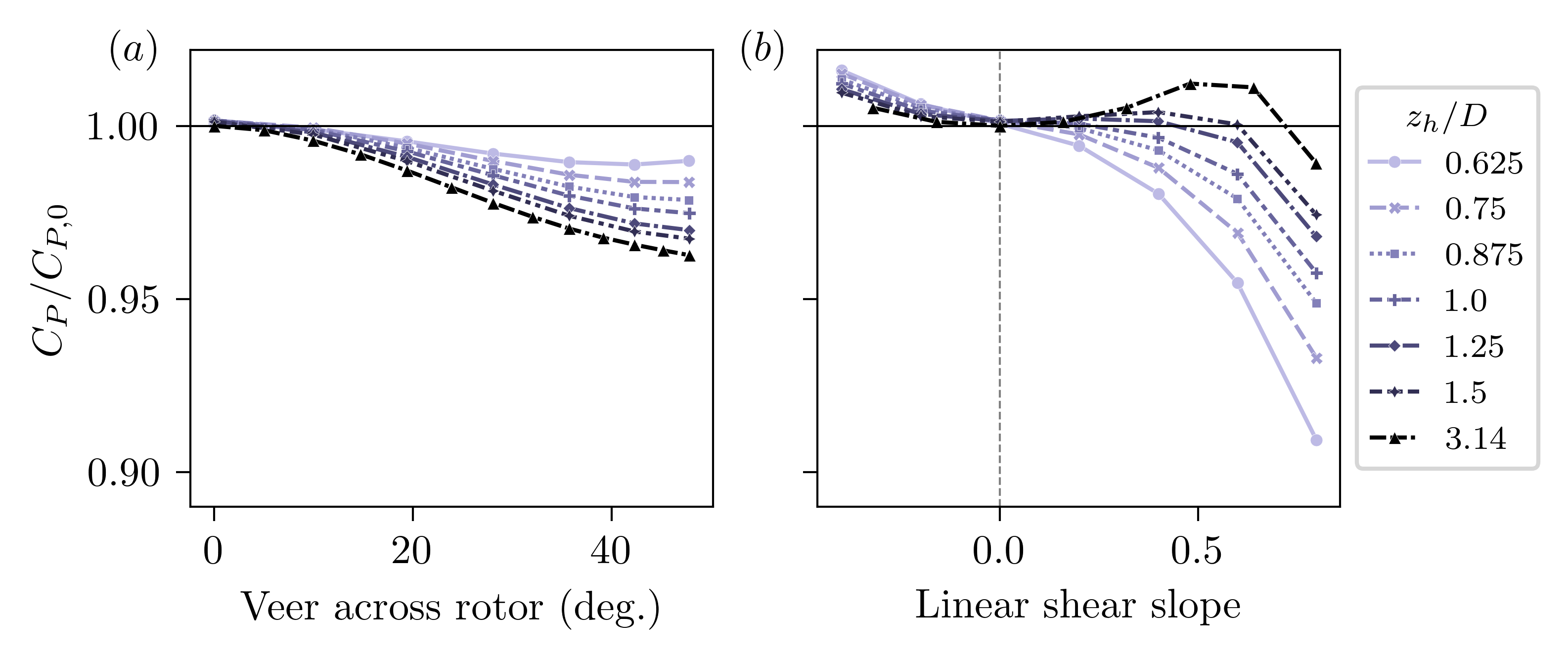}
    \caption{Influence of the wall on the inductive effects of ($a$) $\tanh$ wind shear and ($b$) $\tanh$ wind shear for varying turbine hub heights $\zhub/D$. The `no-wall' case ($\zhub = L_z/2 = \pi D$) is shown in the black triangles. }
    \label{fig:appx_wall}
\end{figure}

In both sets of synthetic inflow large-eddy simulations, we impose laminar inflow and hold the rotor-equivalent wind speed $\rews$ constant. 
That is, the veered inflow simulations use the profiles parameterized by \cref{eq:const_rews_Uphi}, while the sheared inflow simulations use the profiles parameterized by \cref{eq:tanh_shear} and $\vB=0$. 
This isolates the inductive effects of shear and veer from the geometric effects (i.e., simply changing $\rews$). 
For the veered inflow cases, we simulate $\veer \in [0, \SI{60}{\degree}]$ in increments of $\SI{10}{\degree}$ at each non-dimensional hub height for a total of 42 simulations. 
We note that for an irrotational actuator disk-modeled rotor, the effects of wind veer are symmetric across positive and negative veer; thus, we do not simulate $\veer < 0$. 
For the sheared inflow cases, we simulate $\shear \in [-0.4, 0.8]$ in increments of $0.2$ for a total of 42 simulations. 
The resulting turbine power is normalized by the uniform inflow, no-wall case ($P_0 = P(\shear=0, \veer=0)$) as a function of inflow properties and non-dimensional hub height, shown in \cref{fig:appx_wall}.  

In \cref{fig:appx_wall}($a$), we show the inductive response of wind veer on power as a function of wall height and veer magnitude. 
For a fixed amount of wind veer (total wind angle change across the rotor), the rotor power monotonically decreases with increasing turbine hub height $\zhub/D$. 
That is, the mechanism for inductive losses in veered inflow (detailed in \cref{ssec:synthetic_veer_vorticity}) is suppressed by the near-wall proximity and monotonically decreases as $\zhub/D$ decreases. 
By contrast, the `no-wall' wind shear case has the highest rotor power in positive shear ($\shear > 0$, wind speed increasing with increasing height) and vice versa for negative shear, with losses increasing as $\zhub/D$ decreases. 
That is, inductive effects in sheared inflow are amplified by the wall. 
This amplification either monotonically increases rotor power production in negative shear or monotonically decreases rotor power production in positive shear.

\section{Pressure Poisson approach in veered inflow}
\label{appx:prss_veer}

In the inviscid Euler equations, which govern the flow dynamics up to the end of the near-wake region ($x=x_0$, as defined in \cref{ssec:synthetic_veer_dynamics}), the pressure field is determined by the pressure Poisson equation, which can be derived by taking the divergence of the Navier--Stokes equations (\cref{eq:NSE_filtered}): 
\begin{equation}
\label{eq:div_rans}
\frac{\partial}{\partial x_i} \left( {u}_j \frac{\partial {u}_i}{\partial x_j} \right) 
= -\frac{\partial^2 p}{\partial x_i^2} 
+ \frac{\partial {F}_i}{\partial x_i}
\end{equation}
where the transient term $\partial u_i/\partial t$ and all of the forcing terms are lumped into $F_i$ for simplicity. 
Rearranging \cref{eq:div_rans} the pressure Poisson equation and applying continuity yields
\begin{equation}
\frac{\partial^2 p}{\partial x_i^2} 
=
-\frac{\partial u_j}{\partial x_i}\frac{\partial u_i}{\partial x_j} 
-\frac{\partial F_i}{\partial x_i}. 
\end{equation}

\begin{figure}[htb]
    \centering
    \includegraphics[width=0.9\linewidth]{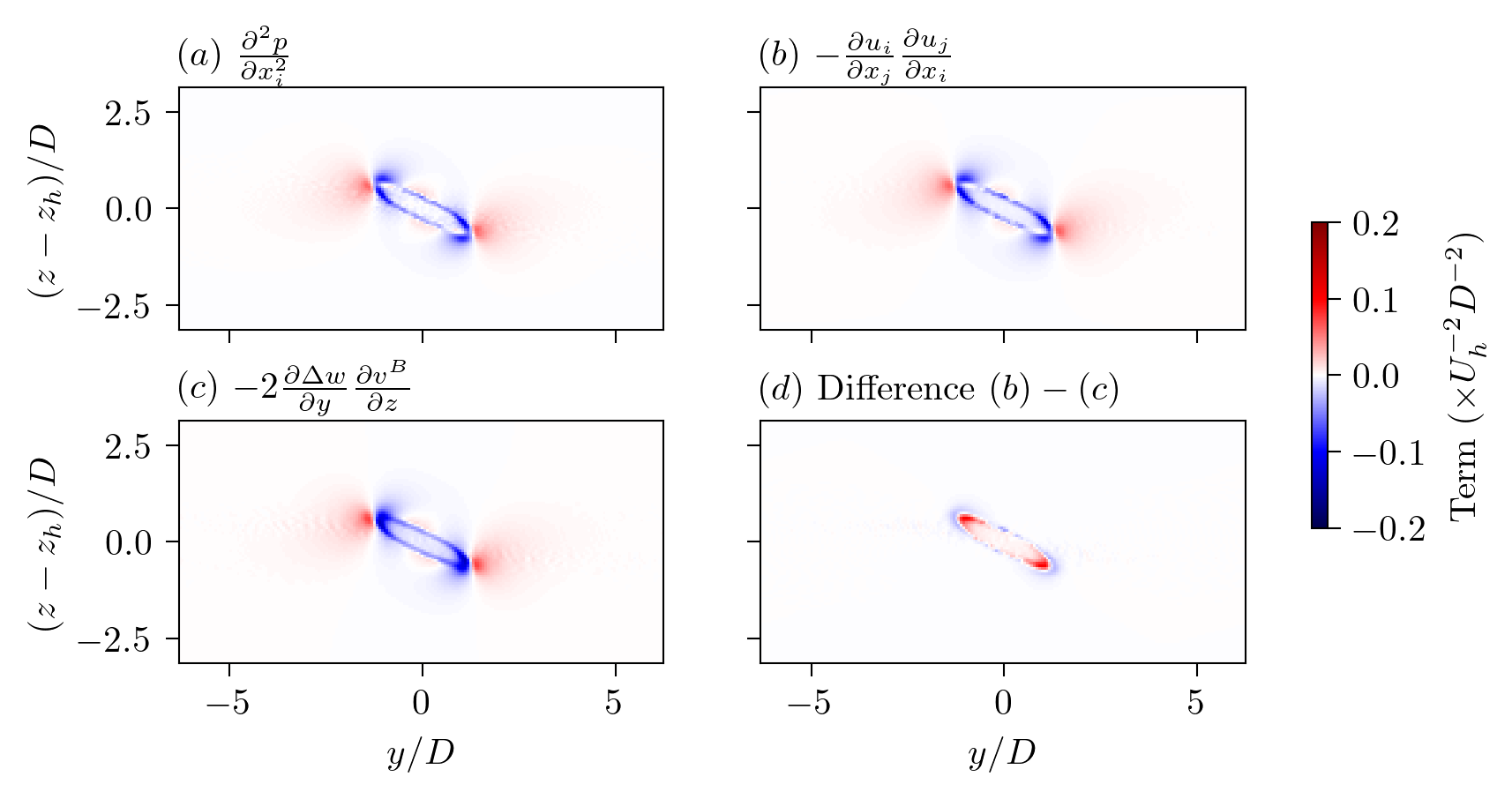}
    \caption{Contours of terms in the Poisson equation for the wake deficit pressure $\Delta p$ (\cref{eq:poisson_inter1}), computed using LES data for $\veer=\SI{30}{\degree}$ at $x=x_0\approx 3.2D$. ($a$) Laplacian of the wake deficit pressure, ($b$) product of velocity gradient terms, ($c$) simplified product term in \cref{eq:poisson_final}, ($d$) difference between ($b$) and ($c$).}
    \label{fig:appx_laplace_pressure}
\end{figure}

We are interested in isolating the effect of wind veer, $\partial \vB/\partial z$, on the wake pressure perturbation. 
To accomplish this, we decompose the velocity and pressure fields into the wake and base flow components, $u_i \equiv u^B + \Delta u$, and likewise for pressure. 
The resulting equation is
\begin{equation}
\frac{\partial^2 (p^B + \Delta p)}{\partial x_i^2} 
= 
-\frac{\partial (u^B + \Delta u)_j}{\partial x_i}\frac{\partial (u^B + \Delta u)_i}{\partial x_j} 
- \frac{\partial (F^B + \Delta F)_i}{\partial x_i}.
\end{equation}
%
Expanding and multiplying terms yields 
\begin{equation}
\frac{\partial^2 (p^B + \Delta p)}{\partial x_i^2} 
= 
-\left( 
\frac{\partial u^B_j}{\partial x_i} \frac{\partial u^B_i}{\partial x_j}
+ \frac{\partial \Delta u_j}{\partial x_i} \frac{\partial u^B_i}{\partial x_j}
+ \frac{\partial u^B_j}{\partial x_i} \frac{\partial \Delta u_i}{\partial x_j}
+ \frac{\partial \Delta u_j}{\partial x_i} \frac{\partial \Delta u_i}{\partial x_j}
\right)
- \frac{\partial (F^B + \Delta F)_i}{\partial x_i}
.
\end{equation}
The indices $i$ and $j$ are dummy indices, so the mixed $\Delta (\cdot)$ and $(\cdot)^B$ velocity gradient terms can be combined. 
After subtracting off the terms corresponding to the base flow pressure $p^B$, we are left with a prognostic equation for the pressure perturbation $\Delta p$ induced by the turbine and wake: 
\begin{equation}
\label{eq:poisson_inter1}
\frac{\partial^2 \Delta p}{\partial x_i^2} 
= 
-\left( 
2\frac{\partial \Delta u_j}{\partial x_i} \frac{\partial u^B_i}{\partial x_j}
+ \frac{\partial \Delta u_j}{\partial x_i} \frac{\partial \Delta u_i}{\partial x_j}
\right)
- \frac{\partial \Delta F_i}{\partial x_i}. 
\end{equation}
To simplify further, we assume that the second-order terms are not relevant (product of wake velocity gradient terms). 
Additionally, in the mixed terms, only vertical gradients of $v$ velocity are present in the veered base flow. 
Furthermore, we neglect streamwise gradients, which we assume to be small in the wake, relative to the cross-stream directions. 
The resulting term from these assumptions is: 
\begin{equation}
\label{eq:poisson_final}
\frac{\partial^2 \Delta p}{\partial x_i^2} 
\approx 
- 2\frac{\partial \Delta w}{\partial y} 
\underbrace{\frac{\partial v^B}{\partial z}}_\text{Veer}; 
\quad i = 2, 3.
\end{equation}
We test the validity of \cref{eq:poisson_final} by computing the left-hand side of \cref{eq:poisson_final} and the successive approximations of the right-hand side in \cref{eq:poisson_inter1} and \eqref{eq:poisson_final} using LES data. 
In \cref{fig:appx_laplace_pressure}, we show contours of the Laplacian of the pressure field and approximations to the right-hand side, computed at the downstream location $x=x_0$ for the ``Constant $\rews$ (no-wall)'' case at $\veer = \SI{30}{\degree}$.

From \cref{fig:appx_laplace_pressure}, it is clear that the simplified wake interaction term between gradients of $\Delta w$ and the background wind veer comprises the majority of the forcing in the wake pressure Poisson equation. 
The challenge this reveals is that the wake deficit pressure is generated by the gradients of the vertical wake velocity, $\Delta w$, which is typically neglected in engineering models and flow physics analysis of wind turbine wakes. 
In the absence of inflow wind veer, $\Delta w = 0$ after the initial wake expansion, but $\Delta w \neq 0$ when inflow wind veer is present because of the evolution of streamwise vorticity in the wake (see \cref{ssec:synthetic_veer_vorticity}). 
Furthermore, spatial information about the $\Delta w$ field is needed to compute lateral gradients (in the $y$ direction). 

\begin{figure}[htb]
    \centering
    \includegraphics[width=1\linewidth]{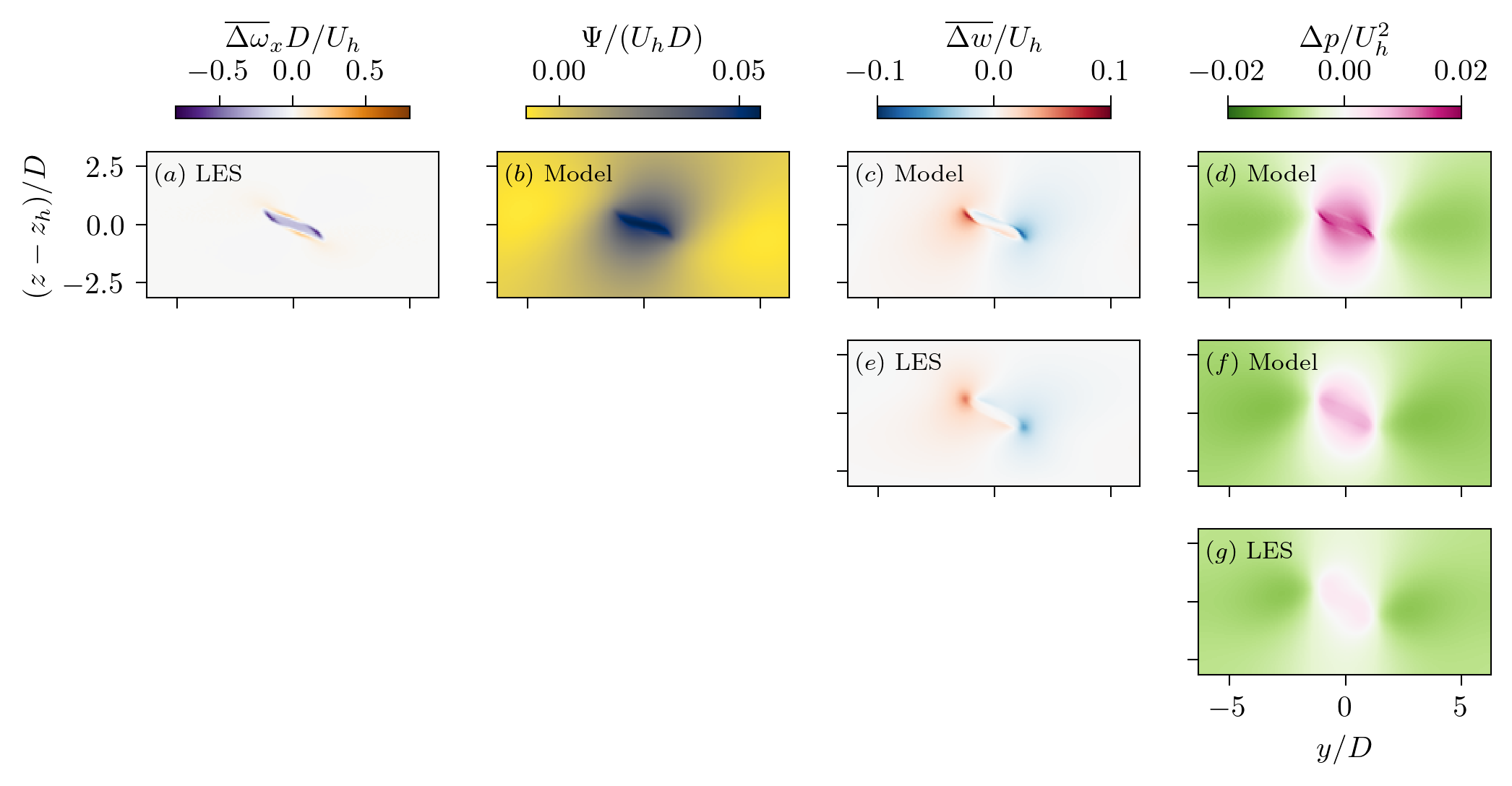}
    \caption{Veer-induced pressure perturbation constructed from ($a$-$d$) the LES vorticity field or ($e$-$f$) the LES velocity field using a Green's function approach. In ($a$-$d$), the double Poisson inversion is used to translate ($a$) $\overline{\Delta \omega}_x$ into the ($b$) streamfunction $\Psi$, then differentiated to the ($c$) vertical velocity $\overline{\Delta w}$, and ($d$) second Poisson inversion for $\Delta p_\text{veer}$ using \cref{eq:poisson_final}. In ($e$), the LES $\bar{w}$ field is used to compute ($f$) an approximate $\Delta p_\text{veer}$. ($g$) LES pressure perturbation due to wind veer $\Delta p_\text{veer} = \Delta p - \Delta p_0$.}
    \label{fig:appx_pressure_model}
\end{figure}

One strategy for deriving an equation for the vertical wake velocity $\Delta w(y, z)$ is to invert the wake-added streamwise vorticity. 
Defining a two-dimensional streamfunction $\Psi$ in the $yz$ plane,
\begin{equation}
\begin{cases}
\Delta w = \cfrac{\partial \Psi}{\partial y}, \\
\Delta v = \cfrac{\partial \Psi}{\partial z}.
\end{cases}
\end{equation}
The wake streamfunction $\Psi$ is therefore related to the wake vorticity $\Delta \omega_x$ with a second Poisson equation: 
\begin{equation}
\frac{\partial^2 \Psi}{\partial x_i^2} = \Delta \omega_x,
\end{equation}
where $i = 2, 3$ only (i.e., the streamfunction is two-dimensional). 
The transformation from the veer-induced wake pressure to the wake vorticity thus constitutes two Poisson inversions, first from $\Delta \omega_x(y, z)$ to $\Psi(y, z)$, and the second from $\Delta w = \partial \Psi/\partial y$ to $\Delta p(y, z)$. 
The boundary conditions for the streamfunction and veer pressure $\Delta p_\text{veer}$ decay to zero as $r \to \infty$. 
Therefore, each Poisson equation can be solved numerically with a Green's function approach using the two-dimensional fundamental solution: 
$$
g(y, z; y', z') = \frac{1}{2\pi} \ln(\sqrt{(y-y')^2 + (z-z')^2}),
$$
which solves the Poisson equation for a delta function located at $(y', z')$.
We show the procedure for computing the veer-induced wake pressure from the streamwise wake vorticity $\Delta \omega_x$ in \cref{fig:appx_pressure_model}.

From \cref{fig:appx_pressure_model}($a$-$d$), we can see that the model framework qualitatively captures the structure of the vertical velocity and pressure fields. 
Furthermore, using the exact LES vertical velocity field, the magnitude of the modeled $\Delta p_\text{veer}$ improves (\cref{fig:appx_pressure_model}($e$-$f$)), indicating that some error is incurred in the two-dimensional flow assumption in the transformation from $\Delta \omega_x$ to $\Delta w$. 
The simplified model indicates that the physical mechanism leading to an adverse pressure gradient in the wake is the interaction between the wake velocity (specifically, through $\Delta w$) and the background wind veer, but that the quantitative response of $\Delta p_\text{veer}$ likely also depends on three-dimensional interactions which are not captured by the two-dimensional Poisson inversion model described here.

\end{appendix}

\bibliography{references_.bib}

@article{albers_influence_2007,
  title = {Influence of Meteorological Variables on Measured Wind Turbine Power Curves},
  author = {Albers, Axel and Jakobi, Tim and Rohden, Rolf and Stoltenjohannes, J{\"u}rgen},
  year = 2007,
  month = jan,
  journal = {European Wind Energy Conference and Exhibition 2007, EWEC 2007},
  volume = {3}
}

@article{antoniou_wind_2009,
  title = {Wind {{Shear}} and {{Uncertainties}} in {{Power Curve Measurement}} and {{Wind Resources}}},
  author = {Antoniou, Ioannis and Pedersen, S{\o}ren Markilde and Enevoldsen, Peder Bay},
  year = 2009,
  month = oct,
  journal = {Wind Engineering},
  volume = {33},
  number = {5},
  pages = {449--468},
  publisher = {SAGE Publications}
}

@misc{bastankhah_generalised_2026,
  title = {Generalised Actuator Disk Theory: Wake Development with Turbulent Entrainment},
  shorttitle = {Generalised Actuator Disk Theory},
  author = {Bastankhah, Majid and Hydon, Peter E. and Shapiro, Carl and Gayme, Dennice F. and Meneveau, Charles},
  year = 2026,
  month = mar,
  number = {arXiv:2510.08213},
  primaryclass = {physics.flu-dyn},
  publisher = {arXiv},
  archiveprefix = {arXiv}
}

@book{burton_wind_2011,
  title = {Wind {{Energy Handbook}}},
  author = {Burton, Tony and Jenkins, Nick and Sharpe, David and Bossanyi, Ervin},
  year = 2011,
  month = may,
  publisher = {John Wiley \& Sons},
  isbn = {978-1-119-99392-6}
}

@article{calaf_large_2010,
  title = {Large Eddy Simulation Study of Fully Developed Wind-Turbine Array Boundary Layers},
  author = {Calaf, Marc and Meneveau, Charles and Meyers, Johan},
  year = 2010,
  month = jan,
  journal = {Physics of Fluids},
  volume = {22},
  number = {1},
  pages = {015110}
}

@article{cheung_greens_2024,
  title = {A {{Green}}'s {{Function Wind Turbine Induction Model That Incorporates Complex Inflow Conditions}}},
  author = {Cheung, Lawrence and Brown, Kenneth and Sakievich, Philip and {deVelder}, Nathaniel and Herges, Thomas and Houck, Daniel and Hsieh, Alan},
  year = 2024,
  journal = {Wind Energy},
  volume = {27},
  number = {12},
  pages = {1526--1544},
  langid = {english}
}

@article{dar_experimental_2023,
  title = {An Experimental and Analytical Study of Wind Turbine Wakes under Pressure Gradient},
  author = {Dar, Arslan Salim and Gertler, Abraham Starbuck and {Port{\'e}-Agel}, Fernando},
  year = 2023,
  month = apr,
  journal = {Physics of Fluids},
  volume = {35},
  number = {4},
  pages = {045140}
}

@article{diaz_review_2020,
  title = {Review of the Current Status, Technology and Future Trends of Offshore Wind Farms},
  author = {D{\'i}az, H. and Guedes Soares, C.},
  year = 2020,
  month = aug,
  journal = {Ocean Engineering},
  volume = {209},
  pages = {107381},
  langid = {english}
}

@article{dorenkamper_atmospheric_2014,
  title = {Atmospheric {{Impacts}} on {{Power Curves}} of {{Multi-Megawatt Offshore Wind Turbines}}},
  author = {D{\"o}renk{\"a}mper, M. and Tambke, J. and Steinfeld, G. and Heinemann, D. and K{\"u}hn, M.},
  year = 2014,
  month = dec,
  journal = {Journal of Physics: Conference Series},
  volume = {555},
  number = {1},
  pages = {012029},
  publisher = {IOP Publishing},
  langid = {english}
}

@article{ekman_influence_1905,
  title = {On the {{Influence}} of the {{Earth}}'s {{Rotation}} on {{Ocean-Currents}}},
  author = {Ekman, Vagn Walfrid},
  year = 1905,
  month = oct,
  journal = {Arkiv f\"or Matematik, Astronomi och Fysik},
  volume = {2},
  number = {11},
  pages = {1--52}
}

@techreport{gaertner_definition_2020,
  title = {Definition of the {{IEA}} 15-{{Megawatt Offshore Reference Wind Turbine}}},
  author = {Gaertner, Evan and Rinker, Jennifer and Sethuraman, Latha and Zahle, Frederik and Anderson, Benjamin and Barter, Garrett and Abbas, Nikhar and Meng, Fanzhong and Bortolotti, Pietro and Skrzypinski, Witold and Scott, George and Feil, Roland and Bredmose, Henrik and Dykes, Katherine and Shields, Matt and Allen, Christopher and Viselli, Anthony},
  year = 2020,
  number = {NREL/TP-5000-75698},
  pages = {54},
  address = {Golden, CO, United States},
  institution = {National Renewable Energy Lab (NREL)},
  langid = {english}
}

@article{gao_effect_2021,
  title = {Effect of Wind Veer on Wind Turbine Power Generation},
  author = {Gao, Linyue and Li, Bochen and Hong, Jiarong},
  year = 2021,
  month = jan,
  journal = {Physics of Fluids},
  volume = {33},
  number = {1},
  pages = {015101}
}

@techreport{glauert_general_1926,
  title = {A {{GENERAL THEORY OF THE AUTOGYRO}}.},
  author = {Glauert, H},
  year = 1926,
  number = {1111},
  pages = {558--593},
  institution = {HM Stationery Office},
  langid = {english}
}

@article{heck_coriolis_2025,
  title = {Coriolis Effects on Wind Turbine Wakes across Neutral Atmospheric Boundary Layer Regimes},
  author = {Heck, Kirby S. and Howland, Michael F.},
  year = 2025,
  month = apr,
  journal = {Journal of Fluid Mechanics},
  volume = {1008},
  pages = {A7},
  langid = {english}
}

@article{heck_modelling_2023,
  title = {Modelling the Induction, Thrust and Power of a Yaw-Misaligned Actuator Disk},
  author = {Heck, K. S. and Johlas, H. M. and Howland, M. F.},
  year = 2023,
  month = mar,
  journal = {Journal of Fluid Mechanics},
  volume = {959},
  pages = {A9},
  langid = {english}
}

@misc{heck_unraveling_2025,
  title = {Unraveling the Effects of Atmospheric Dynamics on Wakes with a Controlled Synthetic Inflow Methodology},
  author = {Heck, Kirby S. and Howland, Michael F.},
  year = 2025,
  month = dec,
  number = {arXiv:2512.18518},
  primaryclass = {physics},
  publisher = {arXiv},
  archiveprefix = {arXiv}
}

@article{howland_influence_2020,
  title = {Influence of Atmospheric Conditions on the Power Production of Utility-Scale Wind Turbines in Yaw Misalignment},
  author = {Howland, Michael F. and Gonz{\'a}lez, Carlos Moral and Mart{\'i}nez, Juan Jos{\'e} Pena and Quesada, Jes{\'u}s Bas and Larra{\~n}aga, Felipe Palou and Yadav, Neeraj K. and Chawla, Jasvipul S. and Dabiri, John O.},
  year = 2020,
  month = nov,
  journal = {Journal of Renewable and Sustainable Energy},
  volume = {12},
  number = {6},
  pages = {063307},
  langid = {english}
}

@techreport{iea_renewables_2025,
  title = {Renewables 2025},
  author = {IEA},
  year = 2025,
  month = oct,
  institution = {International Energy Agency},
  langid = {british}
}

@article{jeong_identification_1995,
  title = {On the Identification of a Vortex},
  author = {Jeong, Jinhee and Hussain, Fazle},
  year = 1995,
  month = feb,
  journal = {Journal of Fluid Mechanics},
  volume = {285},
  pages = {69--94},
  langid = {english}
}

@techreport{jonkman_fast_2005,
  title = {{{FAST User}}'s {{Guide}}},
  author = {Jonkman, Jason and Buhl, Marshall},
  year = 2005,
  month = aug,
  number = {NREL/EL-500-38230},
  institution = {National Renewable Energy Lab (NREL), Golden, CO (United States)},
  langid = {english}
}

@article{jung_properties_2023,
  title = {The Properties of the Global Offshore Wind Turbine Fleet},
  author = {Jung, Christopher and Schindler, Dirk},
  year = 2023,
  month = oct,
  journal = {Renewable and Sustainable Energy Reviews},
  volume = {186},
  pages = {113667},
  langid = {english}
}

@article{kelly_shear_2023,
  title = {From Shear to Veer: Theory, Statistics, and Practical Application},
  shorttitle = {From Shear to Veer},
  author = {Kelly, Mark and {van der Laan}, Maarten Paul},
  year = 2023,
  month = jun,
  journal = {Wind Energy Science},
  volume = {8},
  number = {6},
  pages = {975--998},
  publisher = {Copernicus GmbH},
  langid = {english}
}

@article{klemmer_momentum_2024,
  title = {Momentum Deficit and Wake-Added Turbulence Kinetic Energy Budgets in the Stratified Atmospheric Boundary Layer},
  author = {Klemmer, Kerry S. and Howland, Michael F.},
  year = 2024,
  month = nov,
  journal = {Physical Review Fluids},
  volume = {9},
  number = {11},
  pages = {114607},
  langid = {english}
}

@article{kosovic_impact_2026,
  title = {Impact of Atmospheric Turbulence on Performance and Loads of Wind Turbines: Knowledge Gaps and Research Challenges},
  shorttitle = {Impact of Atmospheric Turbulence on Performance and Loads of Wind Turbines},
  author = {Kosovi{\'c}, Branko and Basu, Sukanta and Berg, Jacob and Berg, Larry K. and Haupt, Sue E. and Lars{\'e}n, Xiaoli G. and Peinke, Joachim and Stevens, Richard J. A. M. and Veers, Paul and Watson, Simon},
  year = 2026,
  month = feb,
  journal = {Wind Energy Science},
  volume = {11},
  number = {2},
  pages = {509--555},
  publisher = {Copernicus GmbH},
  langid = {english}
}

@article{kumar_largeeddy_2006,
  title = {Large-Eddy Simulation of a Diurnal Cycle of the Atmospheric Boundary Layer: {{Atmospheric}} Stability and Scaling Issues},
  shorttitle = {Large-Eddy Simulation of a Diurnal Cycle of the Atmospheric Boundary Layer},
  author = {Kumar, Vijayant and Kleissl, Jan and Meneveau, Charles and Parlange, Marc B.},
  year = 2006,
  journal = {Water Resources Research},
  volume = {42},
  number = {6},
  langid = {english}
}

@article{kustas_wind_1986,
  title = {Wind Profile Constants in a Neutral Atmospheric Boundary Layer over Complex Terrain},
  author = {Kustas, William P. and Brutsaert, Wilfried},
  year = 1986,
  month = jan,
  journal = {Boundary-Layer Meteorology},
  volume = {34},
  number = {1},
  pages = {35--54},
  langid = {english}
}

@article{lele_compact_1992,
  title = {Compact Finite Difference Schemes with Spectral-like Resolution},
  author = {Lele, Sanjiva K.},
  year = 1992,
  month = nov,
  journal = {Journal of Computational Physics},
  volume = {103},
  number = {1},
  pages = {16--42}
}

@article{liew_unified_2024,
  title = {Unified Momentum Model for Rotor Aerodynamics across Operating Regimes},
  author = {Liew, Jaime and Heck, Kirby S. and Howland, Michael F.},
  year = 2024,
  month = aug,
  journal = {Nature Communications},
  volume = {15},
  number = {1},
  pages = {1--12},
  publisher = {Nature Publishing Group},
  copyright = {2024 The Author(s)},
  langid = {english}
}

@article{liu_geostrophic_2021,
  title = {Geostrophic Drag Law for Conventionally Neutral Atmospheric Boundary Layers Revisited},
  author = {Liu, Luoqin and Gadde, Srinidhi N. and Stevens, Richard J.A.M.},
  year = 2021,
  journal = {Quarterly Journal of the Royal Meteorological Society},
  volume = {147},
  number = {735},
  pages = {847--857},
  langid = {english}
}

@incollection{lundquist_wind_2022,
  title = {Wind {{Shear}} and {{Wind Veer Effects}} on {{Wind Turbines}}},
  booktitle = {Handbook of {{Wind Energy Aerodynamics}}},
  author = {Lundquist, Julie K.},
  year = 2022,
  pages = {859--880},
  publisher = {Springer, Cham},
  isbn = {978-3-030-31307-4},
  langid = {english}
}

@article{ma_exploring_2026,
  title = {Exploring the Impact of Wind Veer on the Aerodynamic Performance and Wake Evolution of a Wind Turbine},
  author = {Ma, Jiachen and Feng, Chengdong and Quan, Yong and Yao, Bo and Guo, Zhenshan},
  year = 2026,
  month = aug,
  journal = {Renewable Energy},
  volume = {269},
  pages = {125854},
  langid = {english}
}

@article{madsen_implementation_2020,
  title = {Implementation of the Blade Element Momentum Model on a Polar Grid and Its Aeroelastic Load Impact},
  author = {Madsen, Helge Aagaard and Larsen, Torben Juul and Pirrung, Georg Raimund and Li, Ang and Zahle, Frederik},
  year = 2020,
  month = jan,
  journal = {Wind Energy Science},
  volume = {5},
  number = {1},
  pages = {1--27},
  publisher = {Copernicus GmbH},
  langid = {english}
}

@article{mahrt_stably_2014,
  title = {Stably {{Stratified Atmospheric Boundary Layers}}},
  author = {Mahrt, L.},
  year = 2014,
  month = jan,
  journal = {Annual Review of Fluid Mechanics},
  volume = {46},
  number = {Volume 46, 2014},
  pages = {23--45},
  publisher = {Annual Reviews},
  langid = {english}
}

@article{mata_modeling_2024,
  title = {Modeling the Effect of Wind Speed and Direction Shear on Utility-scale Wind Turbine Power Production},
  author = {Mata, Storm A. and Pena Mart{\'i}nez, Juan Jos{\'e} and Bas Quesada, Jes{\'u}s and Palou Larra{\~n}aga, Felipe and Yadav, Neeraj and Chawla, Jasvipul S. and Sivaram, Varun and Howland, Michael F.},
  year = 2024,
  month = jun,
  journal = {Wind Energy},
  pages = {we.2917},
  langid = {english}
}

@article{meyers_flow_2013,
  title = {Flow Visualization Using Momentum and Energy Transport Tubes and Applications to Turbulent Flow in Wind Farms},
  author = {Meyers, Johan and Meneveau, Charles},
  year = 2013,
  month = jan,
  journal = {Journal of Fluid Mechanics},
  volume = {715},
  pages = {335--358},
  publisher = {Cambridge University Press},
  langid = {english}
}

@article{munoz-esparza_bridging_2014,
  title = {Bridging the {{Transition}} from {{Mesoscale}} to {{Microscale Turbulence}} in {{Numerical Weather Prediction Models}}},
  author = {{Mu{\~n}oz-Esparza}, Domingo and Kosovi{\'c}, Branko and Mirocha, Jeff and {van Beeck}, Jeroen},
  year = 2014,
  month = dec,
  journal = {Boundary-Layer Meteorology},
  volume = {153},
  number = {3},
  pages = {409--440},
  langid = {english}
}

@article{murphy_how_2020,
  title = {How Wind Speed Shear and Directional Veer Affect the Power Production of a Megawatt-Scale Operational Wind Turbine},
  author = {Murphy, Patrick and Lundquist, Julie K. and Fleming, Paul},
  year = 2020,
  month = sep,
  journal = {Wind Energy Science},
  volume = {5},
  number = {3},
  pages = {1169--1190},
  publisher = {Copernicus GmbH},
  langid = {english}
}

@techreport{musial_offshore_2022,
  title = {Offshore {{Wind Market Report}}: 2022 {{Edition}}},
  author = {Musial, Walter and Spitsen, Paul and Duffy, Patrick and Beiter, Philipp and Marquis, Melinda and Hammond, Rob and Shields, Matt},
  year = 2022,
  number = {DOE/GO-102022-5765},
  pages = {126},
  address = {United States},
  institution = {{EERE Publication and Product Library}},
  langid = {english}
}

@article{nicoud_using_2011,
  title = {Using Singular Values to Build a Subgrid-Scale Model for Large Eddy Simulations},
  author = {Nicoud, Franck and Toda, Hubert Baya and Cabrit, Olivier and Bose, Sanjeeb and Lee, Jungil},
  year = 2011,
  month = aug,
  journal = {Physics of Fluids},
  volume = {23},
  number = {8},
  pages = {085106}
}

@article{nordstrom_fringe_1999,
  title = {The {{Fringe Region Technique}} and the {{Fourier Method Used}} in the {{Direct Numerical Simulation}} of {{Spatially Evolving Viscous Flows}}},
  author = {Nordstr{\"o}m, Jan and Nordin, Niklas and Henningson, Dan},
  year = 1999,
  month = jan,
  journal = {SIAM Journal on Scientific Computing},
  volume = {20},
  number = {4},
  pages = {1365--1393},
  langid = {english}
}

@article{parinam_exploring_2024,
  title = {Exploring the Impact of Different Inflow Conditions on Wind Turbine Wakes Using {{Large-Eddy Simulations}}},
  author = {Parinam, Anand and Benard, Pierre and Von Terzi, Dominic and Vir{\'e}, Axelle},
  year = 2024,
  month = jun,
  journal = {Journal of Physics: Conference Series},
  volume = {2767},
  number = {9},
  pages = {092098},
  publisher = {IOP Publishing},
  langid = {english}
}

@article{revaz_effect_2025,
  title = {Effect of {{Turbulence Intensity}} on the {{Induction Factor}} and {{Power Efficiency}} of {{Wind Turbines}}},
  author = {Revaz, Tristan and {Port{\'e}-Agel}, Fernando},
  year = 2025,
  journal = {Wind Energy},
  volume = {28},
  number = {11},
  pages = {e70040},
  langid = {english}
}

@inproceedings{rossby_layer_1935,
  title = {The Layer of Frictional Influence in Wind and Ocean Currents},
  booktitle = {Papers in {{Physical Oceanography}} and {{Meterology}}},
  author = {Rossby, Carl-Gustaf and Montgomery, Raymond B.},
  year = 1935,
  volume = {3},
  pages = {1--101},
  publisher = {MIT Press},
  address = {Cambridge, MA},
  langid = {english}
}

@article{sanchezgomez_effect_2020,
  title = {The Effect of Wind Direction Shear on Turbine Performance in a Wind Farm in Central {{Iowa}}},
  author = {Sanchez Gomez, Miguel and Lundquist, Julie K.},
  year = 2020,
  month = jan,
  journal = {Wind Energy Science},
  volume = {5},
  number = {1},
  pages = {125--139},
  publisher = {Copernicus GmbH},
  langid = {english}
}

@article{sescu_control_2014,
  title = {A Control Algorithm for Statistically Stationary Large-Eddy Simulations of Thermally Stratified Boundary Layers: {{A Control Algorithm}} for {{LES}} of {{Thermally Stratified Boundary Layers}}},
  shorttitle = {A Control Algorithm for Statistically Stationary Large-Eddy Simulations of Thermally Stratified Boundary Layers},
  author = {Sescu, Adrian and Meneveau, Charles},
  year = 2014,
  month = jul,
  journal = {Quarterly Journal of the Royal Meteorological Society},
  volume = {140},
  number = {683},
  pages = {2017--2022},
  langid = {english}
}

@article{shapiro_filtered_2019,
  title = {Filtered Actuator Disks: {{Theory}} and Application to Wind Turbine Models in Large Eddy Simulation},
  shorttitle = {Filtered Actuator Disks},
  author = {Shapiro, Carl R. and Gayme, Dennice F. and Meneveau, Charles},
  year = 2019,
  journal = {Wind Energy},
  volume = {22},
  number = {10},
  pages = {1414--1420},
  langid = {english}
}

@article{shapiro_modelling_2018,
  title = {Modelling Yawed Wind Turbine Wakes: A Lifting Line Approach},
  shorttitle = {Modelling Yawed Wind Turbine Wakes},
  author = {Shapiro, Carl R. and Gayme, Dennice F. and Meneveau, Charles},
  year = 2018,
  month = apr,
  journal = {Journal of Fluid Mechanics},
  volume = {841},
  pages = {R1},
  copyright = {https://www.cambridge.org/core/terms},
  langid = {english}
}

@article{shen_global_2024,
  title = {The Global Properties of Nocturnal Stable Atmospheric Boundary Layers},
  author = {Shen, Zhouxing and Liu, Luoqin and Lu, Xiyun and Stevens, Richard J. A. M.},
  year = 2024,
  month = nov,
  journal = {Journal of Fluid Mechanics},
  volume = {999},
  pages = {A60},
  langid = {english}
}

@article{shin_addressing_2025,
  title = {Addressing {{Grid Convergence}} and {{Log-Layer Mismatch}} in {{Wall Modeled Large Eddy Simulations}} of {{Geophysical Flows Over Rough Surfaces}} and {{Canopies}}},
  author = {Shin, E. Y. and Yang, X. I. A. and Howland, M. F.},
  year = 2025,
  month = aug,
  journal = {Boundary-Layer Meteorology},
  volume = {191},
  number = {9},
  pages = {42},
  langid = {english}
}

@article{st.martin_wind_2016,
  title = {Wind Turbine Power Production and Annual Energy Production Depend on Atmospheric Stability and Turbulence},
  author = {St. Martin, Clara M. and Lundquist, Julie K. and Clifton, Andrew and Poulos, Gregory S. and Schreck, Scott J.},
  year = 2016,
  month = nov,
  journal = {Wind Energy Science},
  volume = {1},
  number = {2},
  pages = {221--236},
  publisher = {Copernicus GmbH},
  langid = {english}
}

@article{steiros_drag_2018,
  title = {Drag on Flat Plates of Arbitrary Porosity},
  author = {Steiros, K. and Hultmark, M.},
  year = 2018,
  month = oct,
  journal = {Journal of Fluid Mechanics},
  volume = {853},
  pages = {R3},
  langid = {english}
}

@article{stevens_concurrent_2014,
  title = {A Concurrent Precursor Inflow Method for {{Large Eddy Simulations}} and Applications to Finite Length Wind Farms},
  author = {Stevens, Richard J.A.M. and Graham, Jason and Meneveau, Charles},
  year = 2014,
  month = aug,
  journal = {Renewable Energy},
  volume = {68},
  pages = {46--50},
  langid = {english}
}

@book{stull_introduction_1988,
  title = {An {{Introduction}} to {{Boundary Layer Meteorology}}},
  editor = {Stull, Roland B.},
  year = 1988,
  publisher = {Springer Netherlands},
  address = {Dordrecht},
  isbn = {978-90-277-2769-5 978-94-009-3027-8}
}

@article{tamaro_power_2024,
  title = {On the Power and Control of a Misaligned Rotor -- beyond the Cosine Law},
  author = {Tamaro, Simone and Campagnolo, Filippo and Bottasso, Carlo L.},
  year = 2024,
  month = jul,
  journal = {Wind Energy Science},
  volume = {9},
  number = {7},
  pages = {1547--1575},
  publisher = {Copernicus GmbH},
  langid = {english}
}

@article{troldborg_actuator_2010,
  title = {Actuator {{Disc Simulations}} of {{Influence}} of {{Wind Shear}} on {{Power Production}} of {{Wind Turbines}}},
  author = {Troldborg, Niels and Gaunaa, Mac and Mikkelsen, Robert},
  year = 2010,
  journal = {Journal of Physics: Conference Series},
  pages = {10},
  langid = {english}
}

@article{tumenbayar_effect_2023,
  title = {An {{Effect}} of {{Wind Veer}} on {{Wind Turbine Performance}}},
  author = {Tumenbayar, Undarmaa and Ko, Kyungnam},
  year = 2023,
  month = jan,
  journal = {International Journal of Renewable Energy Development},
  volume = {12},
  number = {1},
  pages = {111--117},
  langid = {english}
}

@article{turk_dependence_2010,
  title = {The Dependence of Offshore Turbulence Intensity on Wind Speed},
  author = {T{\"u}rk, Matthias and Emeis, Stefan},
  year = 2010,
  month = aug,
  journal = {Journal of Wind Engineering and Industrial Aerodynamics},
  volume = {98},
  number = {8},
  pages = {466--471}
}

@article{vanderwende_modification_2012,
  title = {The Modification of Wind Turbine Performance by Statistically Distinct Atmospheric Regimes},
  author = {Vanderwende, B J and Lundquist, J K},
  year = 2012,
  month = sep,
  journal = {Environmental Research Letters},
  volume = {7},
  number = {3},
  pages = {034035},
  publisher = {IOP Publishing},
  langid = {english}
}

@article{vansark_we_2019,
  title = {Do We Really Need Rotor Equivalent Wind Speed?},
  author = {Van Sark, Wilfried G.J.H.M. and Van Der Velde, Henrik C. and Coelingh, Jan P. and Bierbooms, Wim A.A.M.},
  year = 2019,
  month = jun,
  journal = {Wind Energy},
  volume = {22},
  number = {6},
  pages = {745--763},
  langid = {english}
}

@article{vratsinis_impact_2026,
  title = {Impact of Inflow Conditions and Turbine Placement on the Performance of Offshore Wind Turbines Exceeding 7\&thinsp;{{MW}}},
  author = {Vratsinis, Konstantinos and Marini, Rebeca and Daems, Pieter-Jan and Pauscher, Lukas and {van Beeck}, Jeroen and Helsen, Jan},
  year = 2026,
  month = may,
  journal = {Wind Energy Science},
  volume = {11},
  number = {5},
  pages = {1803--1820},
  publisher = {Copernicus GmbH},
  langid = {english}
}

@article{wagner_accounting_2011,
  title = {Accounting for the Speed Shear in Wind Turbine Power Performance Measurement},
  author = {Wagner, R. and Courtney, M. and Gottschall, J. and {Lindel{\"o}w-Marsden}, P.},
  year = 2011,
  journal = {Wind Energy},
  volume = {14},
  number = {8},
  pages = {993--1004},
  copyright = {Copyright \copyright{} 2011 John Wiley \& Sons, Ltd.},
  langid = {english}
}

@article{wagner_influence_2009,
  title = {The Influence of the Wind Speed Profile on Wind Turbine Performance Measurements},
  author = {Wagner, Rozenn and Antoniou, Ioannis and Pedersen, S{\o}ren M. and Courtney, Michael S. and J{\o}rgensen, Hans E.},
  year = 2009,
  journal = {Wind Energy},
  volume = {12},
  number = {4},
  pages = {348--362},
  langid = {english}
}

@techreport{wagner_simulation_2010,
  title = {Simulation of Shear and Turbulence Impact on Wind Turbine Performance},
  author = {Wagner, Rozenn and Courtney, M and Larsen, T. J.},
  year = 2010,
  number = {1722},
  address = {Ris\o{} Nationallaboratoriet for B\ae redygtig Energi. Denmark.},
  institution = {Danmarks Tekniske Universitet},
  langid = {english}
}

@article{wharton_assessing_2012,
  title = {Assessing Atmospheric Stability and Its Impacts on Rotor-Disk Wind Characteristics at an Onshore Wind Farm},
  author = {Wharton, Sonia and Lundquist, Julie K.},
  year = 2012,
  journal = {Wind Energy},
  volume = {15},
  number = {4},
  pages = {525--546},
  copyright = {Copyright \copyright{} 2011 John Wiley \& Sons, Ltd.},
  langid = {english}
}

@article{wharton_atmospheric_2012,
  title = {Atmospheric Stability Affects Wind Turbine Power Collection},
  author = {Wharton, Sonia and Lundquist, Julie K},
  year = 2012,
  month = jan,
  journal = {Environmental Research Letters},
  volume = {7},
  number = {1},
  pages = {014005},
  publisher = {IOP Publishing},
  langid = {english}
}

@techreport{wilson_applied_1974,
  title = {Applied Aerodynamics of Wind Power Machines},
  author = {Wilson, R. E. and Lissaman, P. B. S.},
  year = 1974,
  month = jul,
  number = {PB-238595},
  institution = {Oregon State Univ., Corvallis (USA)},
  langid = {english}
}

@article{zilitinkevich_further_2007,
  title = {Further Comments on the Equilibrium Height of Neutral and Stable Planetary Boundary Layers},
  author = {Zilitinkevich, Sergej and Esau, Igor and Baklanov, Alexander},
  year = 2007,
  journal = {Quarterly Journal of the Royal Meteorological Society},
  volume = {133},
  number = {622},
  pages = {265--271},
  copyright = {Copyright \copyright{} 2007 Royal Meteorological Society},
  langid = {english}
}

\end{document}